\documentclass[twocolumn,useAMS,usenatbib]{mn2e}
\usepackage{graphicx}
\usepackage{subfigure}
\usepackage{xcolor}
\usepackage{mathtext,bbm,amsmath,amsfonts,amssymb,indentfirst,syntonly,graphicx}
\usepackage{slashbox}
\usepackage[english]{babel}
\usepackage{calc}
\usepackage{tikz}
\usepackage{bm}
\usepackage{epstopdf}
\usepackage{multicol}

\def\bc{\begin{center}}
\def\ec{\end{center}}
\def\be{\begin{eqnarray}}
\def\ee{\end{eqnarray}}

\title[The Dark Matter Profiles in the Milky Way]{The Dark Matter Profiles in the Milky Way}
\author[H.-N. Lin and X. Li]
        {Hai-Nan Lin\thanks{E-mail: linhn@cqu.edu.cn} and
        Xin Li\thanks{E-mail: lixin1981@cqu.edu.cn}\\
Department of Physics, Chongqing University, Chongqing 401331, China\\}
\begin{document}

\date{Accepted xxxx; Received xxxx; in original form xxxx}

\pagerange{\pageref{firstpage}--\pageref{lastpage}} \pubyear{2019}

\maketitle

\label{firstpage}

\begin{abstract}
  We investigate the dark matter profile of the Milky Way using the observed rotation curve data out to 100 kpc. The baryonic matter of the Milky Way is divided into bulge, disk and gas components, and each component is modelled using various possible mass profiles available in literature. The arbitrary combination of seven bulge profiles, four disk profiles and two gas profiles results in fifty-six baryon models. These baryon models are combined with one of the four dark matter profiles: Burkert profile, core-modified profile, pseudo-isothermal profile and NFW profile, to fit the observed rotation curve data. Results show that in general the NFW profile fits the data better than the Burkert profile, while the core-modified profile and the pseudo-isothermal profile are essentially ruled out. The best-fitting NFW model has the scale length $r_0=8.1\pm 0.7$ kpc, and the corresponding local density of dark matter is $\rho_{\rm dm}(R=R_\odot)=0.51\pm0.09~{\rm GeV}/{\rm cm}^3$.
\end{abstract}

\begin{keywords}
Galaxy: kinematics and dynamics -- dark matter
\end{keywords}

\section{Introduction}

In the early 1970s, it was noticed that the luminous mass in some galaxies cannot provides enough gravitational potential to support the observed rotation velocity in the outer of galaxies \citep{Freeman:1970,Whitehurst:1972,Rogstad:1972,Roberts:1973}. This mass missing  problem leads to the hypothesis that there is a large amount of non-luminous matter inside the galaxies which has not been seen yet, i.e. the dark matter hypothesis. Later on, the mass missing problem has been discovered in many more galaxies \citep{Rubin:1978,de Blok:2002,Walter:2008wy,Lelli:2016zqa}. Nowadays it is widely believed that a considerable amount of mass is in the form of dark matter in galaxies or even in the whole Universe. Although the mass missing problem can also been solved in part by modifying the Newtonian dynamics \citep{Milgrom:1983,Begeman:1991,Sanders:2007,Swaters:2010} or modifying the Newtonian gravity \citep{Brownstein:2006zz,Cardone:2010,Lin:2013,Chang:2013}, the dark matter scenario is still the most successful one in accounting for e.g. the formation of galaxies and the cosmic microwave background radiation.

Determining the dark matter distribution in galaxies is of great importance because it affects the dynamical evolution of the galaxies as well as of the Universe. There are various dark matter profiles proposed in literature. For example, \citet{Navarro:1996} proposed the so-called NFW model from the N-body simulations in the standard cold dark matter cosmology. \citet{Burkert:1995} proposed a phenomenological profile (the Burkert profile) to explain the observed rotation curves of dwarf spiral galaxies. \citet{Jimenez:2003} showed that a large sample of galaxy rotation curves can be well fitted by the pseudo-isothermal profile. \citet{Brownstein:2009} showed that the core-modified profile with a constant central density fits excellently well to the rotation curves of both high- and low-surface brightness galaxies. All of these dark matter profiles have two free parameters in each, i.e. the characteristic density $\rho_0$ and the scale length $r_0$. Other dark matter profiles with one or more additional parameters include e.g. the Einasto profile \citep{Navarro:2004,Merritt:2006}, the generalized profile \citep{Zhao:1996a,An:2013}, and so on.

As a typical spiral galaxy, the rotation curve of the Milky Way provides an excellent tool to trace the dark matter distribution because this is the very galaxy which we live in. Thanks to our location in the Milky Ways, it is one of the galaxy whose rotation curve has been measured with high accuracy to far distance. The measurements on the rotation curve of the Milky begin decades ago and is still in progress today, see e.g. \citet{Sofue:2017nhd} for recent review. The observed rotation velocity in the outer galaxy shows obvious deviation from the predicted $r^{-1/2}$ law, which implies the existence of dark matter \citep{Blitz:1979,Clemens:1985,Dehnen:1998}. Later on, the rotation curve has been constructed out to 100 kpc or more and further proves the existence of dark matter \citep{Sofue:2013,Bhattacharjee:2014,Sofue:2015,Huang:2016}. Recently, \citet{Iocco:2015xga} collected a large sample of rotation velocity measurements of the Milky Ways from different dynamical tracers out to 30 kpc, and showed that the observed rotation velocity obviously exceeds the contribution from the baryon mass, thus concluding that dark matter is required even in the inner Milky Way.

In this paper, we investigate the dark matter distribution of the Milky Way using the recent rotation curve data. To this end, the precious measurement of the mass distribution of the Milky way is required. However, despite the great progress on observational technique in recent decades, the exact distribution of baryon mass in the Milky is hard to determine even today. An unambiguous fact is that the Milky Way mainly consists of a stellar disk and the dispersive gas, with an additional bulge in the Galactic center. But how to model the mass distribution of each component is widely debated in literature. To include the maximum possibility, we follow \citet{Iocco:2015xga} and consider various possible mass profiles of each component. Specifically speaking, we consider seven bulge profiles, four disk profiles and two gas profiles, which are described in detail in the next section. The arbitrary combination of the bulge, disk and gas profiles results in fifty-six mass models of the Milky Way. Each mass model is combined to one of the dark matter profiles to fit the rotation curve data. Here we mainly focus on the two-parameter dark matter profiles, i.e. the Burkert profile \citep{Burkert:1995}, the core-modified profile \citep{Brownstein:2009}, the pseudo-isothermal profile \citep{Jimenez:2003}, and the NFW profile \citep{Navarro:1996,Navarro:1997}.

The rest of this paper is organized as follows: In section \ref{sec:mass}, we introduce the mass profiles of the Milky Way and calculate their contribution to the rotation curve. The rotation curve data and the best-fitting results are presented in section \ref{sec:results}. Finally, we conclude our paper in section \ref{sec:conclusions}.

\section{Mass profiles of the Milky Way}\label{sec:mass}

To calculate the theoretical prediction of rotation velocity, the mass distribution of the Milky Way should be known. We divide the mass of the Milky Way into baryonic and dark matter. The former is further divided into three components, i.e., a triaxial bulge in the innermost region, an axis-symmetric stellar disk, and the widely spread gas. The exact distribution of the baryonic matter in the Milky is not clearly known even today. To include the maximum possibility, we follow \citet{Iocco:2015xga} and consider various possible profiles of each component. In detail, we consider seven bulge profiles, four disk profiles and two gas profiles, which are summarized in Table \ref{tab:baryon}. The arbitrary combination of bulge, disk and gas profiles gives in total fifty-six possible baryon models. The details of the bulge, disk and gas profiles are described bellow.

\begin{table}
  \centering
  \caption{The baryon models used in this paper.}\label{tab:baryon}
  \begin{tabular}{lll}
  \hline\hline
  component & label & reference\\
  \hline
  Bulge & B1 & \citet{Stanek:1997} [E2 model]\\
  & B2 & \citet{Stanek:1997} [G2 model]\\
  & B3 & \citet{Zhao:1996b}\\
  & B4 & \citet{Bissantz:2002}\\
  & B5 & \citet{Lopez-Corredoira:2007}\\
  & B6 & \citet{Vanhollebeke:2009}\\
  & B7 & \citet{Robin:2012}\\
  \hline
  Disk & D1 & \citet{Han:2003}\\
  & D2 & \citet{Calchi:2011}\\
  & D3 & \citet{Juric:2008}\\
  & D4 & \citet{Bovy:2013}\\
  \hline
  Gas & G1 & \citet{Ferriere:1998}\\
  & G2 & \citet{Moskalenko:2002}\\
  \hline
  \end{tabular}
\end{table}

For the bulge component, we consider the following seven models: (B1) the second kind exponential profile discussed in \citet{Stanek:1997}; (B2) the second kind gaussian profile discussed in \citet{Stanek:1997}; (B3) the models with a gaussian sharped bar plus an oblate spheroidal nucleus with a steep inner power law and an exponential outer profile given in \citet{Zhao:1996b}; (B4) the truncated power-law bulge given in \citet{Bissantz:2002}; (B5) a bulge with an extra long bar in the Galaxy plane given in \citet{Lopez-Corredoira:2007}; (B6) the truncated power-law bulge given in \citet{Vanhollebeke:2009}; (B7) the double-ellipsoid bulge given in \citet{Robin:2012}.

For the stellar disk component, we consider the following four models: (D1) the thin plus thick exponential disk \citep{Han:2003}; (D2) the standard double exponential disk \citep{Calchi:2011}; (D3) the thin plus thick disk with an extra halo component \citep{Juric:2008}; (D4) the single maximal disk \citep{Bovy:2013}. We rescale the profiles to a fixed sun to Galactic center distance of $R_{\odot}=8.0$ kpc. Note that \citet{Iocco:2015xga} considered a fifth disk profile from \citet{DeJong:2010}. However, the profile of \citet{DeJong:2010} has a negligible difference from the profile of \citet{Calchi:2011} in terms of the predicted rotation velocity, hence we just consider the latter profile in our paper.

The gas in the Milky Way is widely spread and is extremely irregular. We follow \citet{Iocco:2015xga} and separately model the gas in three different regions. In the center ($R<10$ pc) of the Galaxy, the gas is modeled by a point-like mass. In the inner ($R<2$ kpc) region, as is described in detail in \citet{Ferriere:2007yq}, we divide the gas into molecular hydrogen, atomic hydrogen and ionised hydrogen, and model these three components separately. Beyond $R\approx 2$ kpc, two different morphologies are used. In the first model (which we label it as G1), the gas is split into five forms (the molecular gas, the
cold neutral medium, the warm neutral medium, the warm ionized medium, and the hot ionized medium) and each form is modelled separately \citep{Ferriere:1998}. In the second model (which we label it as G2), the gas is divided, as is similar to the inner region, into molecular hydrogen, atomic hydrogen and ionised hydrogen, and each component is modeled separately \citep{Moskalenko:2002}.

For the dark matter models, here we consider four different profiles: (1) the Navarro-Frenk-White (NFW) profile obtained from collisionless N-body numerical simulations \citep{Navarro:1996,Navarro:1997}, (2) the pseudo-isothermal profile which behaves similar to the isothermal spherical profile in the outer region but with a constant density in the inner core \citep{Jimenez:2003}, (3) the Burkert profile which is first introduced to fit the rotation curves of dwarf galaxies \citep{Burkert:1995}, and (4) the core-modified profile with a constant density in the core \citep{Brownstein:2009}. The dark matter profiles and their contributions to the rotation velocity are summarized in Table \ref{tab:dark}.

\begin{table*}
  \centering
  \caption{The dark matter models and their contributions to the rotation velocity. Note: $v_h\equiv(4\pi \rho_0r_0^2G)^{1/2}$.}\label{tab:dark}
  \begin{tabular}{llll}
  \hline\hline
  model & mass profile & rotation velocity & reference\\
  \hline
  Burkert & $\rho_{\rm bur}(r)=\frac{\rho_0r_0^3}{(r+r_0)(r^2+r_0^2)}$ & $v_{\rm bur}(r)=v_h\sqrt{\frac{r_0}{4r} \left[\ln\left(\frac{(r+r_0)^2(r^2+r_0^2)}{r_0^4}\right) - 2\arctan\left(\frac{r}{r_0}\right)\right]}$ & \citet{Burkert:1995}\\
  core-modified & $\rho_{\rm com}(r)=\frac{\rho_0r_0^3}{r^3+r_0^3}$ & $v_{\rm com}(r)=v_h\sqrt{\frac{r_0}{3r}\ln\left(\frac{r^3+r_0^3}{r_0^3}\right)}$ & \citet{Brownstein:2009}\\
  pseudo-isothermal & $\rho_{\rm iso}(r)=\frac{\rho_0}{1+\left(\frac{r}{r_0}\right)^2}$ & $v_{\rm iso}(r)=v_h\sqrt{1-\frac{r_0}{r}\arctan\left(\frac{r}{r_0}\right)}$ & \citet{Jimenez:2003}\\
  NFW & $\rho_{\rm nfw}(r)=\frac{\rho_0}{\frac{r}{r_0}\left(1+\frac{r}{r_0}\right)^2}$ & $v_{\rm nfw}(r)=v_h\sqrt{\frac{r_0}{r} \ln\left(\frac{r+r_0}{r_0}\right)- \frac{r_0}{r+r_0}}$ & \citet{Navarro:1996,Navarro:1997}\\
  \hline
  \end{tabular}
\end{table*}

The rotation velocity contributes from the total mass, according to the Newtonian gravity, is given by
\begin{equation}
  v(R)=\sqrt{v^2_{\rm baryon}(R)+v^2_{\rm dm}(R)},
\end{equation}
where
\begin{equation}
  v_{\rm baryon}(R)=\sqrt{v^2_{\rm bulge}(R)+v^2_{\rm disk}(R)+v^2_{\rm gas}(R)}
\end{equation}
is the contribution from the baryon mass, and $v_{\rm dm}(R)$ is the dark matter contribution. The best-fitting parameters are obtained by minimizing the $\chi^2$,
\begin{equation}
  \chi^2=\sum_i\frac{[v(R_i)-v_i]^2}{\sigma^2_{v_i}},
\end{equation}
where $v(R_i)$ is the theoretical velocity at $R_i$, $v_i$ and $\sigma_{v_i}$ are the observed rotation velocity and its uncertainty, respectively.

We numerically calculate the rotation velocities contributing from bulge, disk and gas. Since the bulge profiles are non-axis-symmetrical, we average the bulge velocity over the azimuth angle in the Galactic plane. The rotation velocity of dark matter can be calculated analytically, as is summarized in Table \ref{tab:dark}. In each baryon+dark matter model, the only two free parameters are the characteristic density $\rho_0$ and the scale length $r_0$ of the dark matter profile. For the sake of convenience, we use the characteristic velocity $v_h\equiv(4\pi \rho_0r_0^2G)^{1/2}$ instead of $\rho_0$ in the fitting.

\section{Data and Results}\label{sec:results}

\begin{figure}
  \centering
  \includegraphics[width=0.48\textwidth]{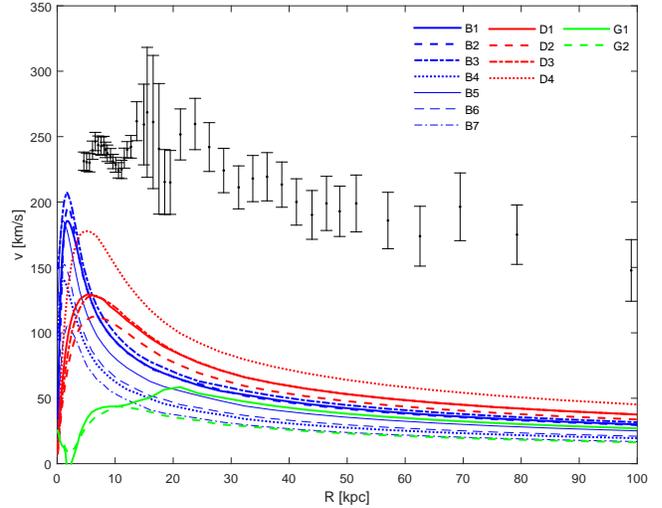}\\
  \caption{The rotation curve data used in our paper. The error bars represent the $1\sigma$ uncertainty. The colored curves represent the contributions from baryon components described in the texts.}\label{fig:data}
\end{figure}

The data used in our analysis is taken from \citet{Huang:2016}. This data set contains in total 43 measurements of Milky Way rotation velocity out to 100 kpc, among which 8 data points are from HI tracer, 12 data points are from primary red clump giant tracer, and 23 data points are from the halo K giant tracer. The data are plotted in Figure \ref{fig:data}. The error bars represent the $1\sigma$ uncertainty. The contributions from baryon components are also shown.

\begin{figure}
  \centering
  \includegraphics[width=0.48\textwidth]{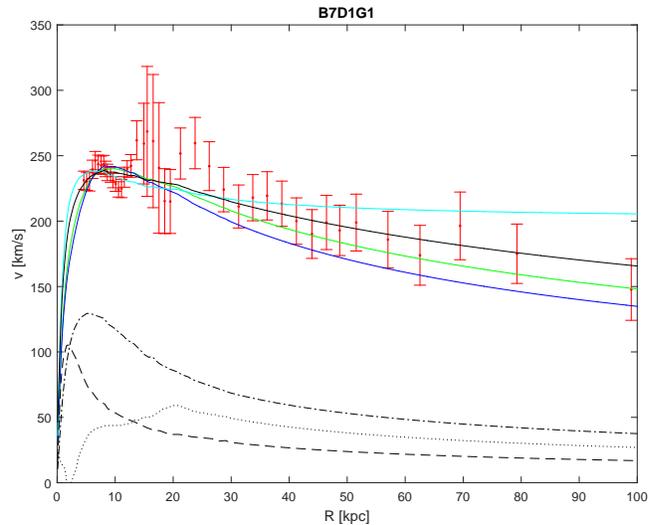}\\
  \caption{An example of the best-fitting results. The baryon profile is taken to be the B7D1G1 model. Green curves for Burkert model; blue curves for core-modified model; cyan curves for pseudo isothermal model; black curves for NFW model. The contributions from bulge (dashed curves), disk (dash-dotted curves) and gas (dotted curves) are also shown.}\label{fig:B7D1G1}
\end{figure}

\begin{figure}
  \centering
  \includegraphics[width=0.48\textwidth]{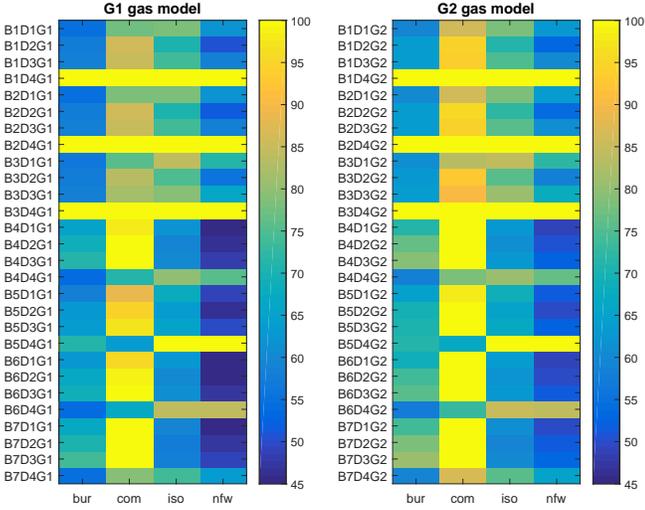}\\
  \caption{The pseudo color map of $\chi^2_{\rm min}$. The smaller (larger) the $\chi^2_{\rm min}$, the bluer (yellower) of the map. Left panel: based on the G1 gas model; right panel: based on the G2 gas model.}\label{fig:colormap}
\end{figure}

The arbitrary combination of 56 baryon models and 4 dark matter models results in 224 mass models of the Milky Way, with two free parameters in each model. The best-fitting parameters and the best-fitting curves of all the models are presented in the {\it supplementary materials}. Figure \ref{fig:B7D1G1} shows an example of the best-fitting results. In Table \ref{tab:chi2}, we list the $\chi^2_{\rm min}$ value of each model. To be more intuitive, in Figure \ref{fig:colormap} we plot the pseudo color map of $\chi^2_{\rm min}$. In this figure, the left panel is based on the G1 gas model, and the right panel is based on the G2 gas model. The vertical axis represents the baryon model and the horizon axis represents the dark matter model. The smaller (larger) the $\chi^2_{\rm min}$, the bluer (yellower) of the map. From Figure \ref{fig:colormap}, we see that the core-modified profile is not a good model. In general, the NFW profile fits the data better than the other dark matter profiles, except in the D4 disk model case. If the disk profile is modelled as D4, most of the mass models couldn't fit the data well. The similarity between the left and right panels of Figure \ref{fig:colormap} implies that the gas model does not significantly affect the results.

\begin{table*} 
\centering 
\caption{The $\chi^2_{\rm min}$ value of each model.}\label{tab:chi2} 
\begin{tabular}{llllp{40pt}lllll} 
\hline\hline 
models&bur&com&iso&nfw&models&bur&com&iso&nfw\\
\hline 
B1D1G1&54.7&77.4&77.7&61.5&B1D1G2&59.4&85.6&78.3&63.0\\
B1D2G1&57.3&85.7&70.4&51.0&B1D2G2&63.1&94.8&71.5&53.5\\
B1D3G1&58.0&84.8&73.8&58.3&B1D3G2&63.4&93.6&75.1&60.4\\
B1D4G1&182.4&139.2&233.1&319.0&B1D4G2&169.1&129.8&218.5&300.5\\
B2D1G1&55.0&77.9&77.8&61.6&B2D1G2&59.7&86.0&78.4&63.1\\
B2D2G1&57.7&86.2&70.7&51.2&B2D2G2&63.5&95.3&71.7&53.8\\
B2D3G1&58.5&85.3&74.1&58.5&B2D3G2&63.8&94.2&75.3&60.6\\
B2D4G1&177.9&135.5&228.5&312.9&B2D4G2&165.4&126.8&214.7&295.3\\
B3D1G1&56.4&75.2&84.0&71.1&B3D1G2&60.6&82.8&84.4&72.1\\
B3D2G1&57.3&83.6&74.5&55.9&B3D2G2&62.7&92.5&75.4&58.2\\
B3D3G1&58.5&81.7&79.2&66.1&B3D3G2&63.4&90.1&80.2&67.8\\
B3D4G1&482.0&408.4&533.2&694.5&B3D4G2&438.1&369.4&488.8&641.9\\
B4D1G1&64.8&98.1&61.3&45.6&B4D1G2&71.6&108.1&62.9&49.3\\
B4D2G1&68.8&101.6&59.1&46.4&B4D2G2&76.1&111.8&60.7&50.6\\
B4D3G1&71.3&106.8&60.2&48.3&B4D3G2&78.7&117.3&62.2&52.4\\
B4D4G1&54.2&71.2&80.2&75.9&B4D4G2&58.0&78.3&81.0&76.5\\
B5D1G1&58.1&88.2&67.4&49.0&B5D1G2&64.0&97.5&68.6&51.7\\
B5D2G1&62.5&94.5&63.5&46.0&B5D2G2&69.1&104.3&64.9&49.5\\
B5D3G1&63.5&96.7&65.3&49.3&B5D3G2&70.0&106.6&67.1&52.5\\
B5D4G1&71.4&63.3&113.0&137.3&B5D4G2&71.6&66.7&111.0&134.1\\
B6D1G1&62.4&95.0&62.4&45.4&B6D1G2&69.0&104.9&63.8&48.8\\
B6D2G1&66.6&99.1&59.9&45.5&B6D2G2&73.7&109.2&61.4&49.6\\
B6D3G1&68.7&103.7&61.0&47.6&B6D3G2&75.9&114.0&62.9&51.5\\
B6D4G1&54.3&66.5&84.4&84.3&B6D4G2&57.5&73.0&84.8&84.3\\
B7D1G1&66.7&100.2&59.3&45.4&B7D1G2&73.8&110.4&60.9&49.3\\
B7D2G1&70.3&102.7&57.5&46.9&B7D2G2&77.8&113.1&59.1&51.4\\
B7D3G1&73.4&108.8&58.3&48.6&B7D3G2&81.0&119.5&60.4&53.0\\
B7D4G1&54.5&78.7&73.7&63.9&B7D4G2&59.2&86.8&75.1&65.3\\
\hline 
\end{tabular} 
\end{table*}

\begin{figure}
  \centering
  \includegraphics[width=0.45\textwidth]{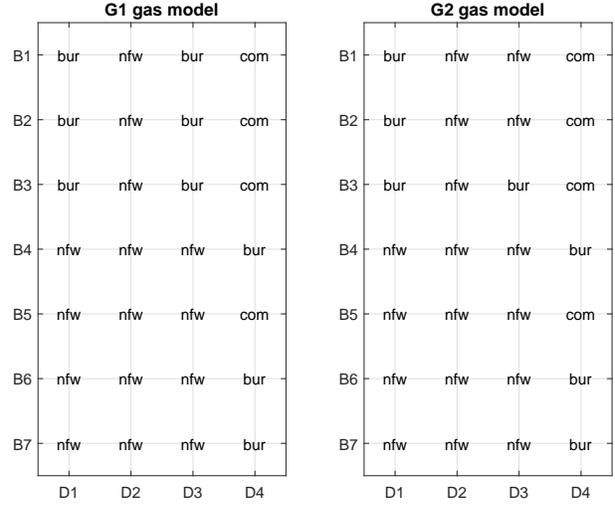}\\
  \caption{The dark matter model which best fits the data given a specific baryon model. Left panel: based on the G1 gas model; right panel: based on the G2 gas model.}\label{fig:bestmodel}
\end{figure}

In Figure \ref{fig:bestmodel}, we list the dark matter model which best fits the data (i.e. which has the minimal $\chi^2_{\rm min}$ value) given a specific baryon model. For example, given the B1D1G1 baryon model, the Burkert profile fits the data best, and so on. From this figure, we see that in 32 out of the 56 baryon models, the NFW profile fits the data best. The Burkert profile and the core-modified profile fit the data best in 16 and 8 baryon models, respectively. Interestingly, non of the baryon model combined with the pseudo-isothermal profile can fit the data best. If the D2 disk model is adopted, the NFW profile fits the data better than the rest three dark matter profiles, in regardless of the bulge and disk models. However, if the D4 disk model is adopted, the NFW profile never fits the data better than the core-modified profile and Burkert profile. The core-modified profile fits best only if the D4 disk model is adopted. However, as is seen from Table \ref{tab:chi2} and Figure \ref{fig:colormap}, within the core-modified profile the $\chi^2_{\rm min}$ values are usually to large to be acceptable. Comparing the left and right panels of Figure \ref{fig:bestmodel}, we can also see that the gas model does not significantly affect the results. In summary, the pseudo-isothermal and core-modified profiles are essentially not supported, and the NFW profile seems to be better than the Burkert profile.

The baryon+NFW models have the weighted average parameters (ignoring the D4 disk model case) $v_h=428.2~{\rm km}~{\rm s}^{-1}$, and $r_0=9.4$ kpc. Among all the mass models, the best one is the B7D1G1+NFW model, which has $\chi_{\rm min}^2=45.4$. The best-fitting parameters of this model are $v_h=430.4\pm5.4~{\rm km}~{\rm s}^{-1}$, and $r_0=8.1\pm 0.7$ kpc, which corresponds to the local density of dark matter $\rho_{\rm dm}(R=R_\odot)=0.013\pm0.002~M_{\odot}/{\rm pc}^3=0.51\pm0.09~{\rm GeV}/{\rm cm}^3$, and the dark matter mass interior the solar position $M_{\rm dm}(R<R_\odot)=(6.6\pm 1.2)\times 10^{10}~M_\odot$. If the dark matter is modeled by the Burkert profile, then the best model is the B4D4G1+Burkert model, in which the best fitting parameters are $v_h=414.0\pm8.5~{\rm km}~{\rm s}^{-1}$, and $r_0=7.8\pm 0.4$ kpc, and $\chi_{\rm min}=54.2$. This corresponding to the local density of dark matter $\rho_{\rm dm}(R=R_\odot)=0.012\pm0.001~M_{\odot}/{\rm pc}^3=0.48\pm0.05~{\rm GeV}/{\rm cm}^3$, and the dark matter mass interior the solar position $M_{\rm dm}(R<R_\odot)=(4.1\pm 0.5)\times 10^{10}~M_\odot$. Although the B4D4G1+Burkert model and B7D1G1+NFW model have a consistent local density of dark matter, the $\chi_{\rm min}$ value of the former is larger than that of the latter by $\sim 9$. On the other hand, there are in total 28 baryon profiles combined with the NFW model fit better than the B4D4G1+Burkert model, while non of the baryon profile combined with the core-modified model or pseudo-isothermal model fits better than the B4D4G1+Burkert model. Therefore, among the four dark matter profiles considered here, the NFW profile maybe the best one in mimicking the dark matter distribution in the Milky Way.

\section{Discussion and Conclusions}\label{sec:conclusions}

In this paper, we have investigated the dark matter distribution in the Milky Way using the recent rotation curve data. The baryon matter of the Milky Way is divided into bulge, disk and gas components, and each component is modelled using various possible mass profiles appearing in literature. The baryon mass models are then combined with one of the following dark matter profiles: Burkert profile, core-modified profile, pseudo-isothermal profile and NFW profile, to fit the observed rotation curved data. Results shows that the core-modified profile and pseudo-isothermal profile couldn't well fit the data, and the NFW profile generally fits the data better than the Burkert profile. This conclusion is not significantly affected by the gas model, but strongly depends on the disk model.

The best (B7D1G1+NFW) model has local density of dark matter $\rho_{\rm dm}(R=R_\odot)=0.51\pm0.09~{\rm GeV}/{\rm cm}^3$, and scale length $r_0=8.1\pm 0.7$ kpc, with $\chi_{\rm min}^2=45.4$. The local dark matter density derived here is at $\sim 1\sigma$ level consistent with the $0.40\pm0.04~{\rm GeV}/{\rm cm}^3$ value given by \cite{Sofue:2017nhd}, but it is a little larger than the $0.32\pm0.02~{\rm GeV}/{\rm cm}^3$ value given by \citet{Huang:2016}. One reason leading to the discrepancy is the different baryon mass profiles used in our paper from that used in \cite{Huang:2016}. Nevertheless, all the three values are consistent with each other within $\sim 2\sigma$ uncertainty.

Finally, it should be noticed that there are additional uncertainties arising from the modelling of the baryon profiles. The baryon profiles considered in our paper have not taken into account the rings, the spiral arms, and other irregular structures, which may also affect the rotation velocity. In addition, due to historical reasons some of the baryon profiles may not be consistent with the present knowledge on our Galaxy. More precise observations on the baryon mass of the Milky Way are required to tightly constrain the dark matter distribution.

\section*{Acknowledgments}
We thanks Dr. Li Tang for useful discussions. This work has been supported by the National Natural Science Fund of China under grant Nos. 11603005 and 11775038.

\section*{Supplementary Materials}
The best-fitting parameters and the best-fitting curves are presented in the supplementary materials.

\bsp

\label{lastpage}

\end{document}


\title{Supplementary Materials}
\author{Hai-Nan Lin and Xin Li}
\date{\today}
\maketitle

\section{The best-fitting parameters}
Table 1 summarises the best-fitting parameters of each dark matter model given a specific baryonic profile listed in the first column. For each dark matter model, the free parameters are $v_h$ and $r_0$.

\section{The best-fitting curves}
Figure 1 shows the best-fitting curves of each dark matter model given a specific baryonic profile, together with the contribution from each component of baryon and the observational data. Green curves for Burkert model; blue curves for core-modified model; cyan curves for pseudo isothermal model; black curves for NFW model. Dashed curves: contribution from bulge; dash-dotted curves: contribution from disk; Dotted curves: contribution from gas. The red error bars are the observational data with $1\sigma$ uncertainties.

\begin{table} 
\centering 
\caption{The best-fitting parameters of each model.} 
\begin{tabular}{l|ll|ll|ll|ll} 
\hline\hline 
models&bur& &com& &iso& &nfw\\&$v_h$ [km/s] &$r_0$ [kpc] &$v_h$ [km/s] &$r_0$ [kpc] &$v_h$ [km/s] &$r_0$ [kpc] &$v_h$ [km/s] &$r_0$ [kpc]\\
\hline 
B1D1G1&$419.5\pm7.3$&$6.2\pm0.3$&$316.7\pm6.0$&$6.6\pm0.3$&$206.1\pm6.2$&$2.2\pm0.3$&$416.5\pm9.2$&$13.4\pm1.4$\\ 
B1D1G2&$423.4\pm7.7$&$6.3\pm0.3$&$319.4\pm6.3$&$6.7\pm0.3$&$208.0\pm6.3$&$2.2\pm0.3$&$421.0\pm9.5$&$13.7\pm1.4$\\ 
B1D2G1&$424.8\pm6.6$&$5.4\pm0.3$&$320.7\pm5.5$&$5.8\pm0.3$&$204.3\pm5.6$&$1.6\pm0.3$&$421.6\pm7.2$&$11.0\pm1.0$\\ 
B1D2G2&$428.3\pm7.0$&$5.5\pm0.3$&$323.1\pm5.9$&$5.9\pm0.3$&$206.1\pm5.7$&$1.6\pm0.3$&$425.7\pm7.6$&$11.3\pm1.1$\\ 
B1D3G1&$416.8\pm7.4$&$6.2\pm0.3$&$314.4\pm6.2$&$6.5\pm0.3$&$203.9\pm6.0$&$2.1\pm0.3$&$414.3\pm8.9$&$13.2\pm1.3$\\ 
B1D3G2&$420.7\pm7.8$&$6.2\pm0.4$&$317.1\pm6.5$&$6.6\pm0.4$&$205.8\pm6.1$&$2.1\pm0.3$&$418.8\pm9.2$&$13.6\pm1.4$\\ 
B1D4G1&$421.2\pm19.9$&$11.5\pm1.1$&$318.6\pm12.0$&$11.2\pm0.7$&$227.1\pm15.1$&$6.3\pm0.9$&$430.9\pm42.3$&$37.0\pm10.9$\\ 
B1D4G2&$425.9\pm19.2$&$11.5\pm1.0$&$322.4\pm11.7$&$11.3\pm0.7$&$229.1\pm14.6$&$6.3\pm0.8$&$436.3\pm41.2$&$37.2\pm10.5$\\ 
B2D1G1&$419.8\pm7.3$&$6.2\pm0.3$&$316.9\pm6.0$&$6.6\pm0.3$&$206.2\pm6.2$&$2.2\pm0.3$&$416.8\pm9.2$&$13.4\pm1.4$\\ 
B2D1G2&$423.7\pm7.7$&$6.3\pm0.3$&$319.6\pm6.3$&$6.7\pm0.3$&$208.1\pm6.3$&$2.2\pm0.3$&$421.3\pm9.5$&$13.7\pm1.4$\\ 
B2D2G1&$425.1\pm6.7$&$5.4\pm0.3$&$320.9\pm5.6$&$5.8\pm0.3$&$204.4\pm5.6$&$1.6\pm0.3$&$421.9\pm7.3$&$10.9\pm1.0$\\ 
B2D2G2&$428.6\pm7.0$&$5.5\pm0.3$&$323.3\pm5.9$&$5.9\pm0.3$&$206.2\pm5.7$&$1.6\pm0.3$&$426.0\pm7.6$&$11.2\pm1.1$\\ 
B2D3G1&$417.1\pm7.4$&$6.1\pm0.3$&$314.6\pm6.2$&$6.5\pm0.3$&$204.0\pm6.0$&$2.1\pm0.3$&$414.6\pm8.9$&$13.2\pm1.3$\\ 
B2D3G2&$420.9\pm7.8$&$6.2\pm0.4$&$317.3\pm6.5$&$6.6\pm0.4$&$205.9\pm6.1$&$2.1\pm0.3$&$419.1\pm9.2$&$13.5\pm1.4$\\ 
B2D4G1&$421.3\pm19.6$&$11.5\pm1.0$&$318.8\pm11.9$&$11.2\pm0.7$&$227.0\pm14.9$&$6.3\pm0.9$&$430.9\pm41.7$&$36.8\pm10.7$\\ 
B2D4G2&$426.1\pm19.0$&$11.5\pm1.0$&$322.6\pm11.5$&$11.2\pm0.6$&$229.0\pm14.4$&$6.3\pm0.8$&$436.3\pm40.7$&$37.0\pm10.3$\\ 
B3D1G1&$418.0\pm7.8$&$6.7\pm0.4$&$315.7\pm6.2$&$7.0\pm0.3$&$207.3\pm6.7$&$2.5\pm0.3$&$415.1\pm10.6$&$14.7\pm1.6$\\ 
B3D1G2&$422.1\pm8.1$&$6.7\pm0.4$&$318.5\pm6.6$&$7.1\pm0.3$&$209.2\pm6.7$&$2.6\pm0.3$&$419.7\pm10.8$&$15.1\pm1.7$\\ 
B3D2G1&$422.4\pm7.0$&$5.7\pm0.3$&$318.8\pm5.8$&$6.2\pm0.3$&$205.0\pm5.9$&$1.8\pm0.3$&$419.2\pm8.1$&$11.9\pm1.2$\\ 
B3D2G2&$426.0\pm7.4$&$5.8\pm0.3$&$321.3\pm6.1$&$6.2\pm0.3$&$206.9\pm6.0$&$1.9\pm0.3$&$423.5\pm8.4$&$12.2\pm1.2$\\ 
B3D3G1&$415.2\pm7.8$&$6.6\pm0.4$&$313.3\pm6.4$&$6.9\pm0.3$&$205.0\pm6.4$&$2.4\pm0.3$&$412.9\pm10.1$&$14.5\pm1.6$\\ 
B3D3G2&$419.3\pm8.2$&$6.6\pm0.4$&$316.2\pm6.8$&$7.0\pm0.4$&$207.0\pm6.5$&$2.5\pm0.3$&$417.6\pm10.4$&$14.9\pm1.6$\\ 
B3D4G1&$427.8\pm35.0$&$13.2\pm1.9$&$320.9\pm21.8$&$12.4\pm1.3$&$236.8\pm24.9$&$7.7\pm1.5$&$450.4\pm78.6$&$49.7\pm24.1$\\ 
B3D4G2&$432.4\pm33.3$&$13.1\pm1.8$&$324.8\pm20.7$&$12.4\pm1.2$&$238.5\pm23.8$&$7.6\pm1.5$&$454.9\pm74.9$&$49.2\pm22.6$\\ 
B4D1G1&$430.1\pm6.1$&$4.6\pm0.3$&$325.3\pm5.2$&$5.1\pm0.3$&$201.1\pm5.0$&$0.9\pm0.3$&$427.0\pm5.8$&$8.8\pm0.8$\\ 
B4D1G2&$433.3\pm6.5$&$4.7\pm0.3$&$327.4\pm5.5$&$5.1\pm0.3$&$202.9\pm5.1$&$0.9\pm0.3$&$430.7\pm6.1$&$9.1\pm0.8$\\ 
B4D2G1&$439.6\pm5.6$&$4.1\pm0.3$&$333.3\pm4.7$&$4.6\pm0.3$&$201.0\pm4.8$&$0.5\pm0.2$&$435.9\pm5.0$&$7.4\pm0.7$\\ 
B4D2G2&$442.4\pm6.0$&$4.1\pm0.3$&$335.2\pm5.0$&$4.6\pm0.3$&$202.8\pm4.8$&$0.5\pm0.3$&$439.1\pm5.4$&$7.6\pm0.7$\\ 
B4D3G1&$427.8\pm6.3$&$4.5\pm0.3$&$323.5\pm5.3$&$5.0\pm0.3$&$199.2\pm4.9$&$0.8\pm0.3$&$425.1\pm5.9$&$8.7\pm0.8$\\ 
B4D3G2&$430.9\pm6.7$&$4.6\pm0.3$&$325.5\pm5.6$&$5.0\pm0.3$&$200.9\pm5.0$&$0.8\pm0.3$&$428.8\pm6.3$&$8.9\pm0.8$\\ 
B4D4G1&$414.0\pm8.5$&$7.8\pm0.4$&$313.6\pm6.9$&$8.1\pm0.4$&$208.3\pm7.0$&$3.3\pm0.4$&$412.9\pm12.9$&$18.9\pm2.3$\\ 
B4D4G2&$418.4\pm8.9$&$7.9\pm0.4$&$316.9\pm7.3$&$8.2\pm0.4$&$210.3\pm7.0$&$3.3\pm0.4$&$418.0\pm13.2$&$19.3\pm2.3$\\ 
B5D1G1&$423.4\pm6.6$&$5.4\pm0.3$&$319.6\pm5.6$&$5.8\pm0.3$&$203.2\pm5.5$&$1.5\pm0.3$&$420.4\pm7.1$&$10.9\pm1.0$\\ 
B5D1G2&$426.9\pm7.0$&$5.5\pm0.3$&$322.0\pm5.9$&$5.9\pm0.3$&$205.1\pm5.5$&$1.6\pm0.3$&$424.6\pm7.4$&$11.2\pm1.0$\\ 
B5D2G1&$430.7\pm6.1$&$4.7\pm0.3$&$325.7\pm5.2$&$5.2\pm0.3$&$202.4\pm5.1$&$1.0\pm0.3$&$427.4\pm5.9$&$9.1\pm0.8$\\ 
B5D2G2&$433.9\pm6.5$&$4.8\pm0.3$&$327.8\pm5.5$&$5.2\pm0.3$&$204.2\pm5.2$&$1.0\pm0.3$&$431.2\pm6.3$&$9.3\pm0.8$\\ 
B5D3G1&$420.8\pm6.8$&$5.3\pm0.3$&$317.4\pm5.8$&$5.7\pm0.3$&$201.1\pm5.4$&$1.4\pm0.3$&$418.4\pm7.0$&$10.8\pm1.0$\\ 
B5D3G2&$424.3\pm7.3$&$5.4\pm0.3$&$319.8\pm6.1$&$5.8\pm0.3$&$202.9\pm5.4$&$1.5\pm0.3$&$422.5\pm7.4$&$11.1\pm1.0$\\ 
B5D4G1&$416.2\pm11.1$&$9.4\pm0.6$&$316.1\pm7.4$&$9.6\pm0.4$&$216.2\pm9.3$&$4.6\pm0.5$&$417.6\pm21.4$&$25.7\pm4.4$\\ 
B5D4G2&$420.9\pm11.1$&$9.5\pm0.5$&$319.7\pm7.6$&$9.7\pm0.4$&$218.3\pm9.2$&$4.6\pm0.5$&$423.1\pm21.4$&$26.1\pm4.4$\\ 
B6D1G1&$428.8\pm6.2$&$4.8\pm0.3$&$324.2\pm5.2$&$5.2\pm0.3$&$201.7\pm5.1$&$1.0\pm0.3$&$425.7\pm6.0$&$9.2\pm0.8$\\ 
B6D1G2&$432.1\pm6.6$&$4.8\pm0.3$&$326.3\pm5.5$&$5.3\pm0.3$&$203.4\pm5.2$&$1.1\pm0.3$&$429.4\pm6.3$&$9.5\pm0.8$\\ 
B6D2G1&$437.9\pm5.7$&$4.2\pm0.3$&$331.8\pm4.8$&$4.7\pm0.3$&$201.4\pm4.8$&$0.6\pm0.3$&$434.2\pm5.2$&$7.7\pm0.7$\\ 
B6D2G2&$440.8\pm6.0$&$4.3\pm0.3$&$333.8\pm5.1$&$4.7\pm0.3$&$203.2\pm4.9$&$0.6\pm0.3$&$437.6\pm5.5$&$8.0\pm0.7$\\ 
B6D3G1&$426.4\pm6.4$&$4.7\pm0.3$&$322.3\pm5.3$&$5.1\pm0.3$&$199.7\pm5.0$&$0.9\pm0.3$&$423.7\pm6.0$&$9.1\pm0.8$\\ 
B6D3G2&$429.6\pm6.8$&$4.7\pm0.3$&$324.4\pm5.7$&$5.2\pm0.3$&$201.4\pm5.1$&$1.0\pm0.3$&$427.5\pm6.4$&$9.3\pm0.9$\\ 
B6D4G1&$414.4\pm8.8$&$8.1\pm0.4$&$314.2\pm6.8$&$8.4\pm0.4$&$209.8\pm7.3$&$3.5\pm0.4$&$413.5\pm14.2$&$20.0\pm2.6$\\ 
B6D4G2&$418.9\pm9.1$&$8.2\pm0.4$&$317.6\pm7.2$&$8.5\pm0.4$&$211.9\pm7.3$&$3.6\pm0.4$&$418.7\pm14.3$&$20.4\pm2.6$\\ 
B7D1G1&$433.7\pm5.9$&$4.3\pm0.3$&$328.4\pm5.0$&$4.8\pm0.3$&$200.6\pm4.8$&$0.7\pm0.3$&$430.4\pm5.4$&$8.1\pm0.7$\\ 
B7D1G2&$436.7\pm6.3$&$4.4\pm0.3$&$330.4\pm5.2$&$4.9\pm0.3$&$202.4\pm4.9$&$0.7\pm0.3$&$433.8\pm5.7$&$8.3\pm0.8$\\ 
B7D2G1&$443.9\pm5.4$&$3.8\pm0.3$&$337.0\pm4.6$&$4.4\pm0.3$&$200.8\pm4.7$&$0.3\pm0.2$&$440.0\pm4.7$&$6.9\pm0.6$\\ 
B7D2G2&$446.6\pm5.7$&$3.9\pm0.3$&$338.8\pm4.8$&$4.4\pm0.3$&$202.5\pm4.7$&$0.3\pm0.2$&$443.1\pm5.0$&$7.1\pm0.7$\\ 
B7D3G1&$431.5\pm6.0$&$4.3\pm0.3$&$326.7\pm5.1$&$4.7\pm0.3$&$198.7\pm4.8$&$0.6\pm0.3$&$428.5\pm5.5$&$8.0\pm0.7$\\ 
B7D3G2&$434.5\pm6.4$&$4.3\pm0.3$&$328.6\pm5.3$&$4.8\pm0.3$&$200.5\pm4.9$&$0.6\pm0.3$&$432.0\pm5.9$&$8.2\pm0.8$\\ 
B7D4G1&$413.9\pm8.1$&$7.2\pm0.4$&$313.0\pm6.8$&$7.6\pm0.4$&$205.8\pm6.4$&$2.9\pm0.3$&$412.5\pm11.0$&$16.9\pm1.8$\\ 
B7D4G2&$418.2\pm8.5$&$7.3\pm0.4$&$316.1\pm7.2$&$7.7\pm0.4$&$207.8\pm6.5$&$2.9\pm0.3$&$417.5\pm11.3$&$17.3\pm1.9$\\ 
\hline 
\end{tabular} 
\end{table} 

\begin{figure}
\centering
\includegraphics[width=0.4\textwidth]{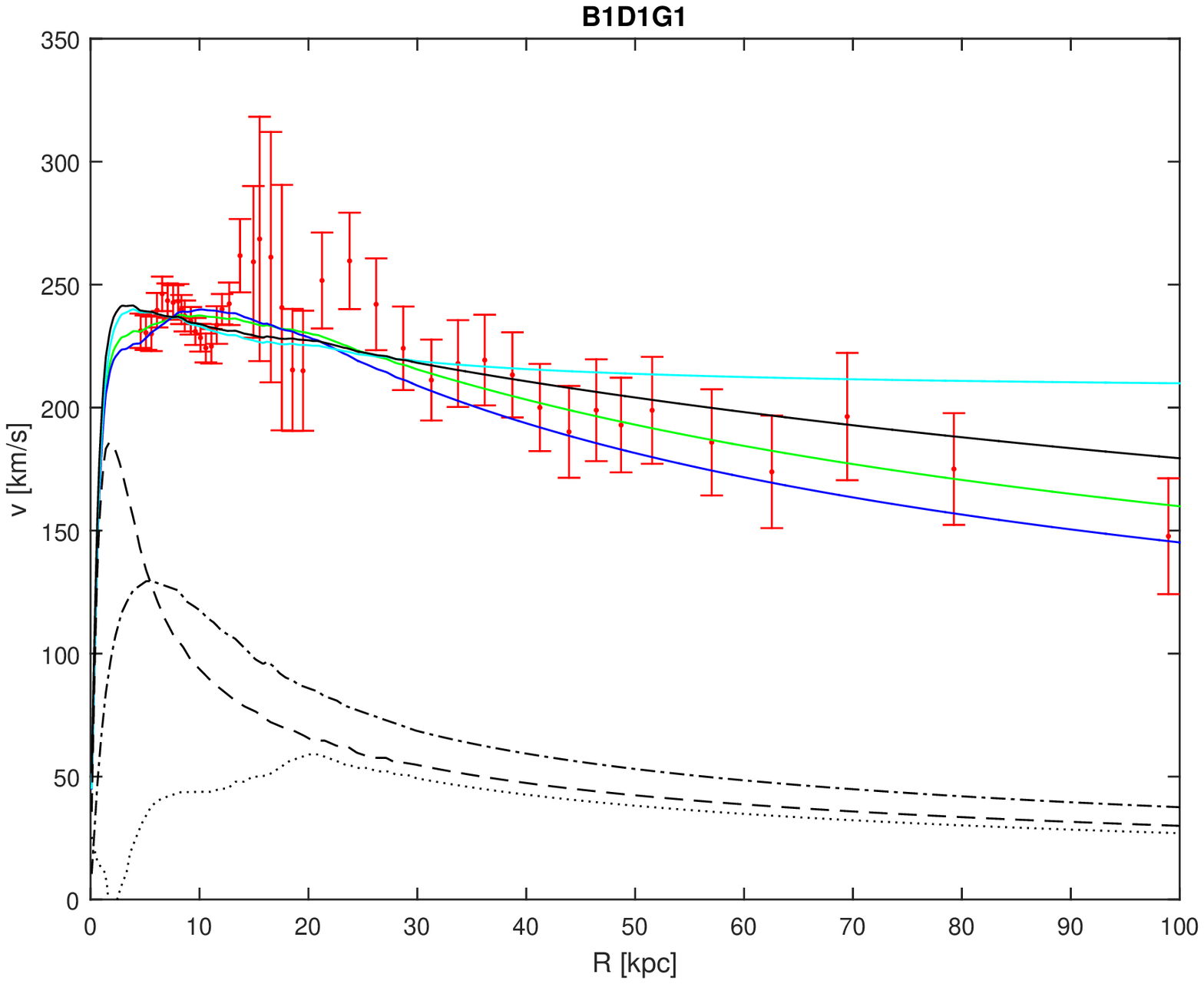}\hspace{0.5cm}
\includegraphics[width=0.4\textwidth]{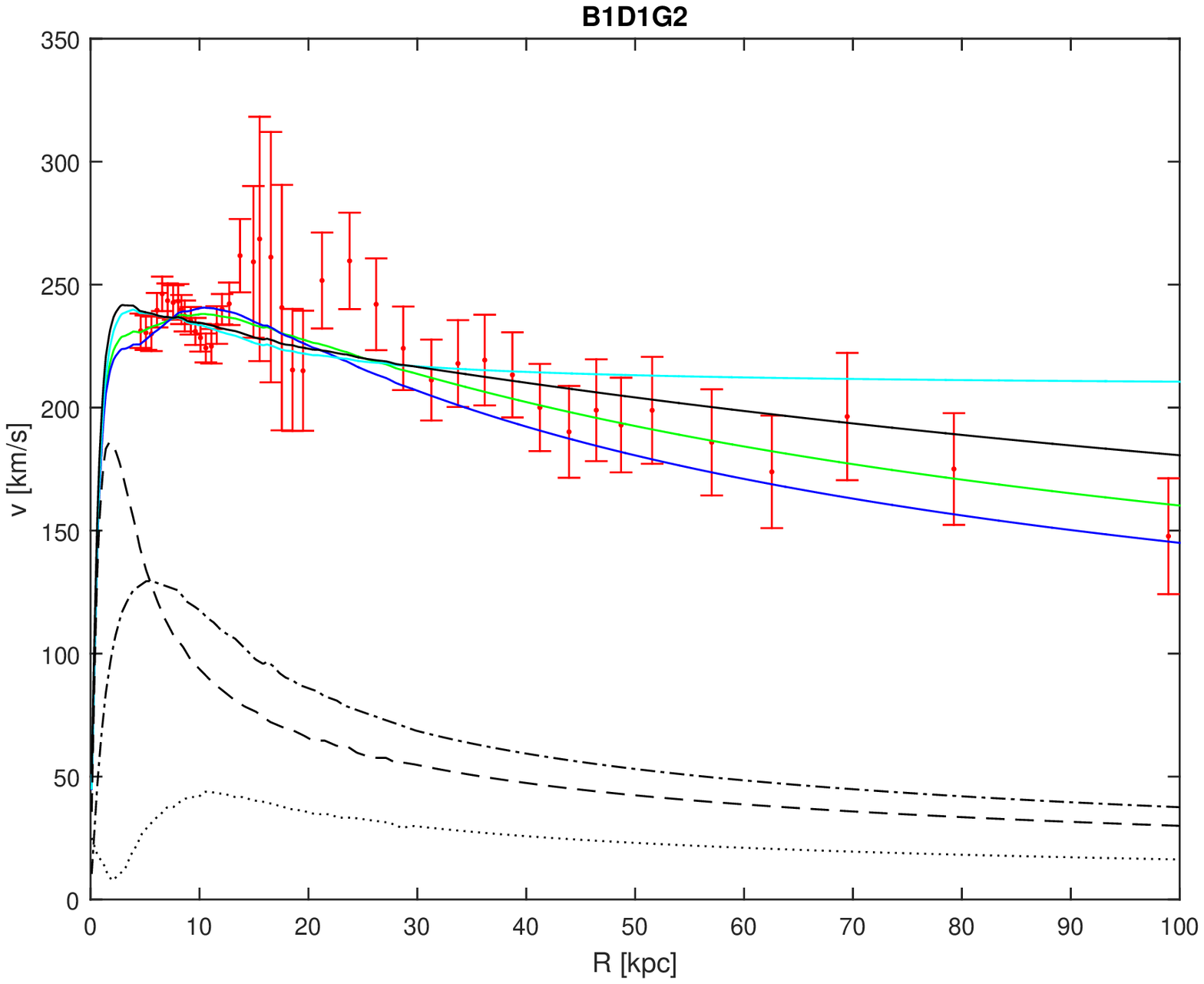}
\includegraphics[width=0.4\textwidth]{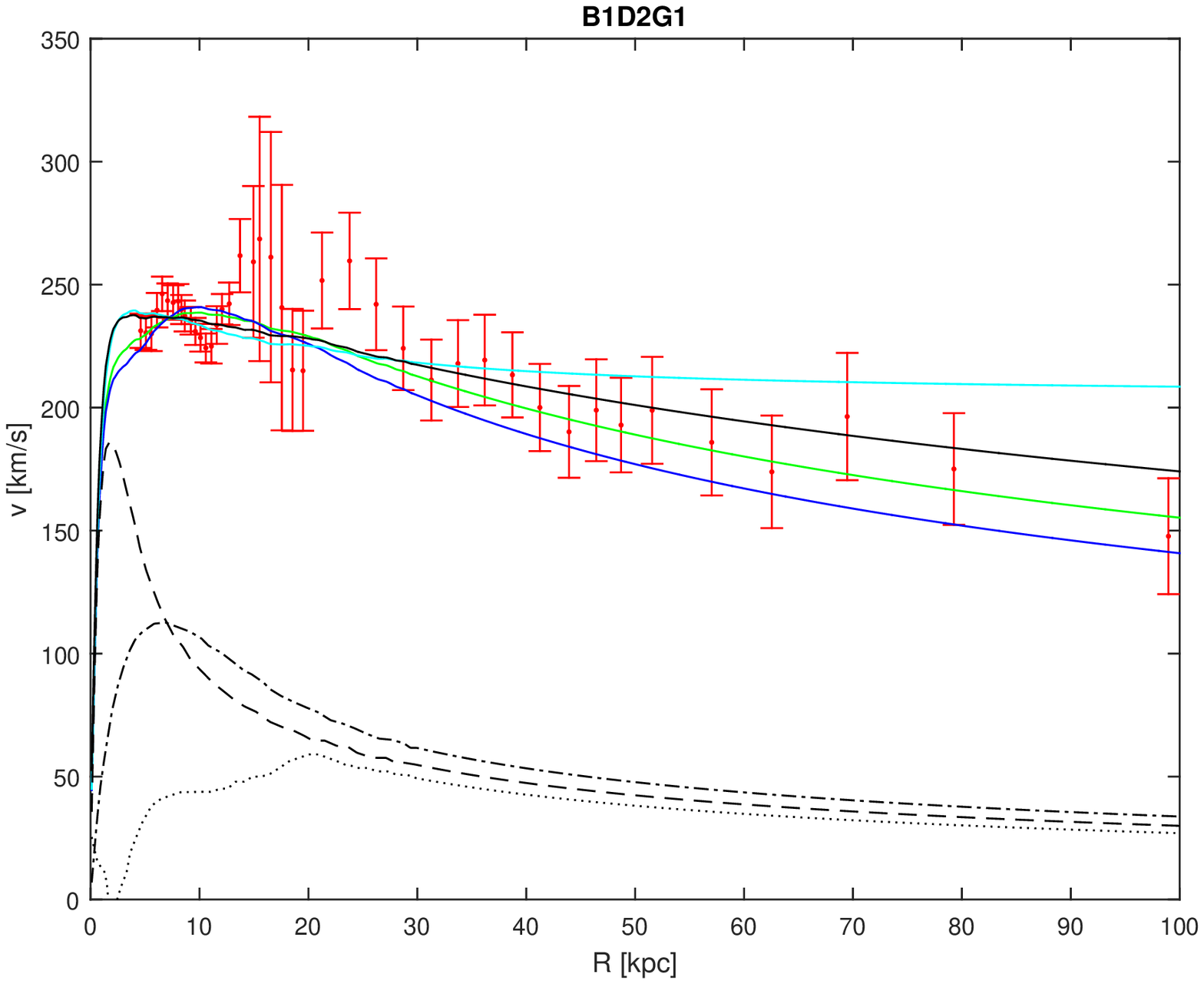}\hspace{0.5cm}
\includegraphics[width=0.4\textwidth]{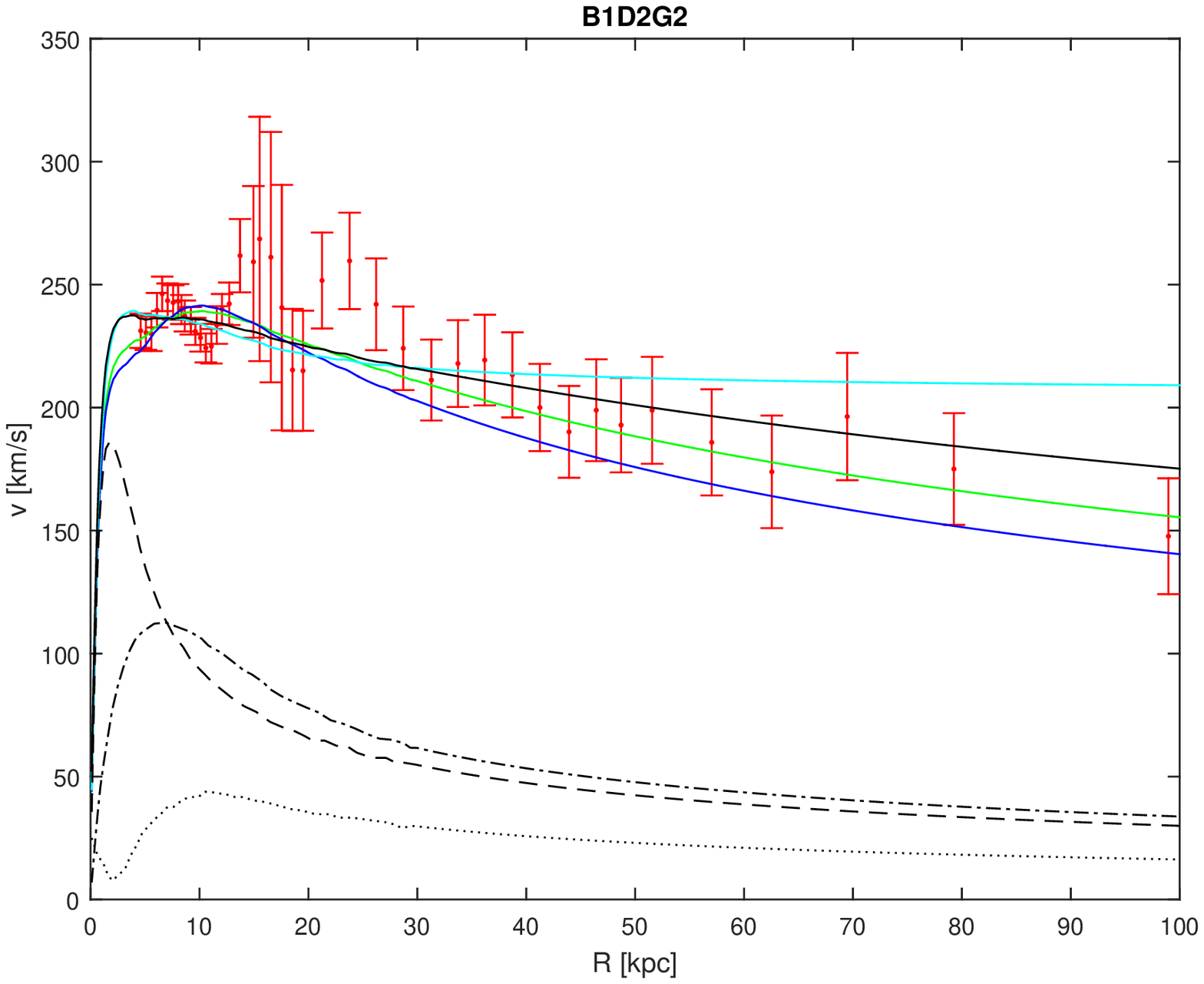}
\includegraphics[width=0.4\textwidth]{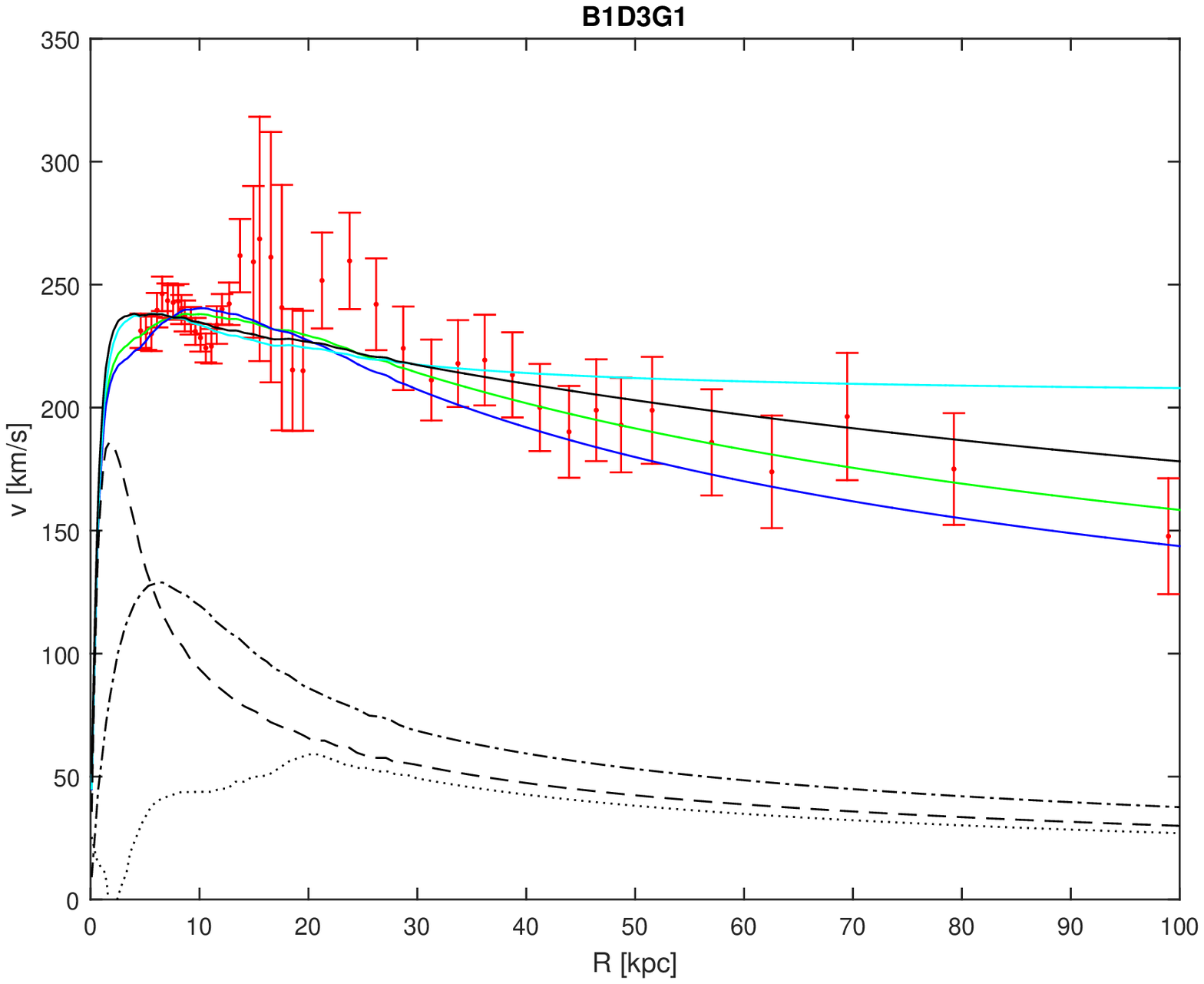}\hspace{0.5cm}
\includegraphics[width=0.4\textwidth]{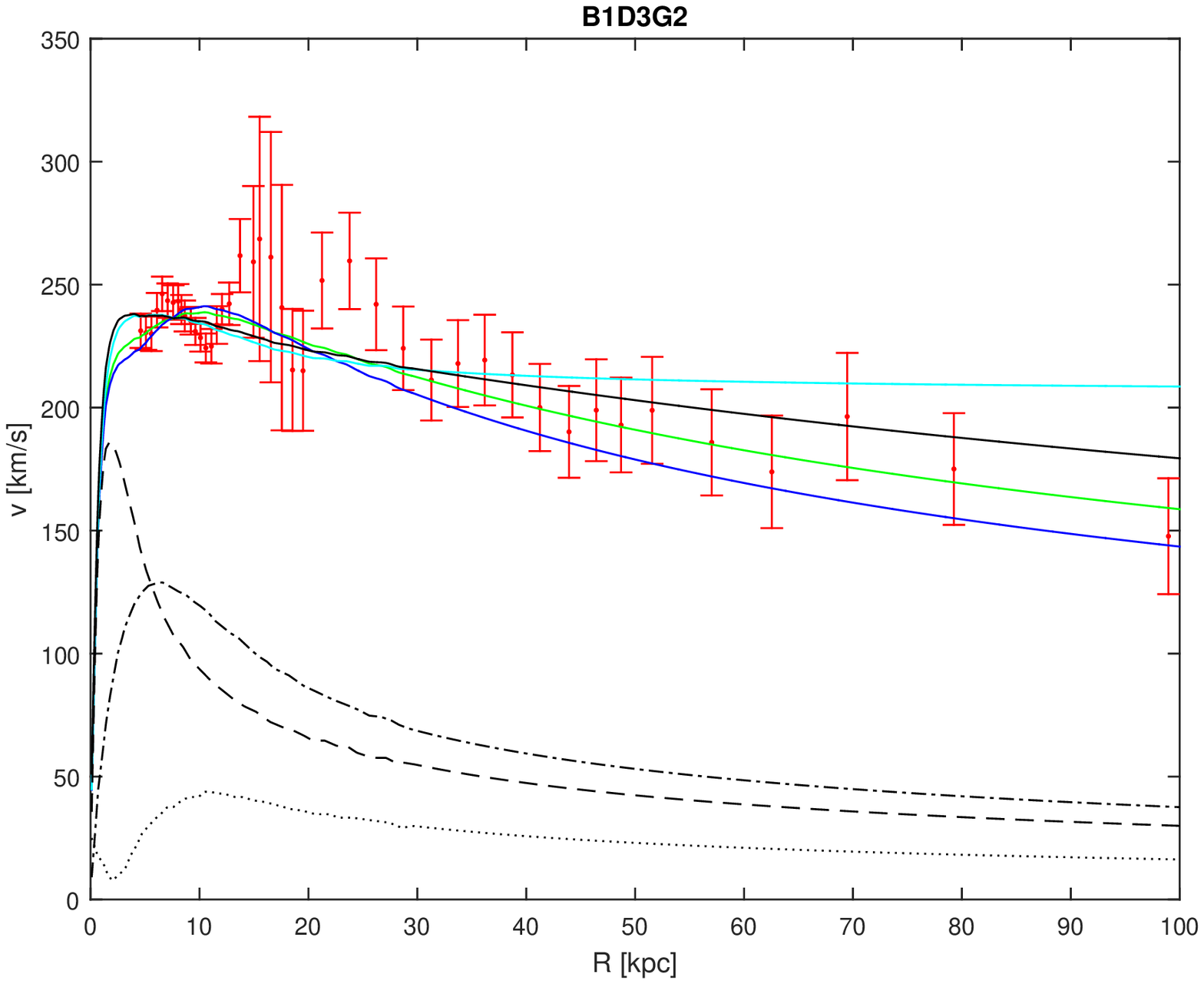}
\includegraphics[width=0.4\textwidth]{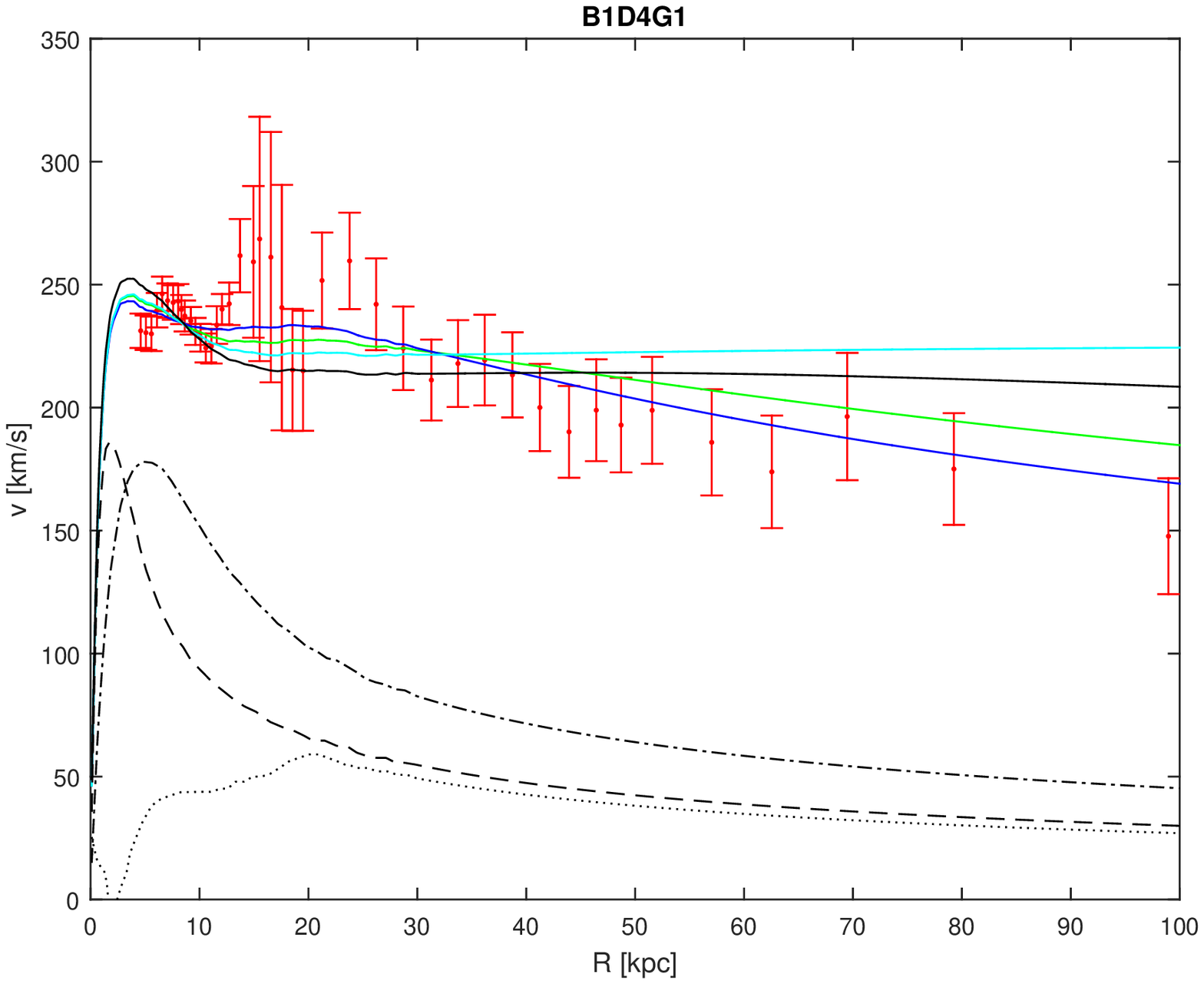}\hspace{0.5cm}
\includegraphics[width=0.4\textwidth]{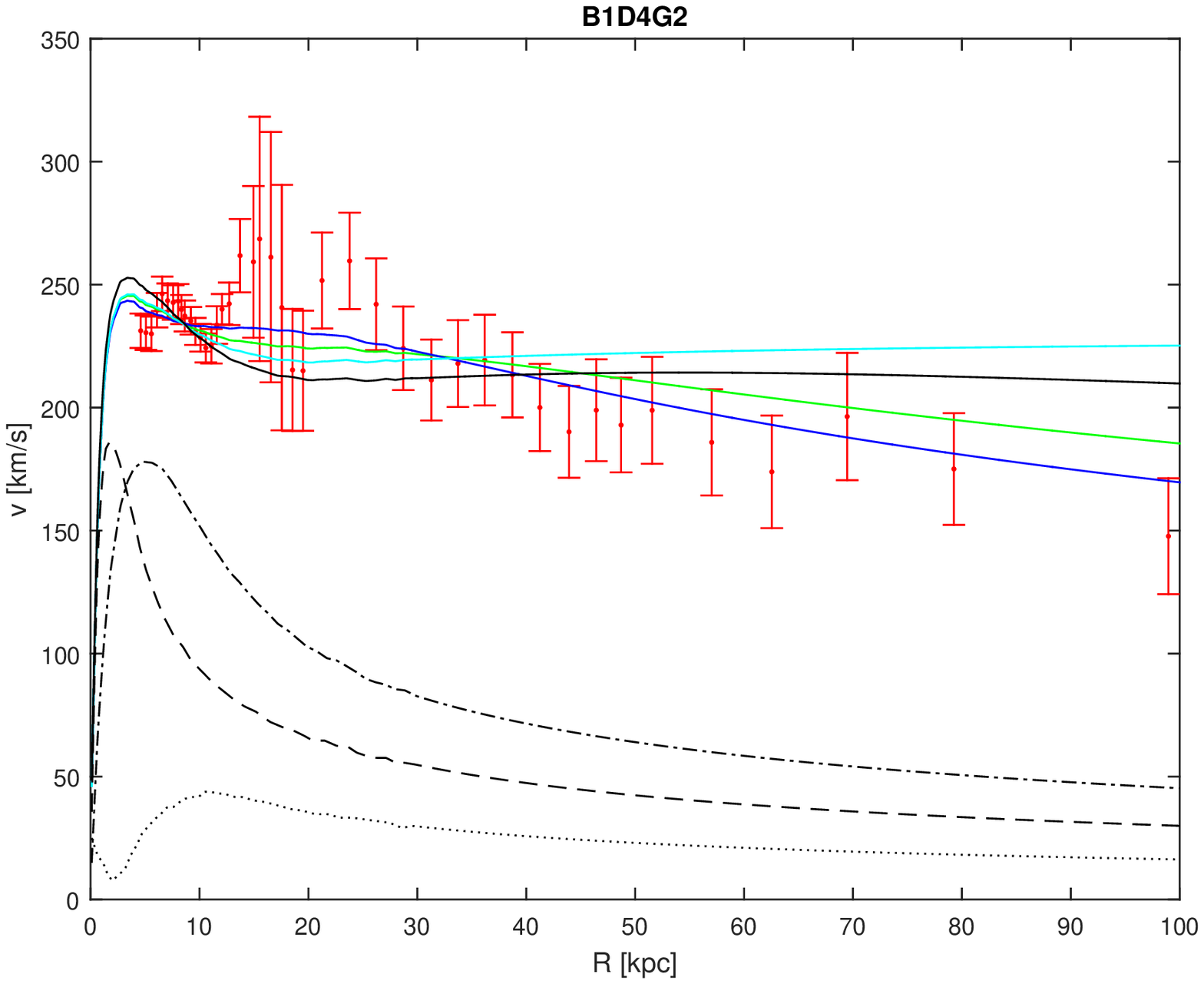}
\caption{Best fits to the rotation curves of the Milky Way. Green curves for Burkert model; blue curves for core-modified model; cyan curves for pseudo isothermal model; black curves for NFW model. The contributions from bulge (dashed curves), disk (dash-dotted curves) and gas (dotted curves) are also shown.}
\end{figure}

\begin{figure}
\centering
\includegraphics[width=0.4\textwidth]{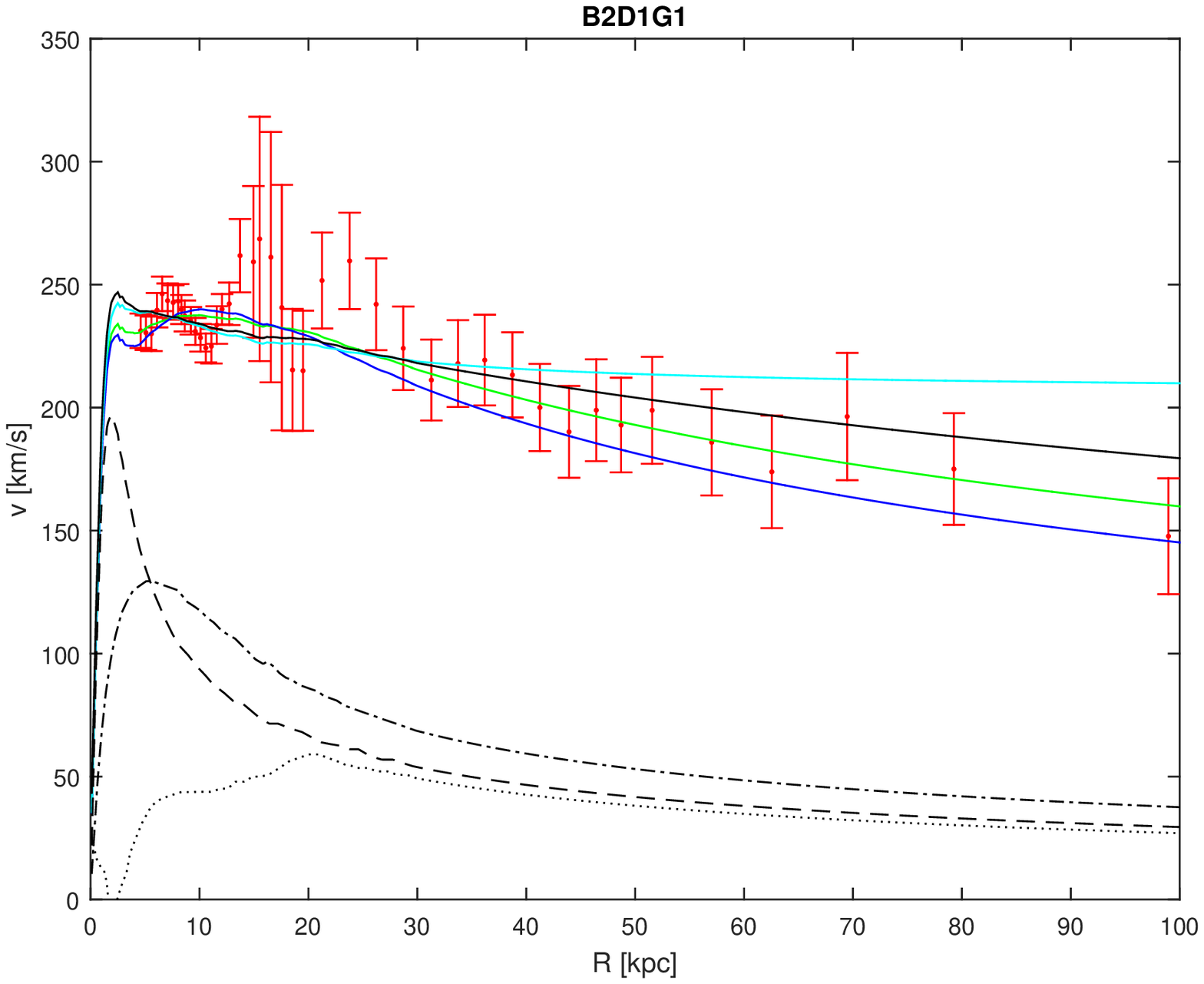}\hspace{0.5cm}
\includegraphics[width=0.4\textwidth]{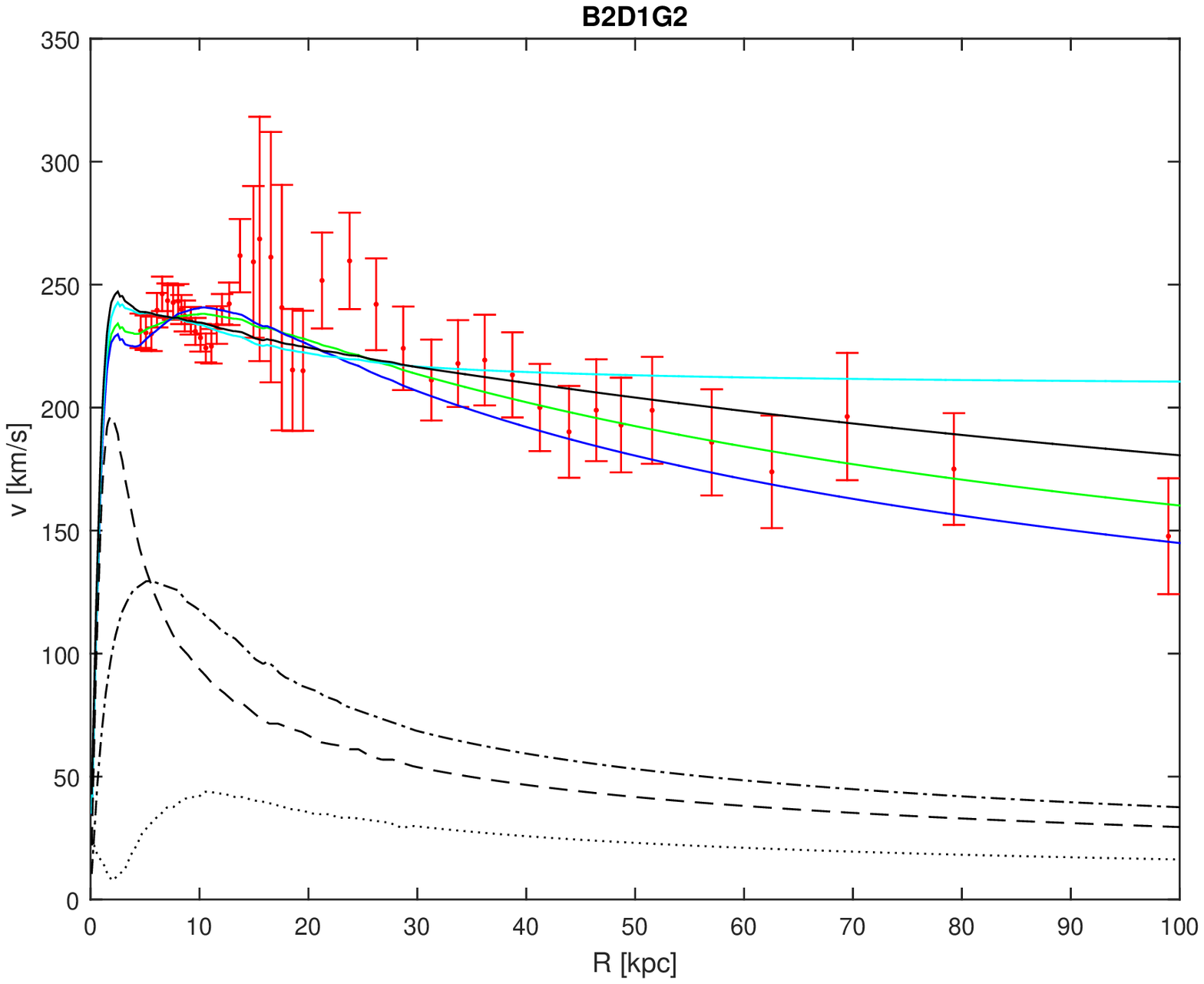}
\includegraphics[width=0.4\textwidth]{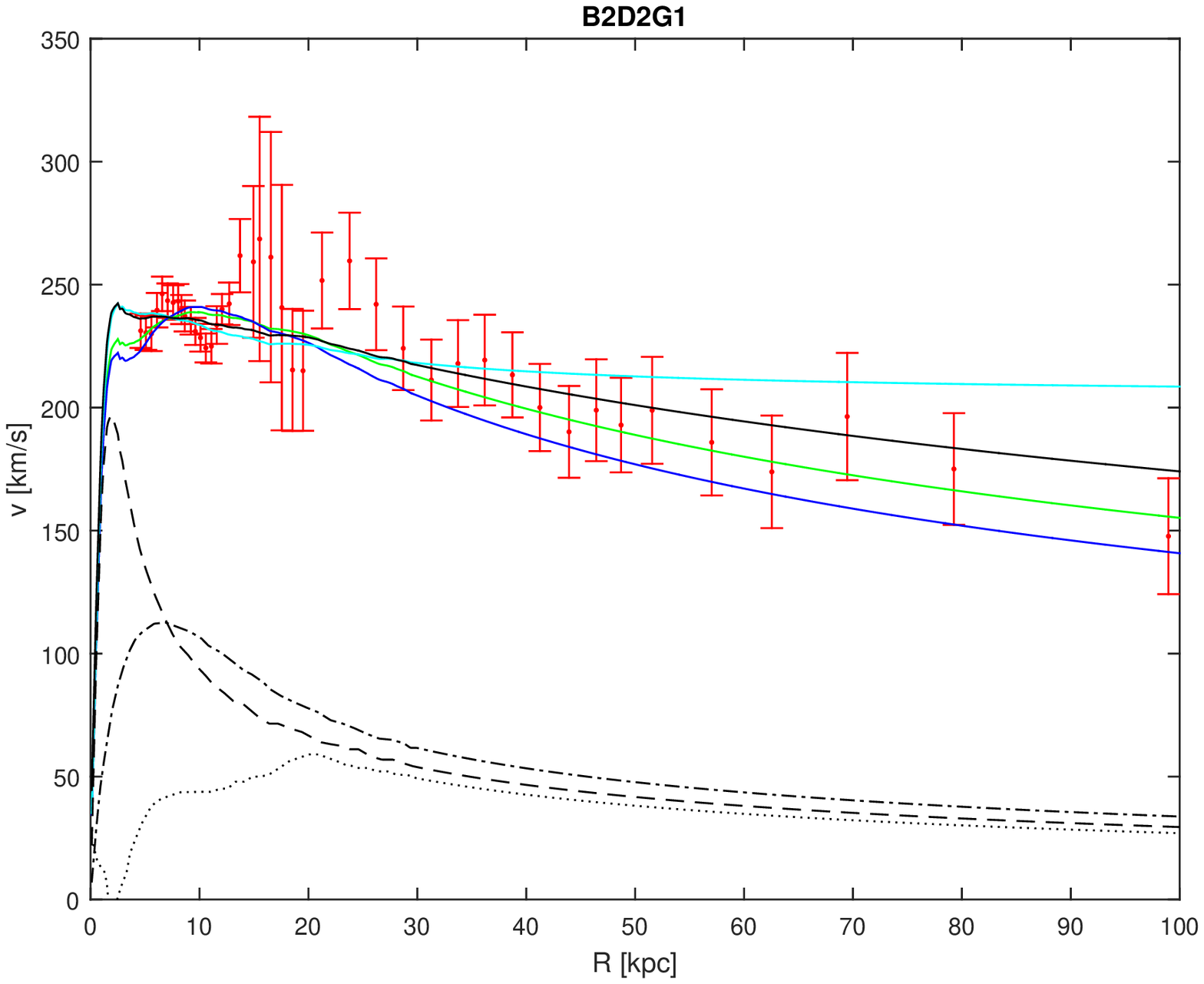}\hspace{0.5cm}
\includegraphics[width=0.4\textwidth]{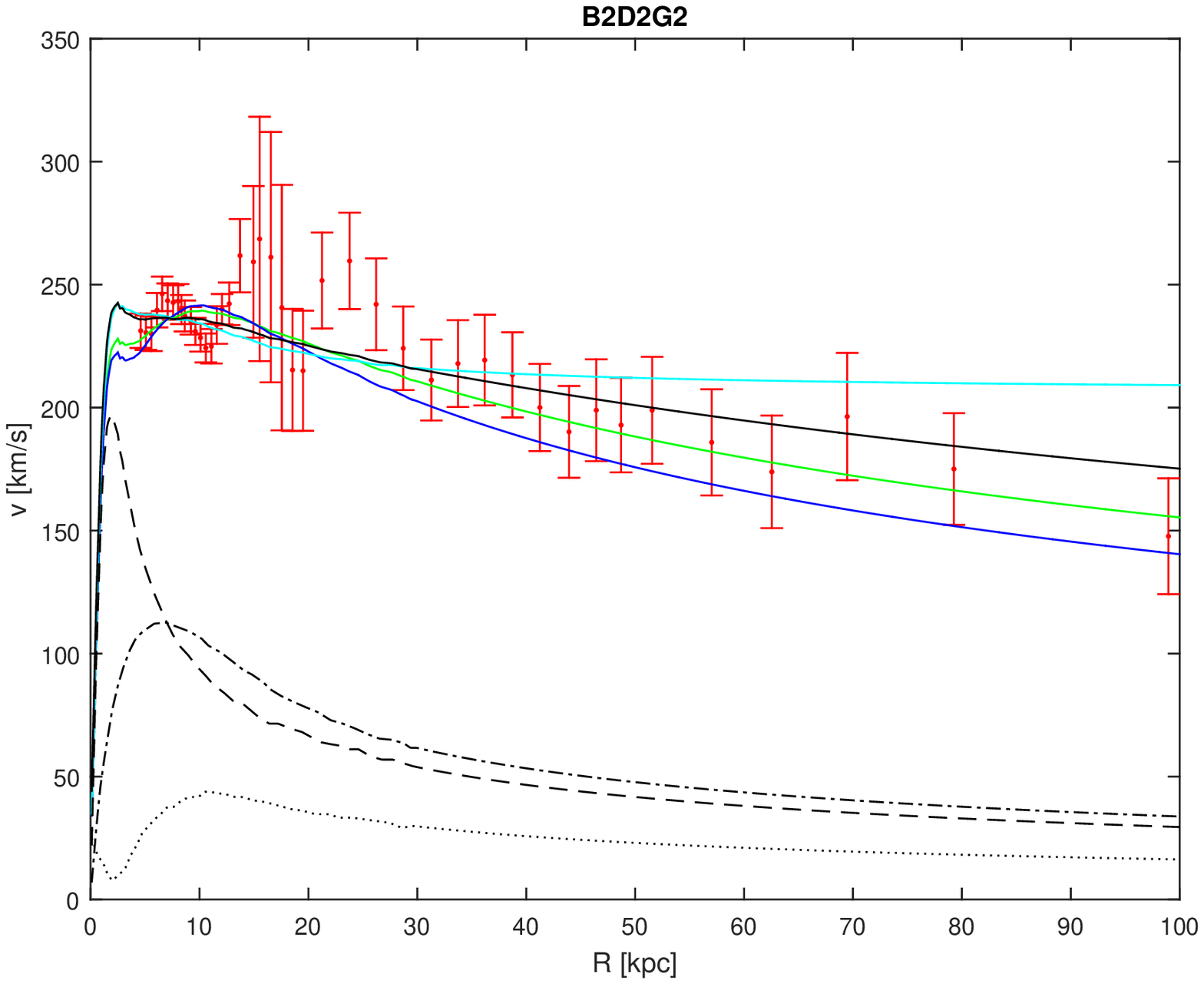}
\includegraphics[width=0.4\textwidth]{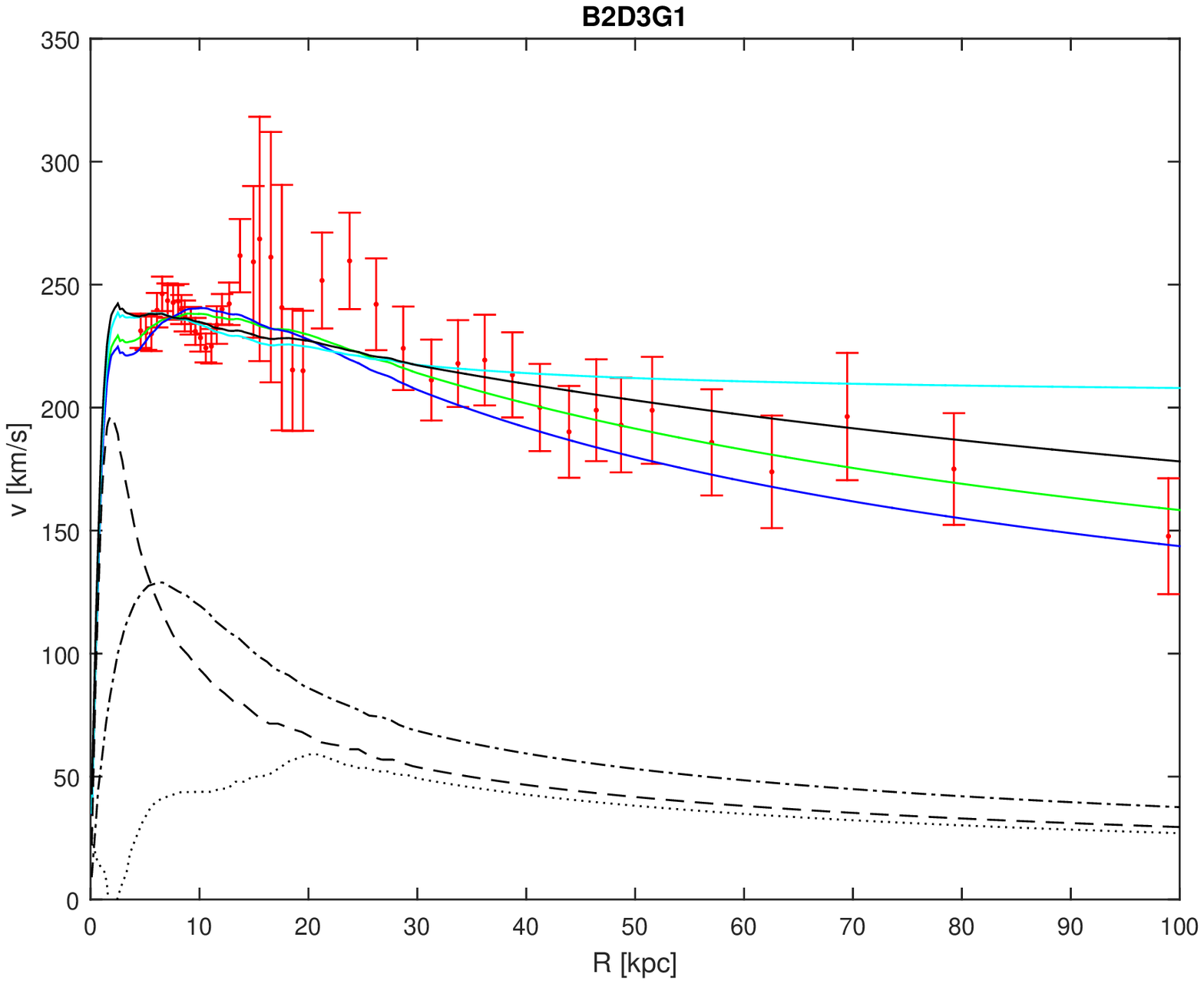}\hspace{0.5cm}
\includegraphics[width=0.4\textwidth]{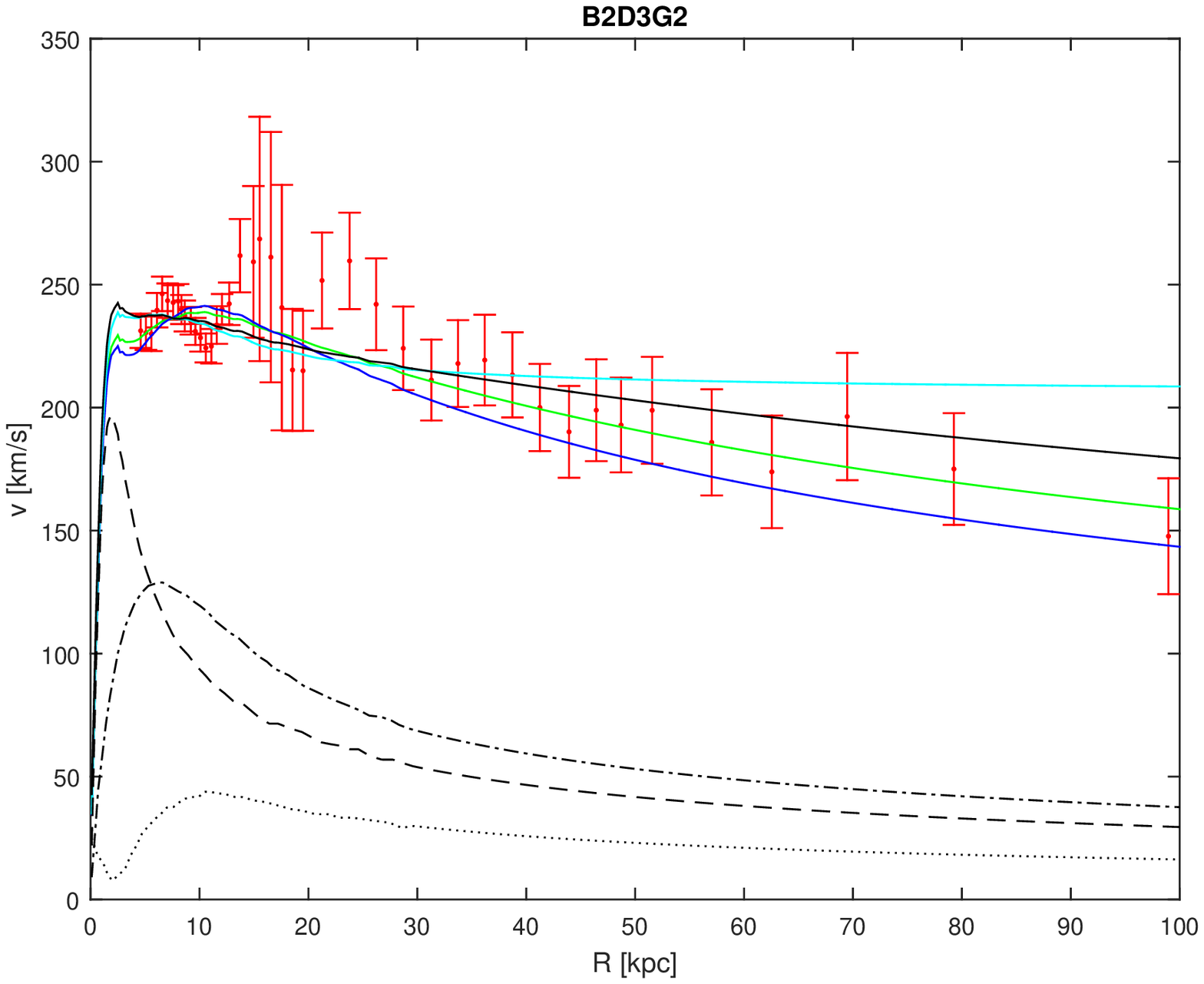}
\includegraphics[width=0.4\textwidth]{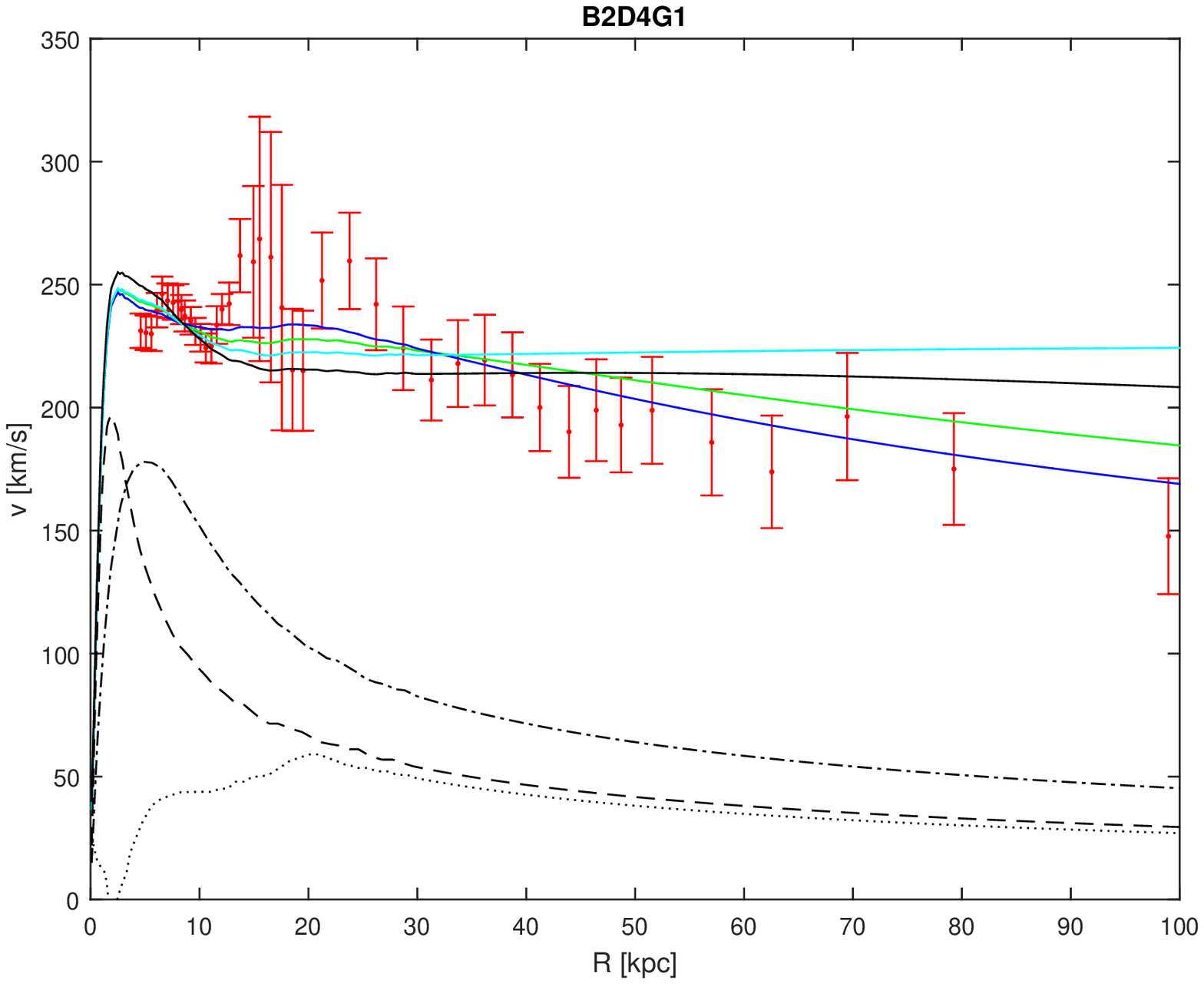}\hspace{0.5cm}
\includegraphics[width=0.4\textwidth]{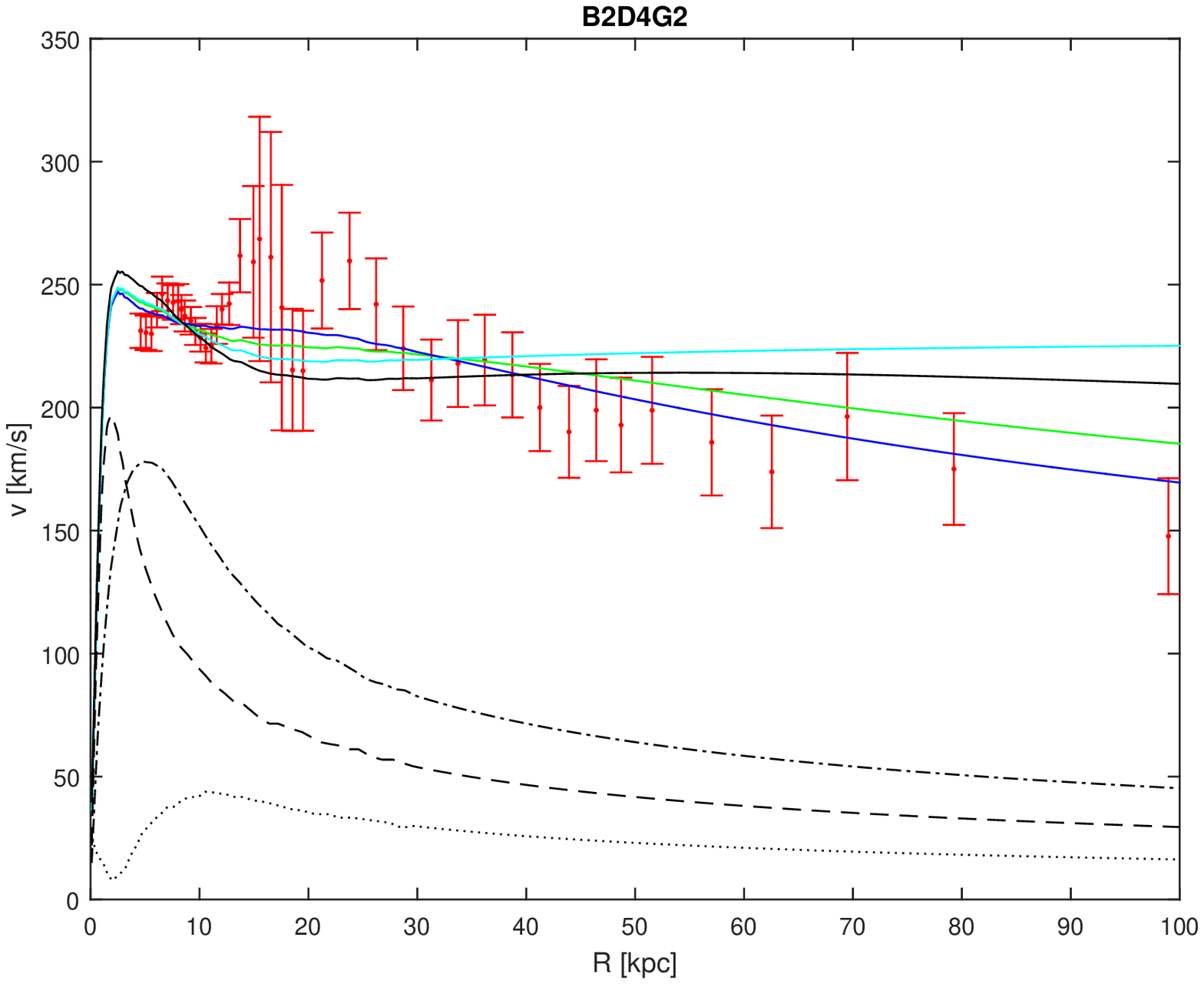}
\addtocounter{figure}{-1}
\caption{--continued}
\end{figure}

\begin{figure}
\centering
\includegraphics[width=0.4\textwidth]{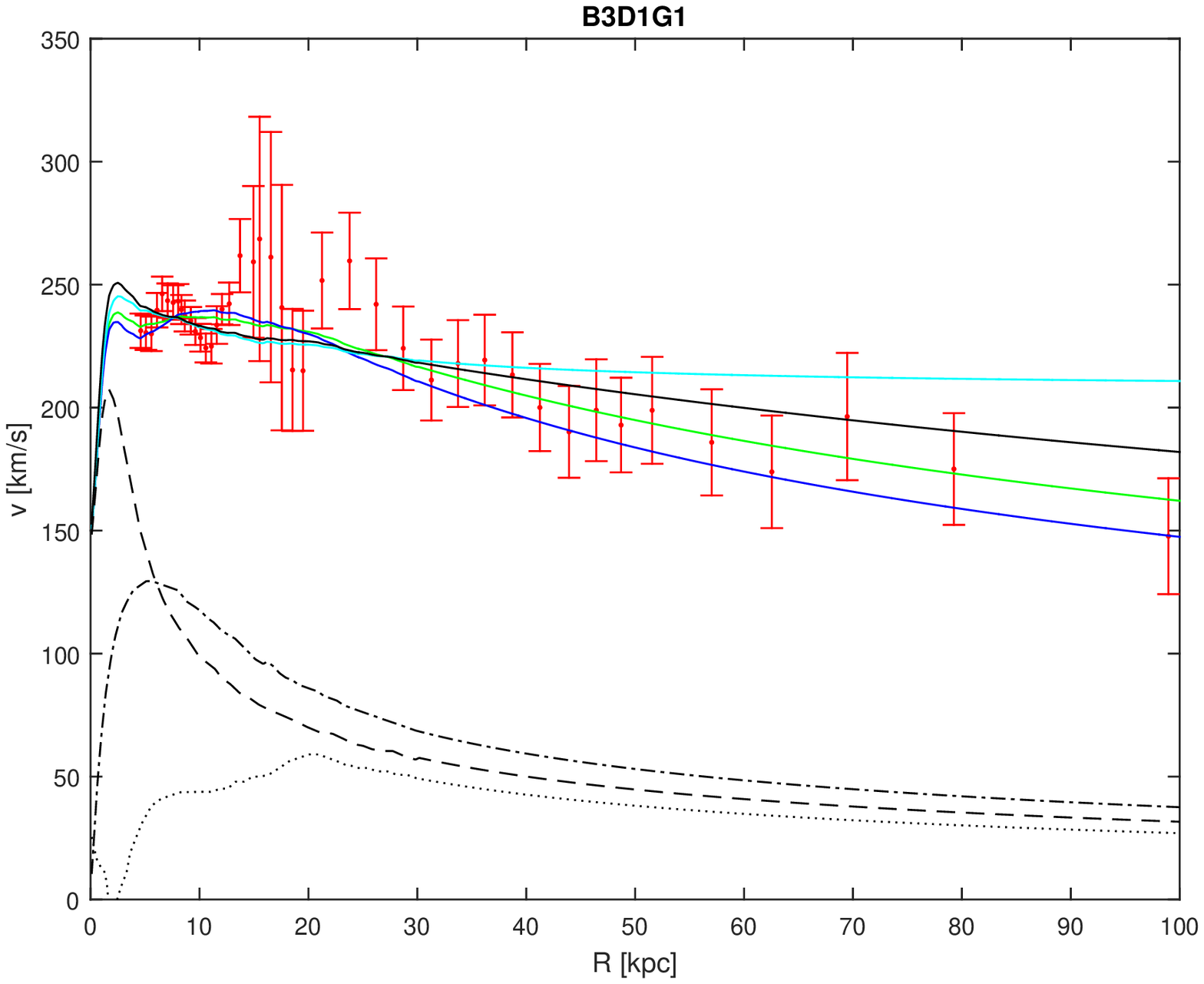}\hspace{0.5cm}
\includegraphics[width=0.4\textwidth]{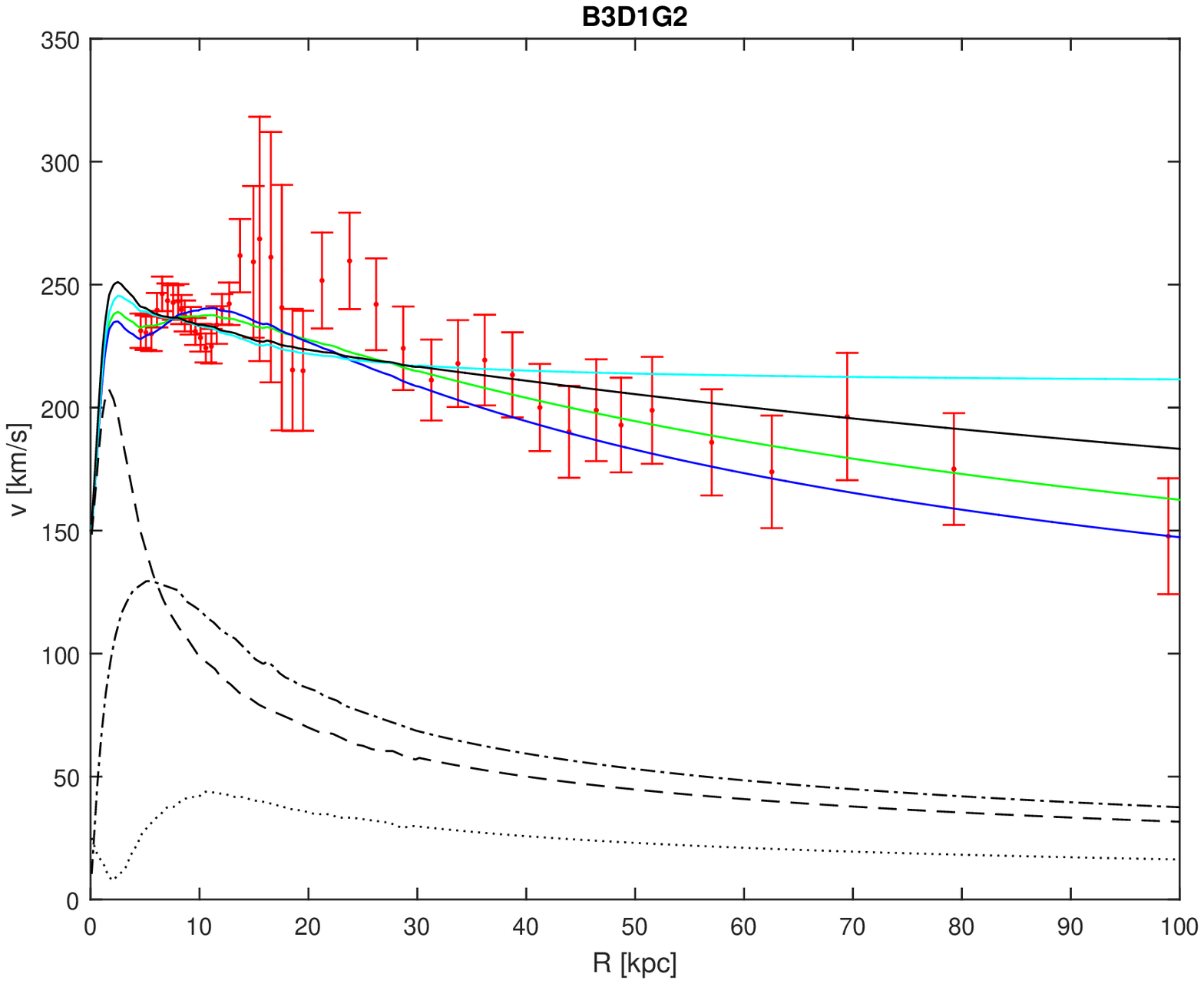}
\includegraphics[width=0.4\textwidth]{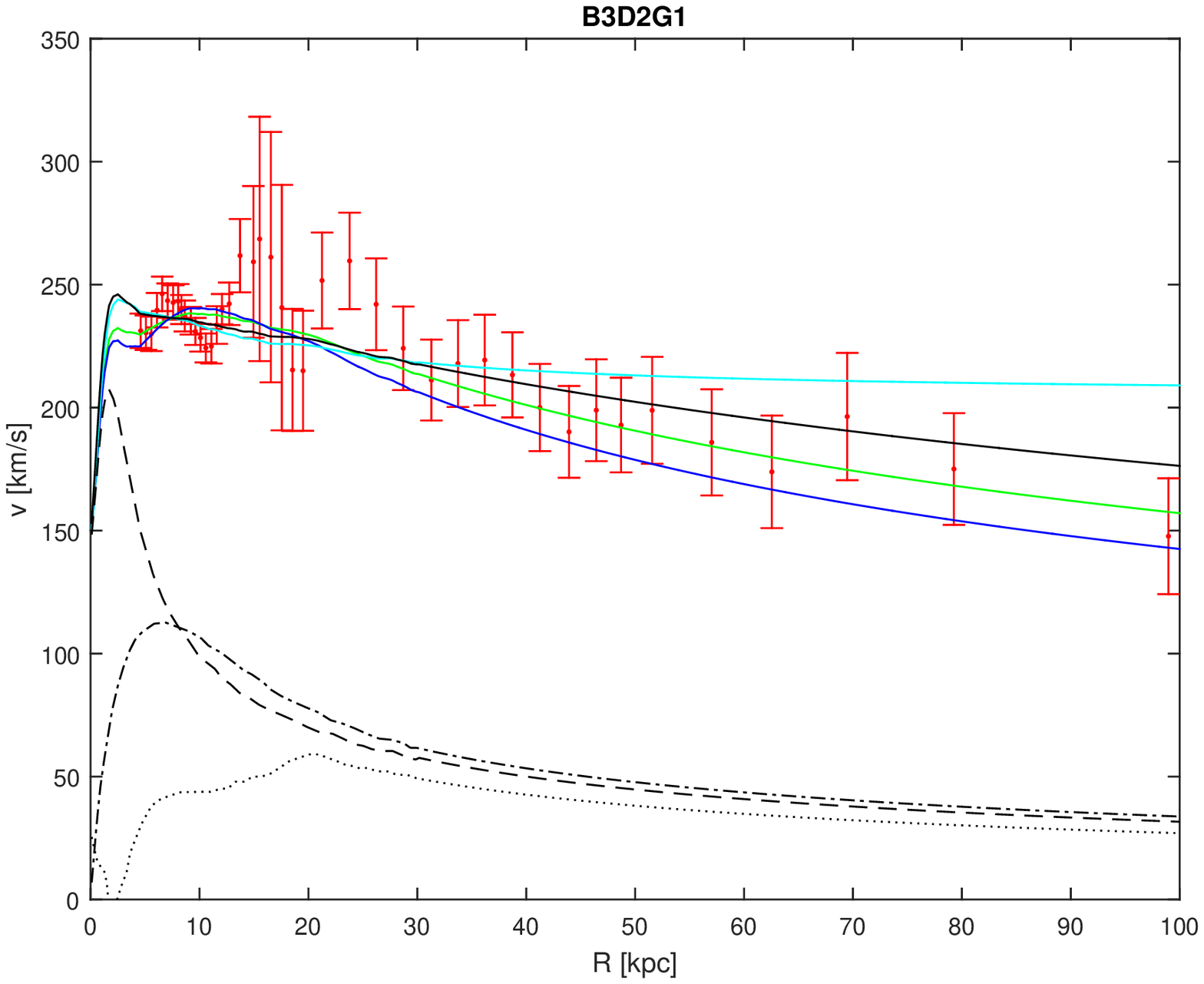}\hspace{0.5cm}
\includegraphics[width=0.4\textwidth]{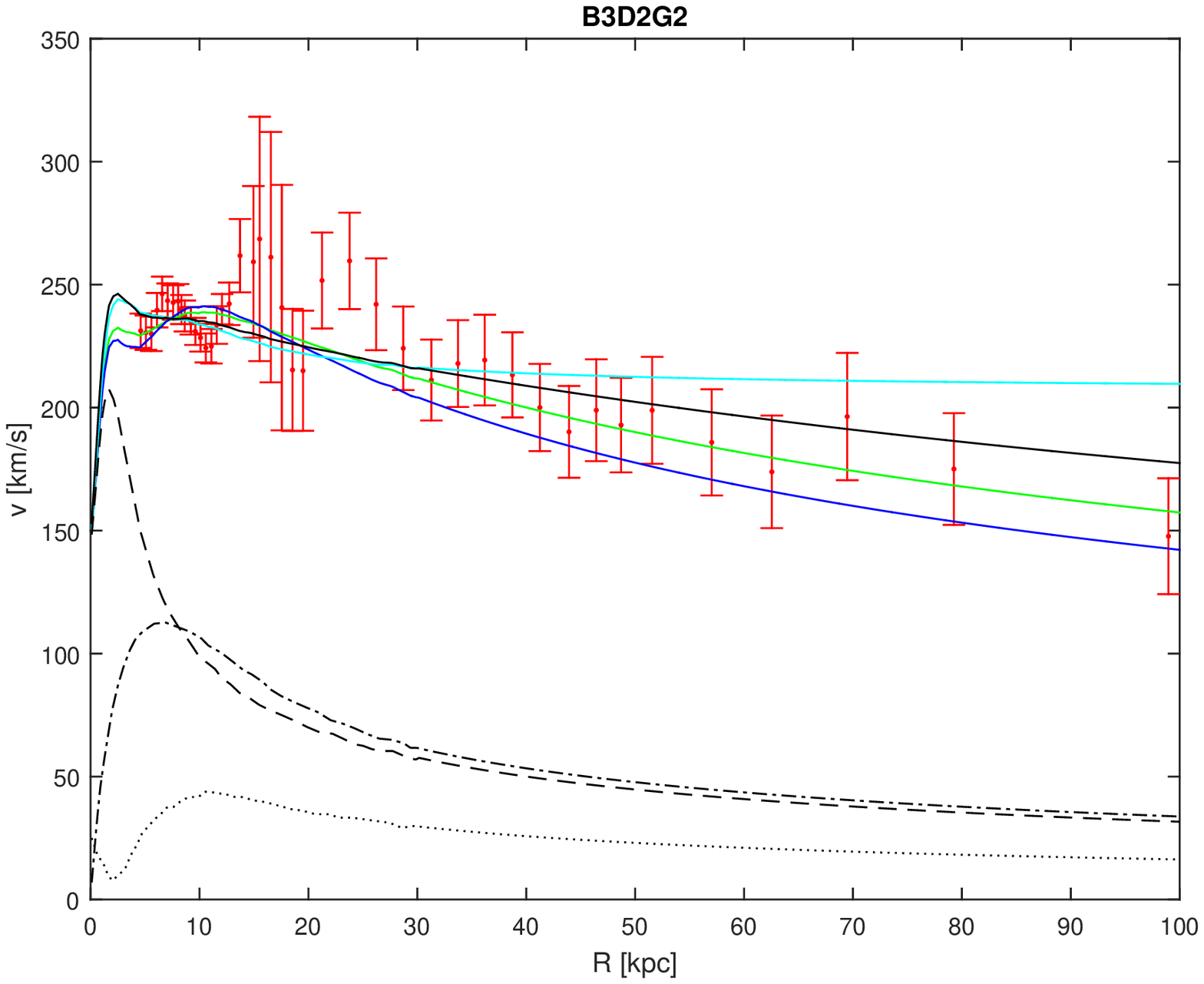}
\includegraphics[width=0.4\textwidth]{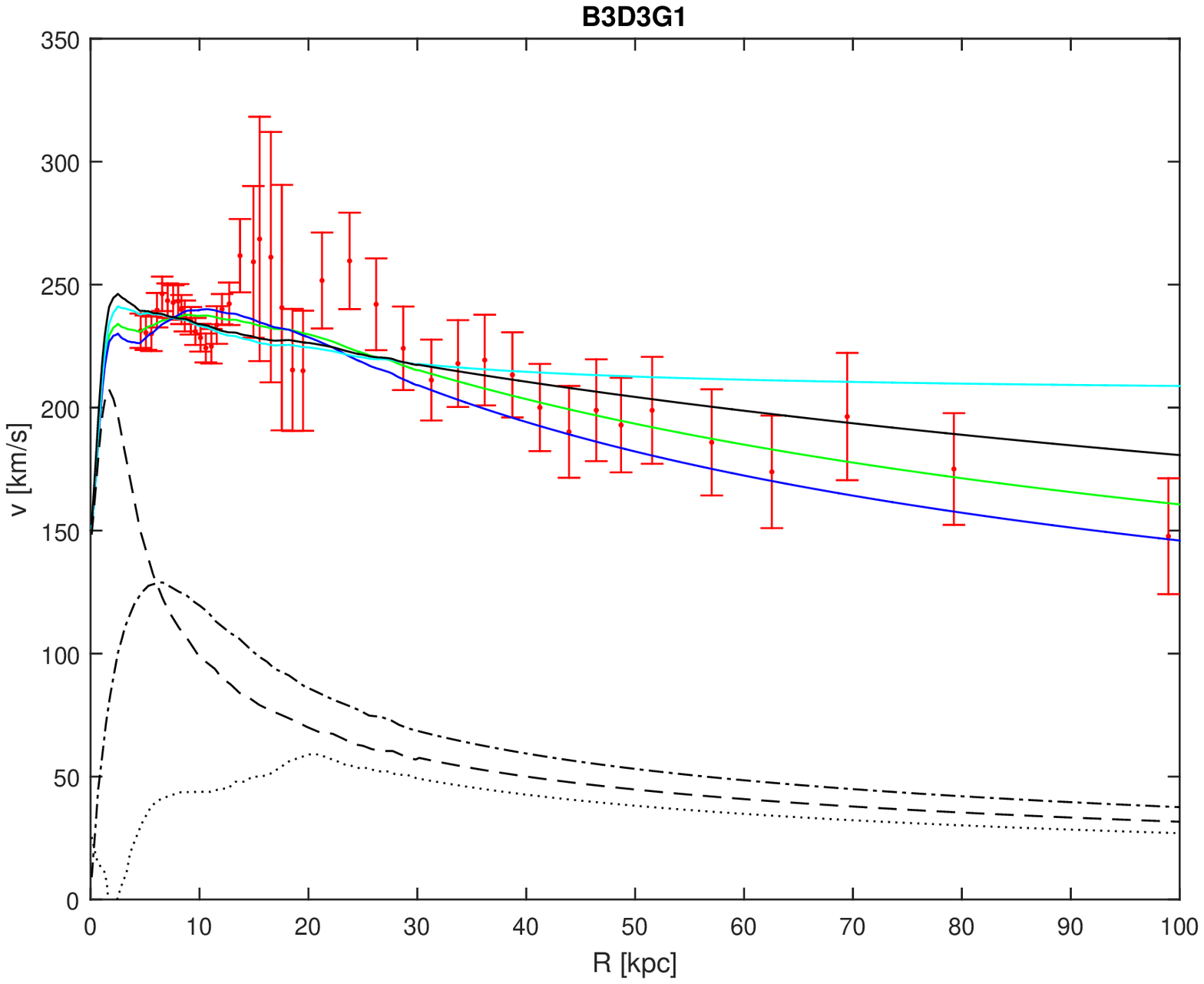}\hspace{0.5cm}
\includegraphics[width=0.4\textwidth]{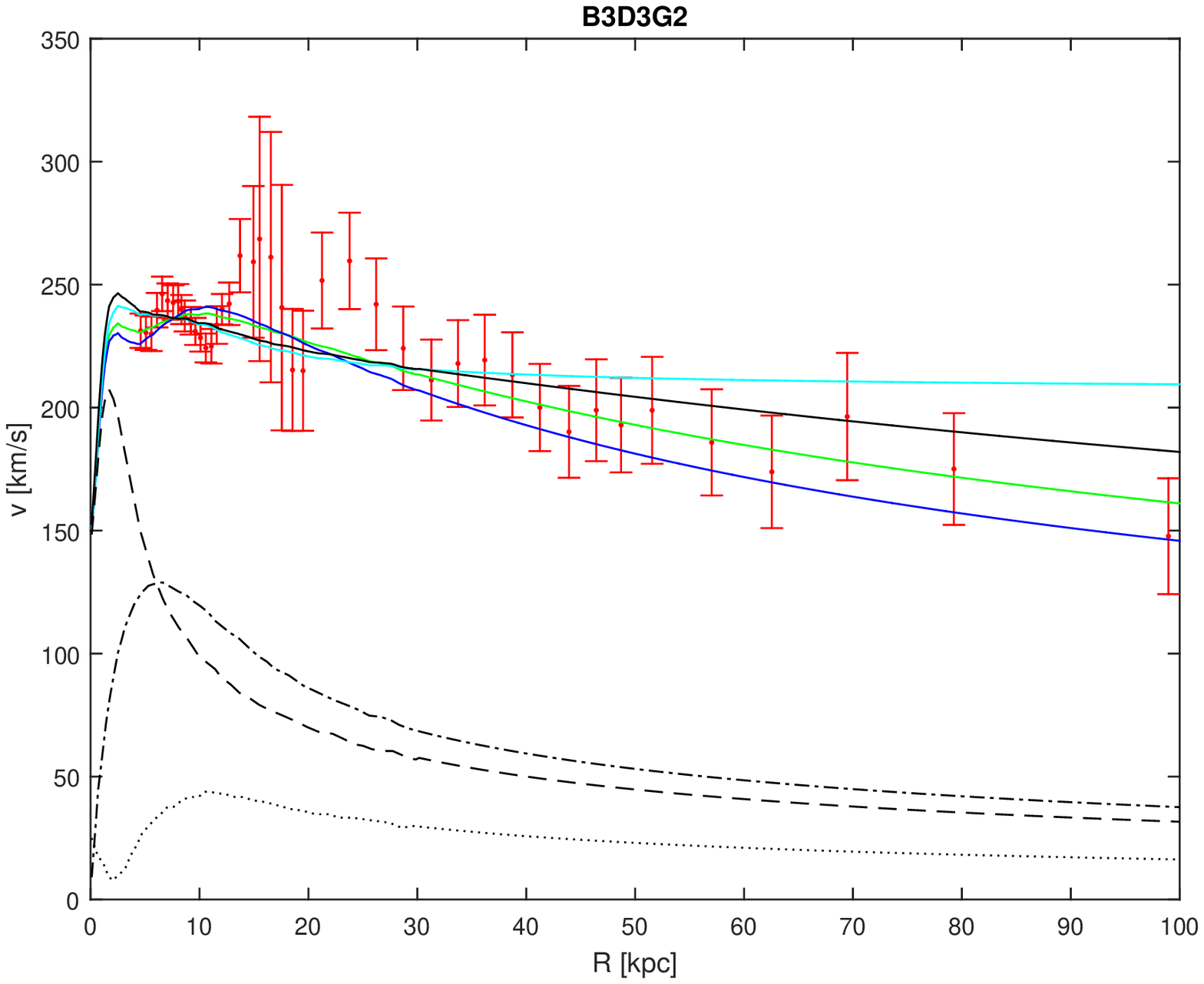}
\includegraphics[width=0.4\textwidth]{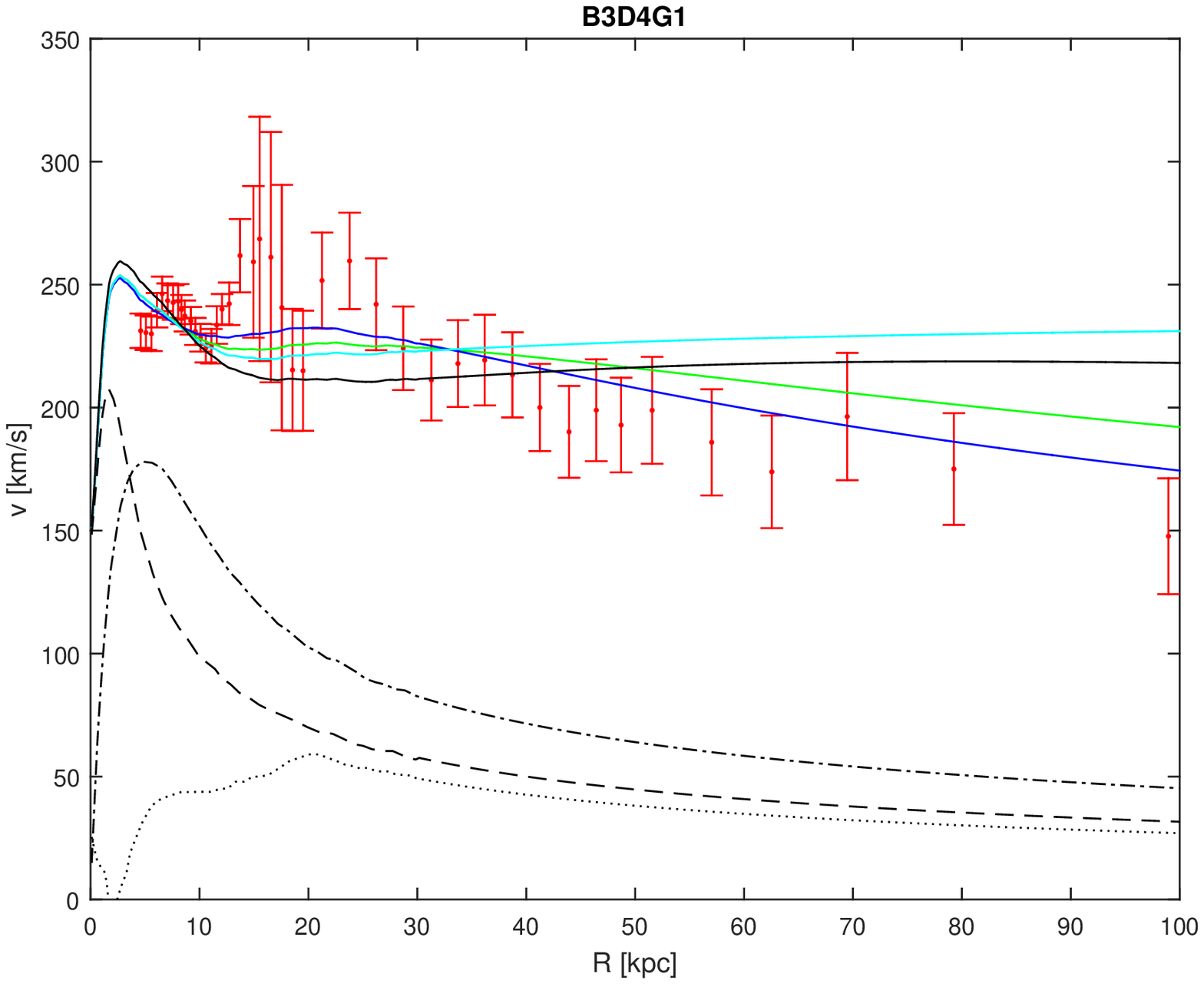}\hspace{0.5cm}
\includegraphics[width=0.4\textwidth]{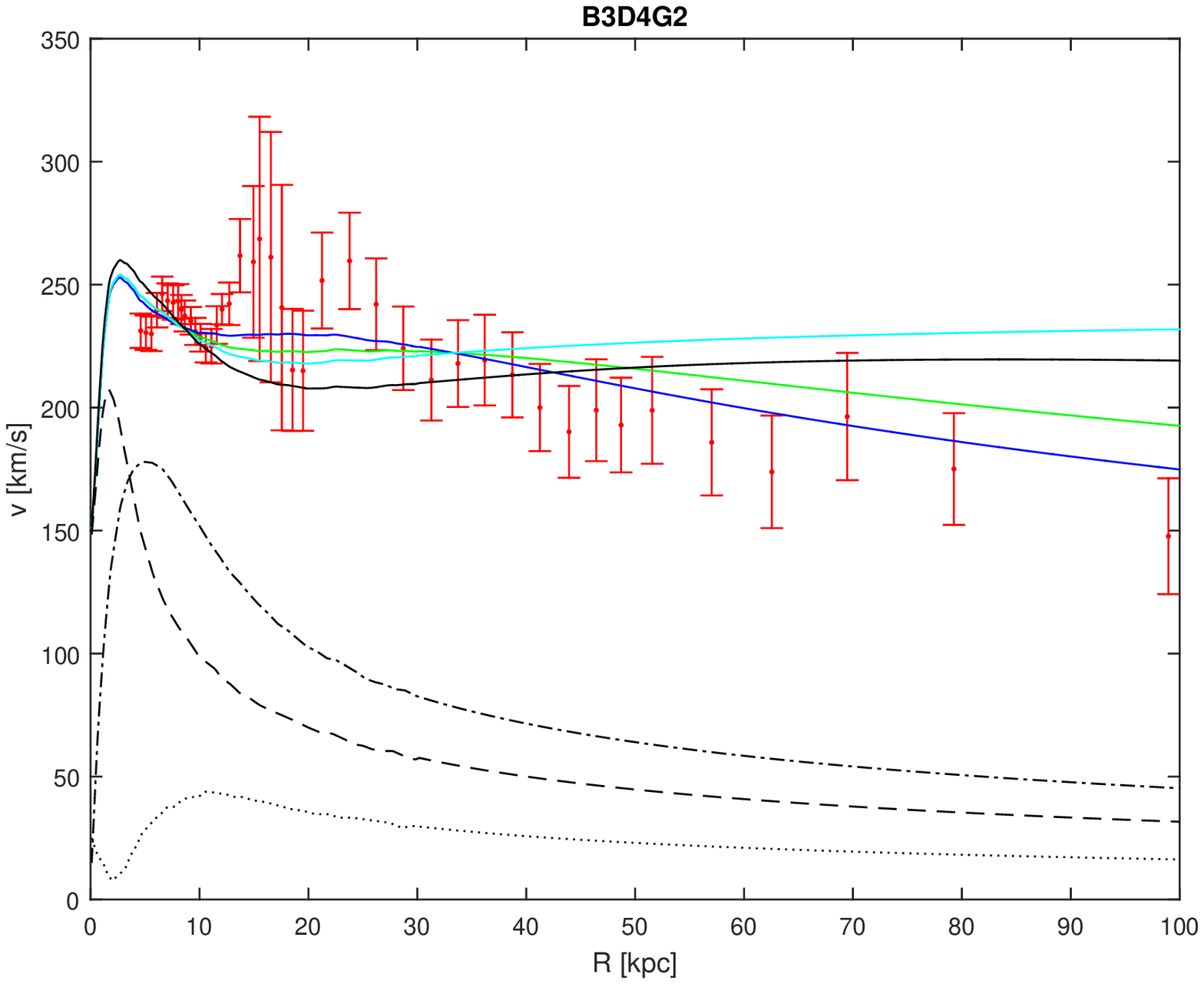}
\addtocounter{figure}{-1}
\caption{--continued}
\end{figure}

\begin{figure}
\centering
\includegraphics[width=0.4\textwidth]{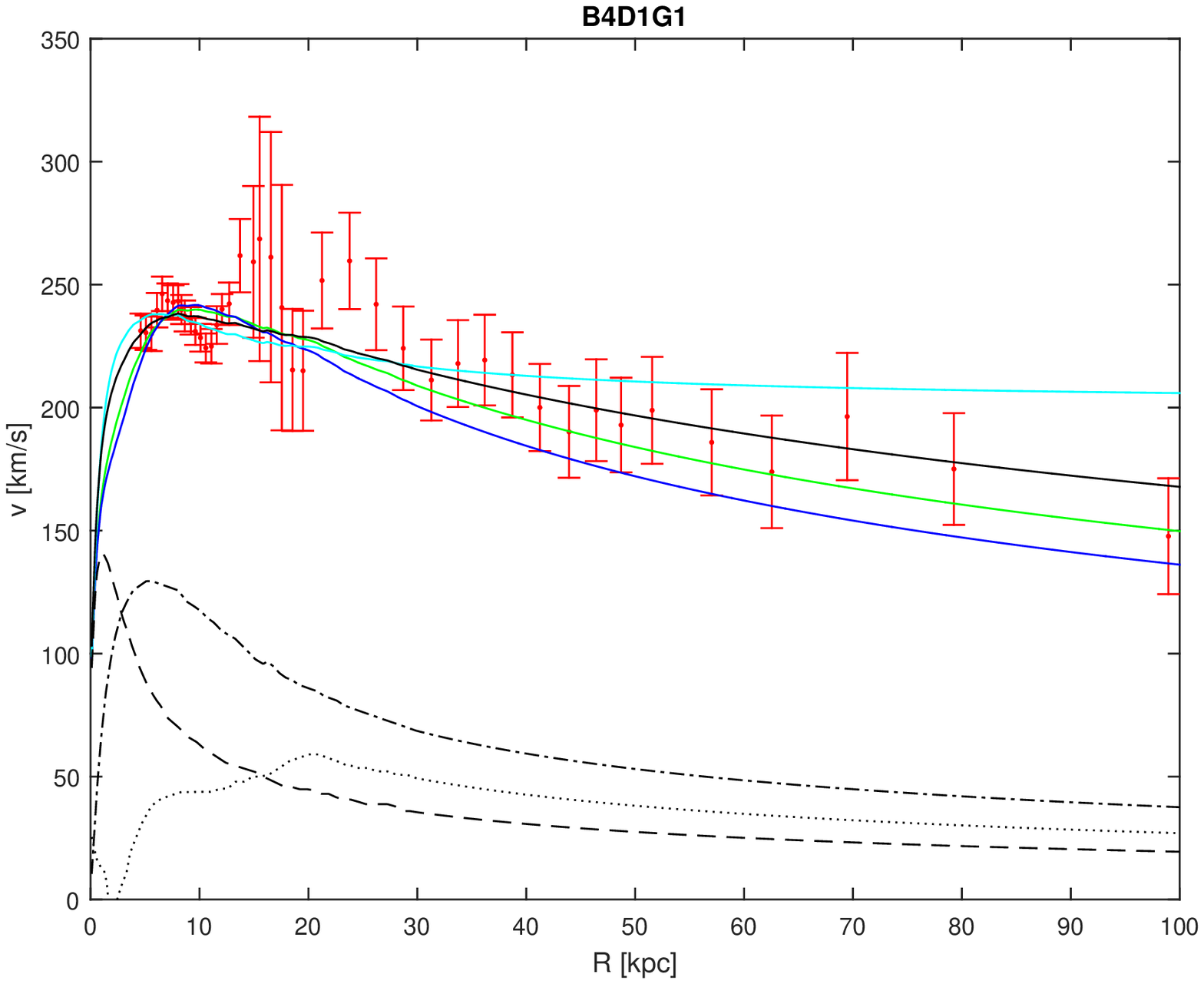}\hspace{0.5cm}
\includegraphics[width=0.4\textwidth]{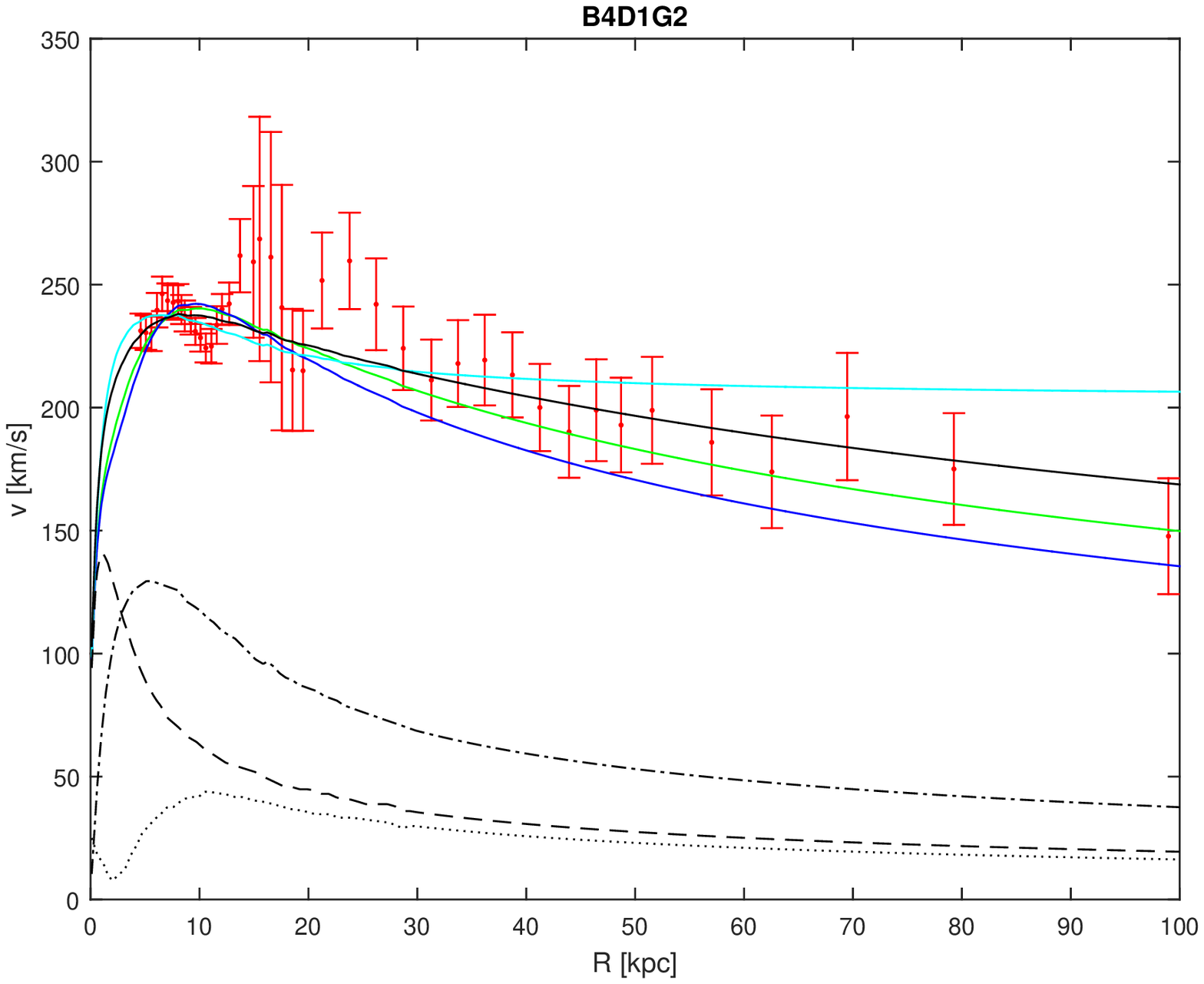}
\includegraphics[width=0.4\textwidth]{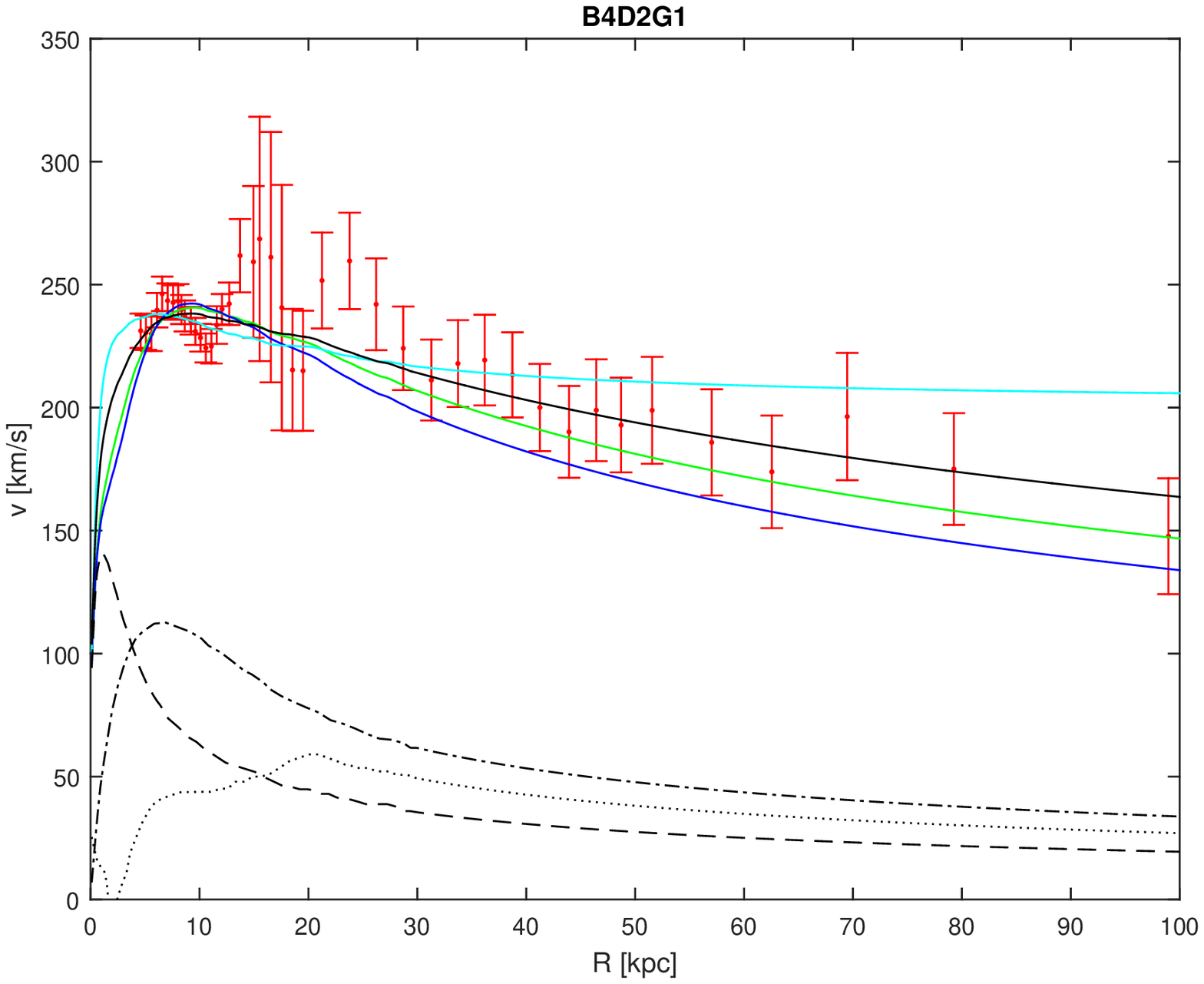}\hspace{0.5cm}
\includegraphics[width=0.4\textwidth]{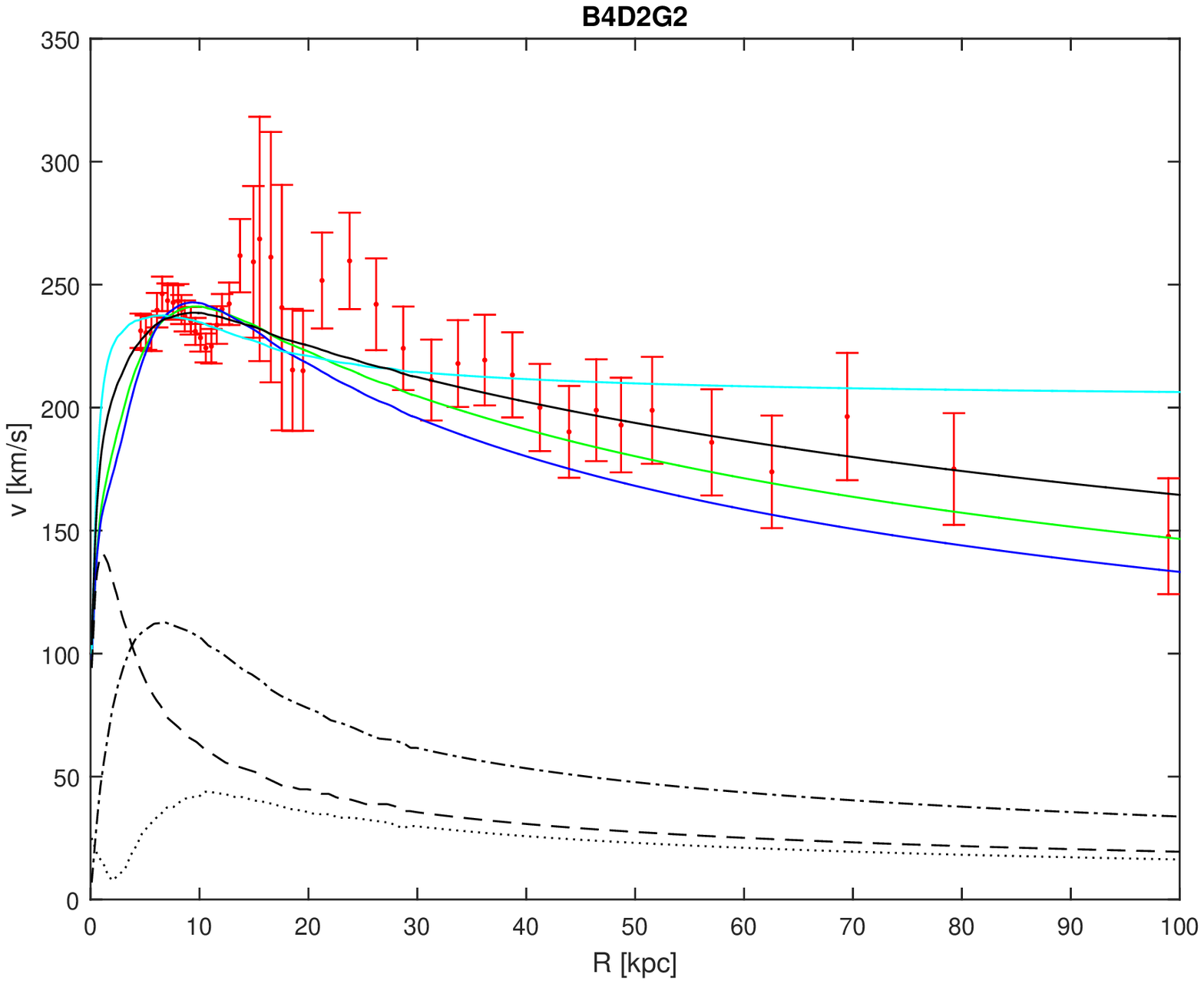}
\includegraphics[width=0.4\textwidth]{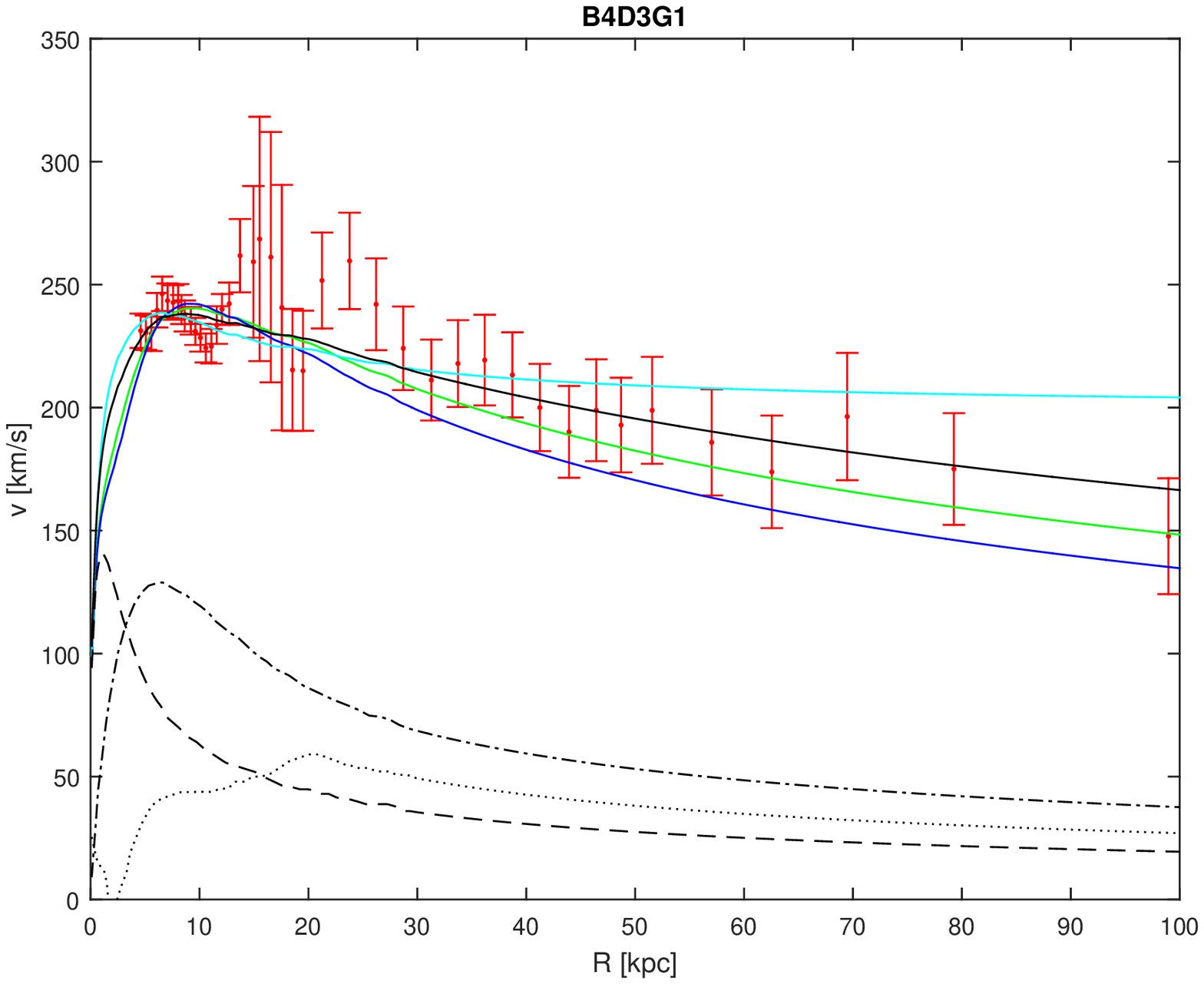}\hspace{0.5cm}
\includegraphics[width=0.4\textwidth]{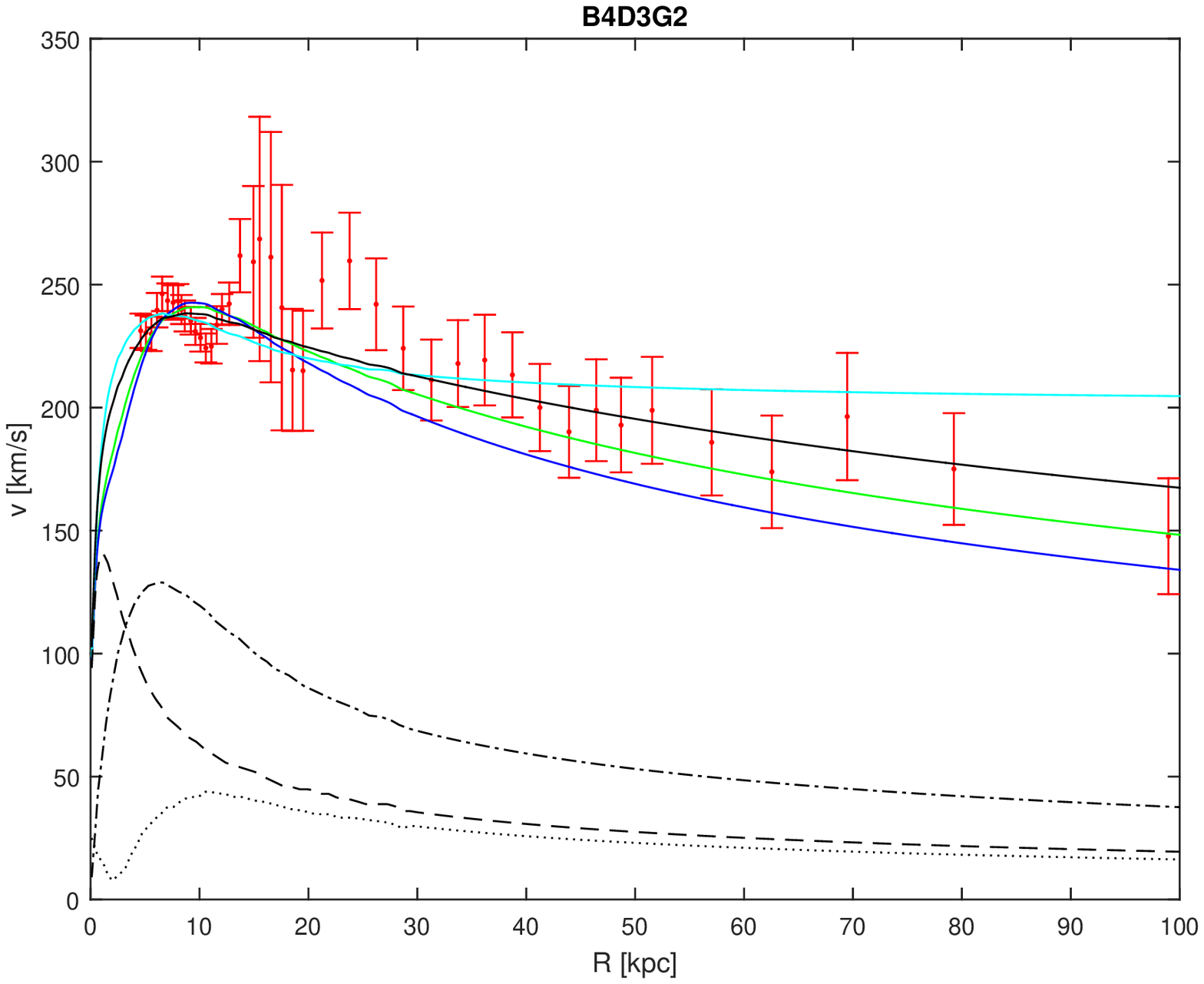}
\includegraphics[width=0.4\textwidth]{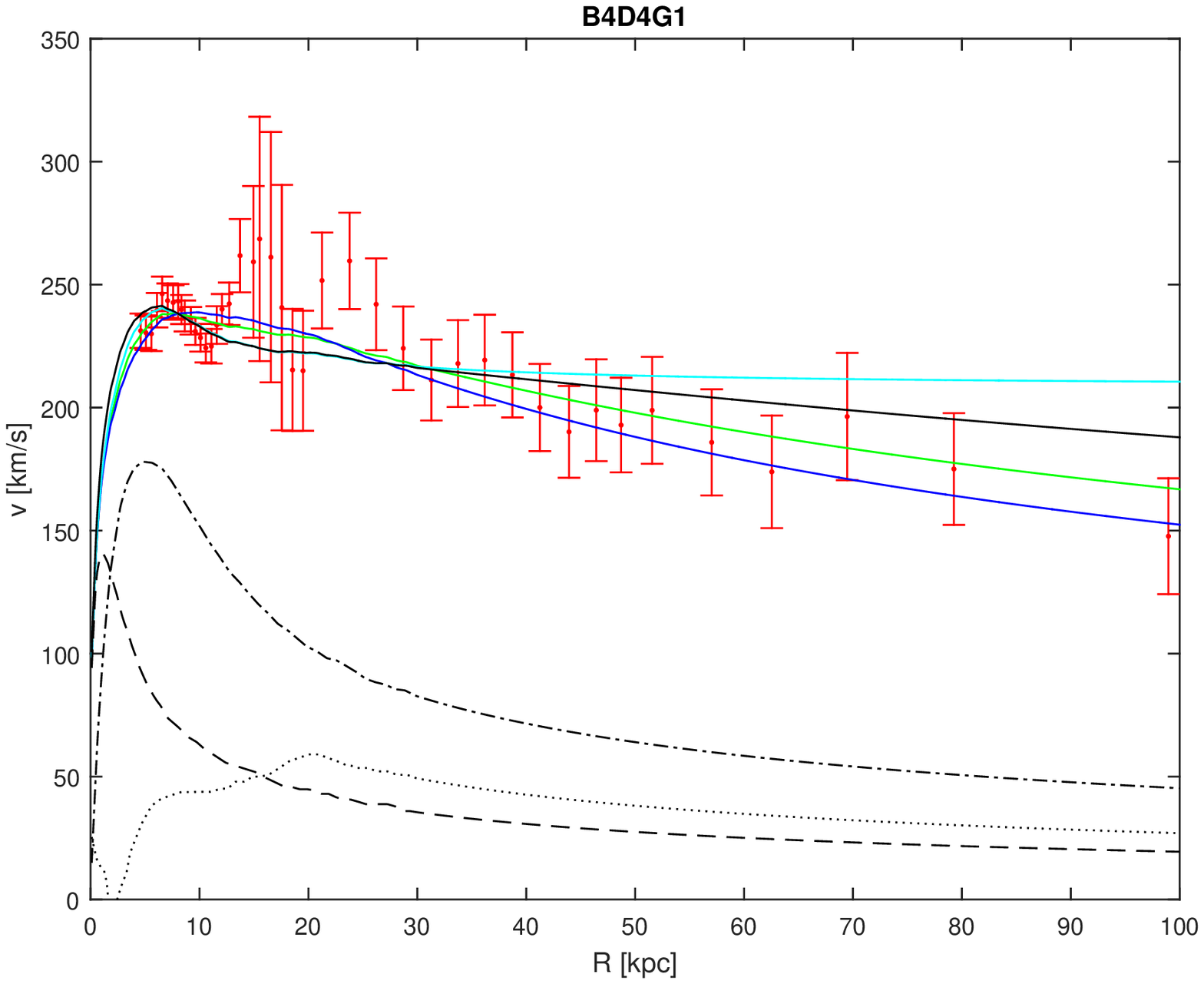}\hspace{0.5cm}
\includegraphics[width=0.4\textwidth]{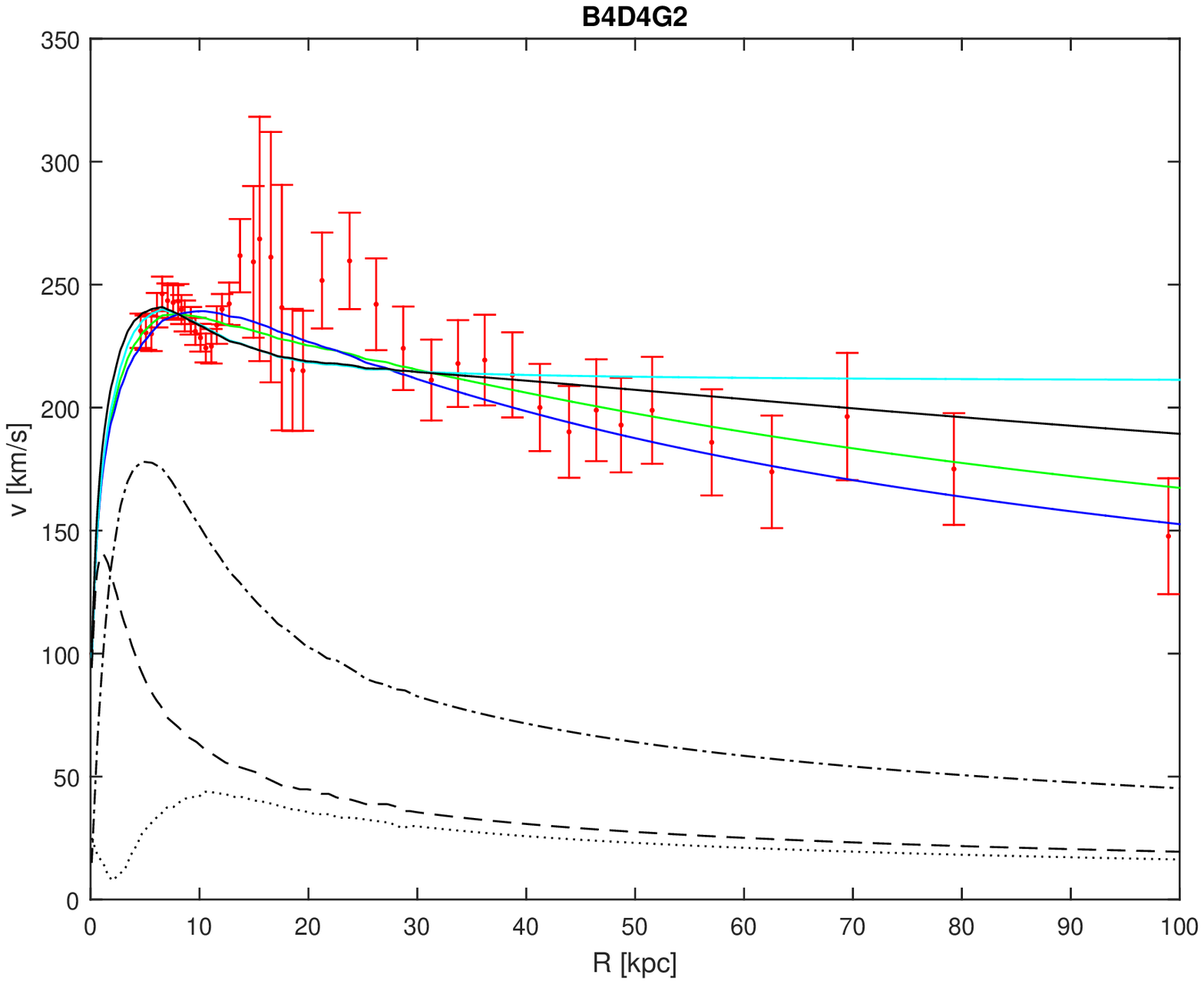}
\addtocounter{figure}{-1}
\caption{--continued}
\end{figure}

\begin{figure}
\centering
\includegraphics[width=0.4\textwidth]{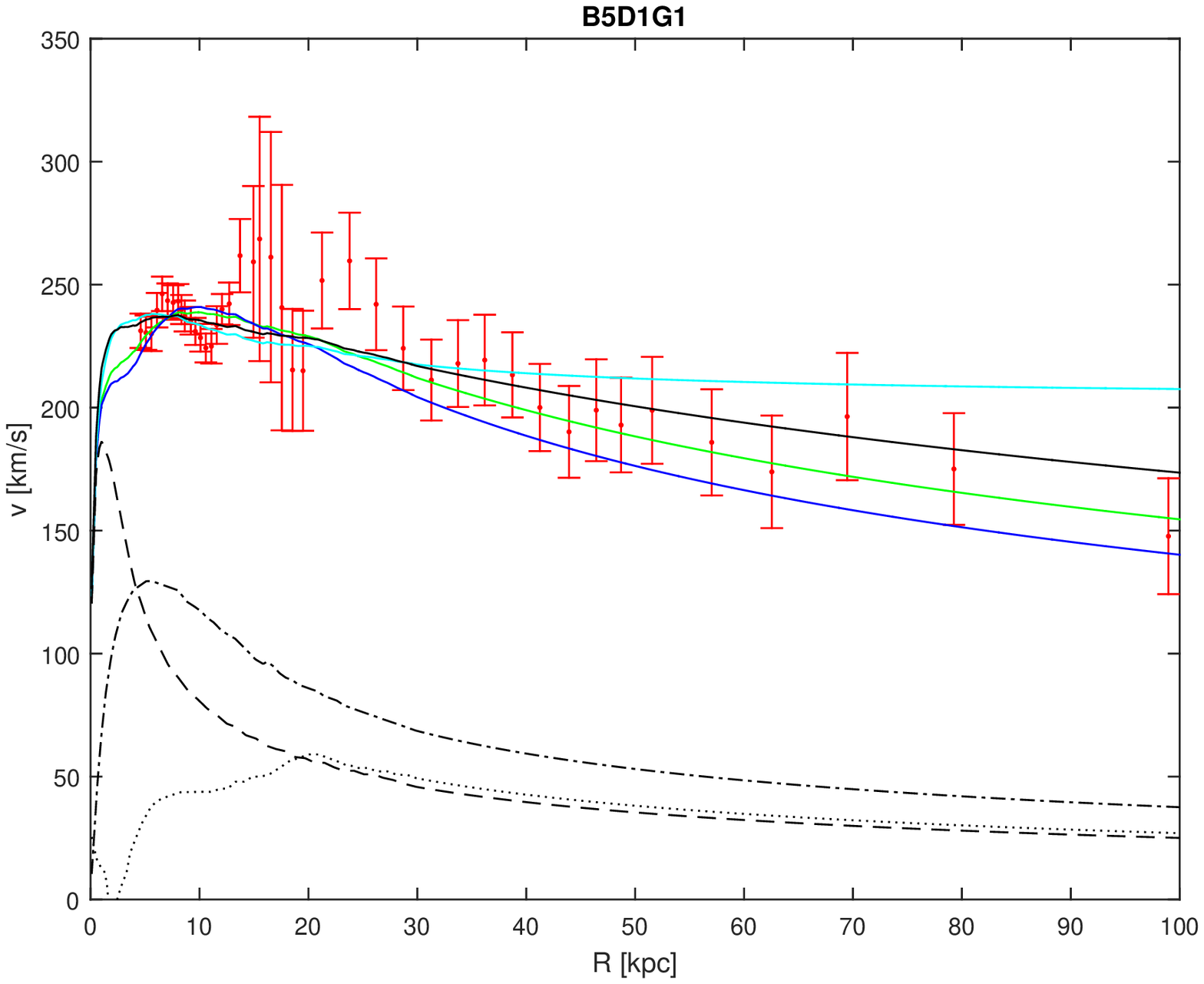}\hspace{0.5cm}
\includegraphics[width=0.4\textwidth]{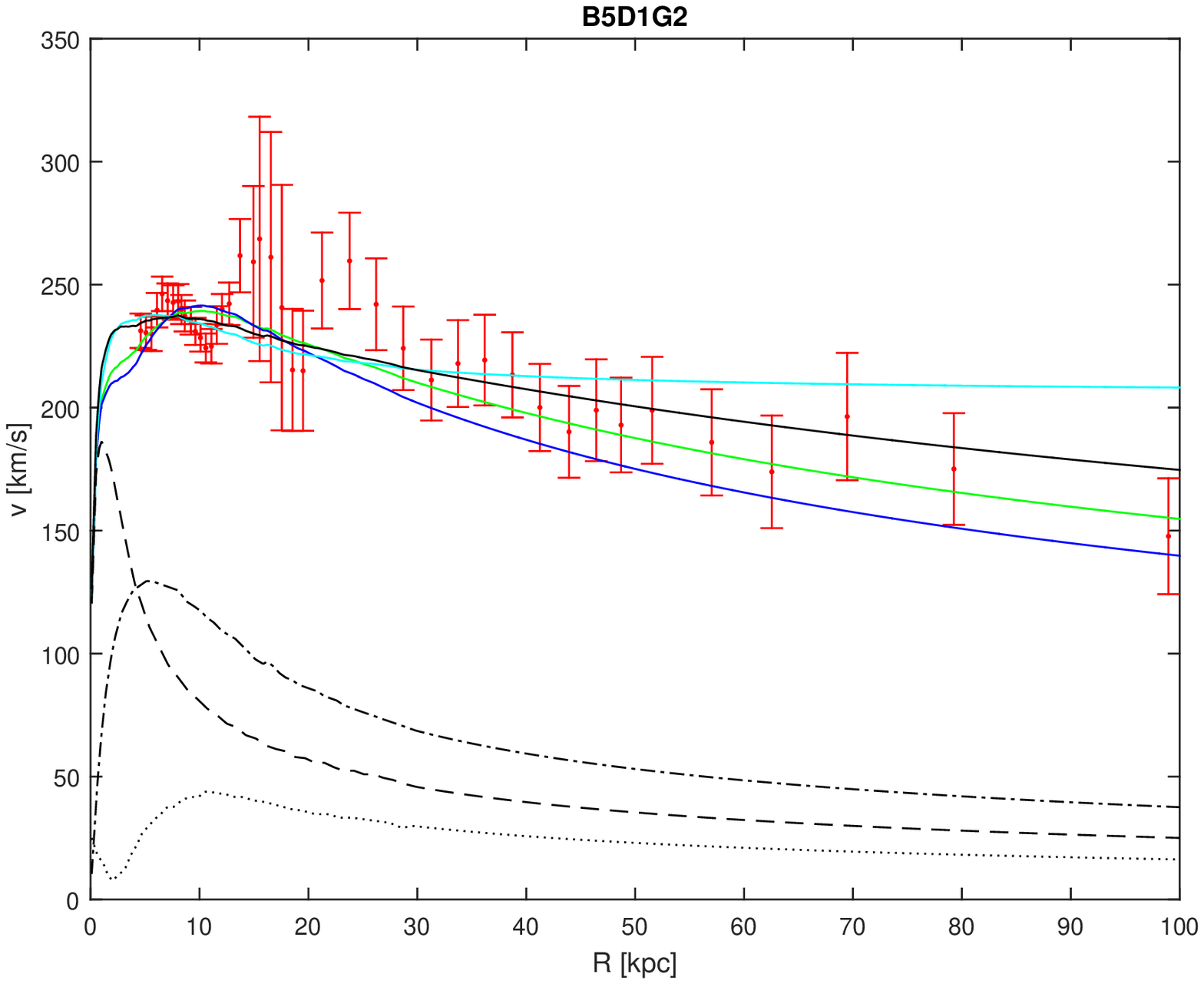}
\includegraphics[width=0.4\textwidth]{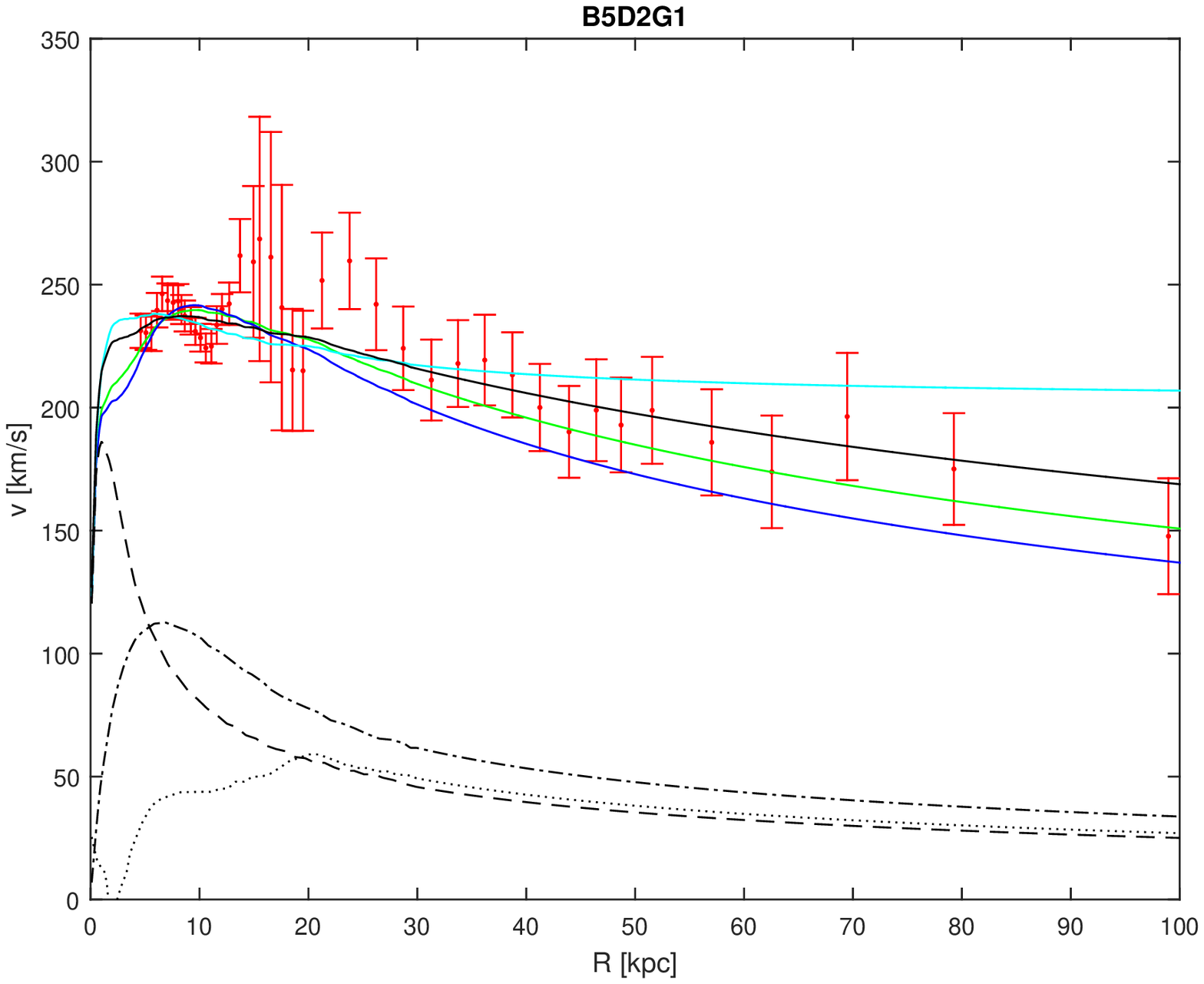}\hspace{0.5cm}
\includegraphics[width=0.4\textwidth]{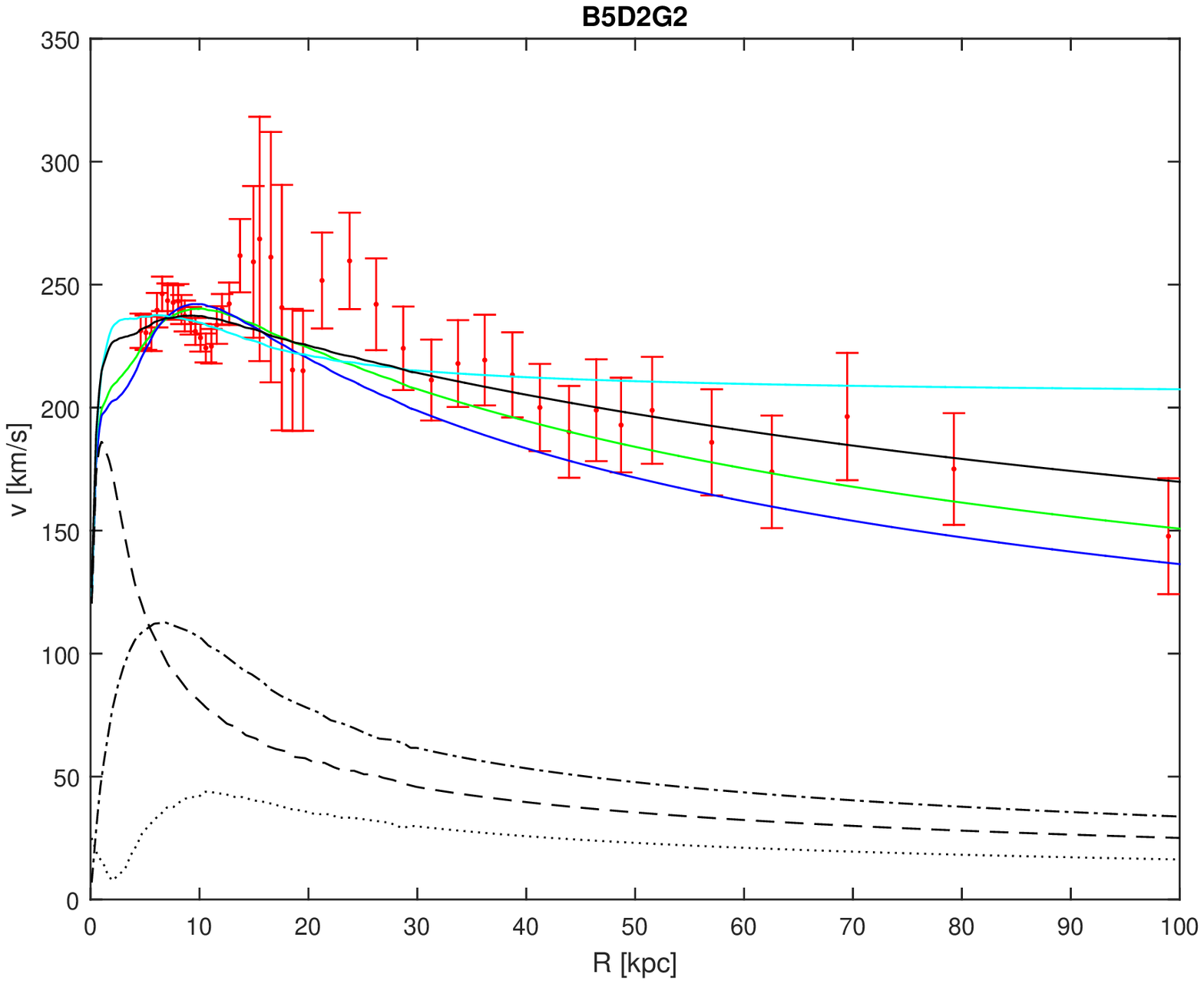}
\includegraphics[width=0.4\textwidth]{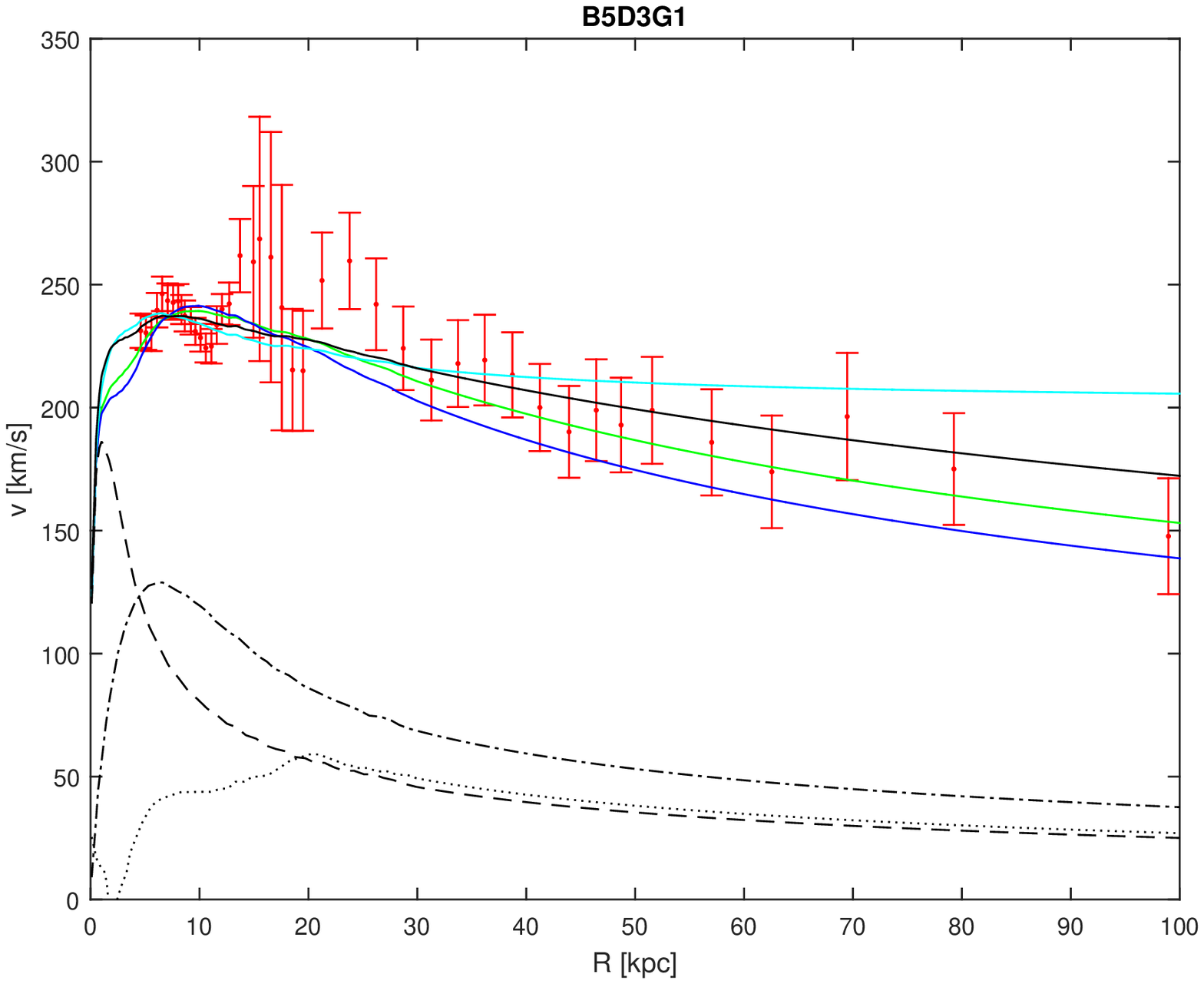}\hspace{0.5cm}
\includegraphics[width=0.4\textwidth]{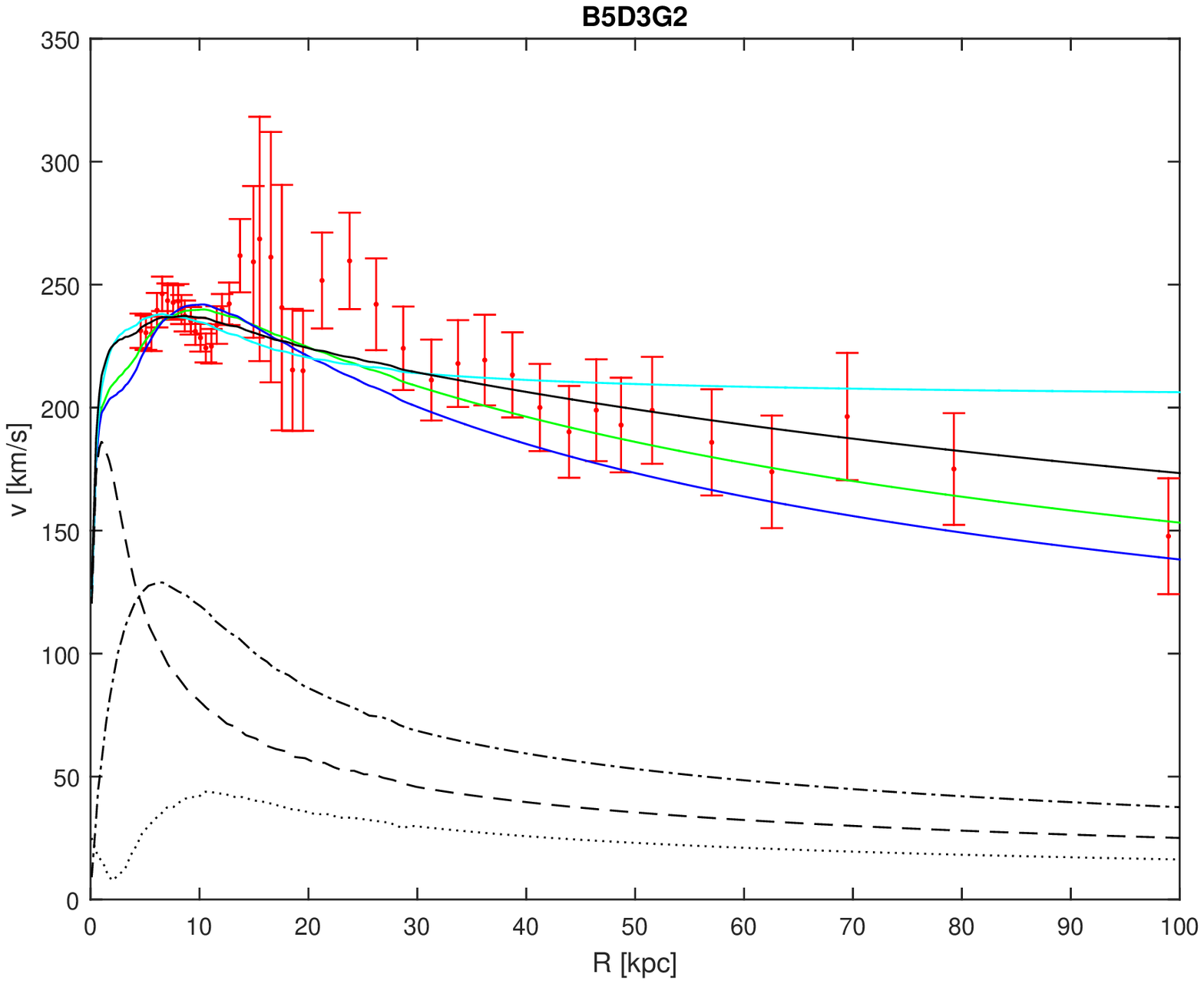}
\includegraphics[width=0.4\textwidth]{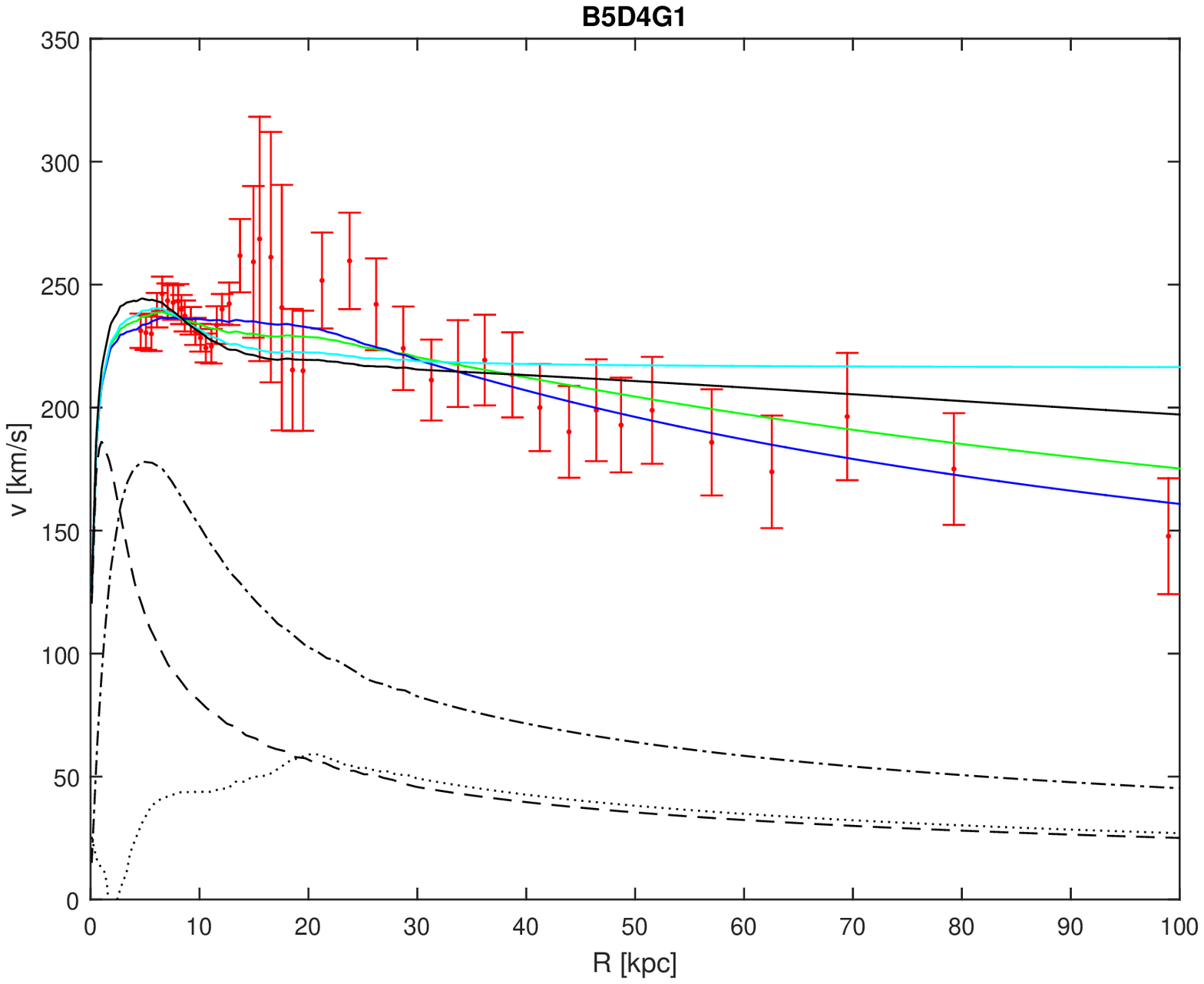}\hspace{0.5cm}
\includegraphics[width=0.4\textwidth]{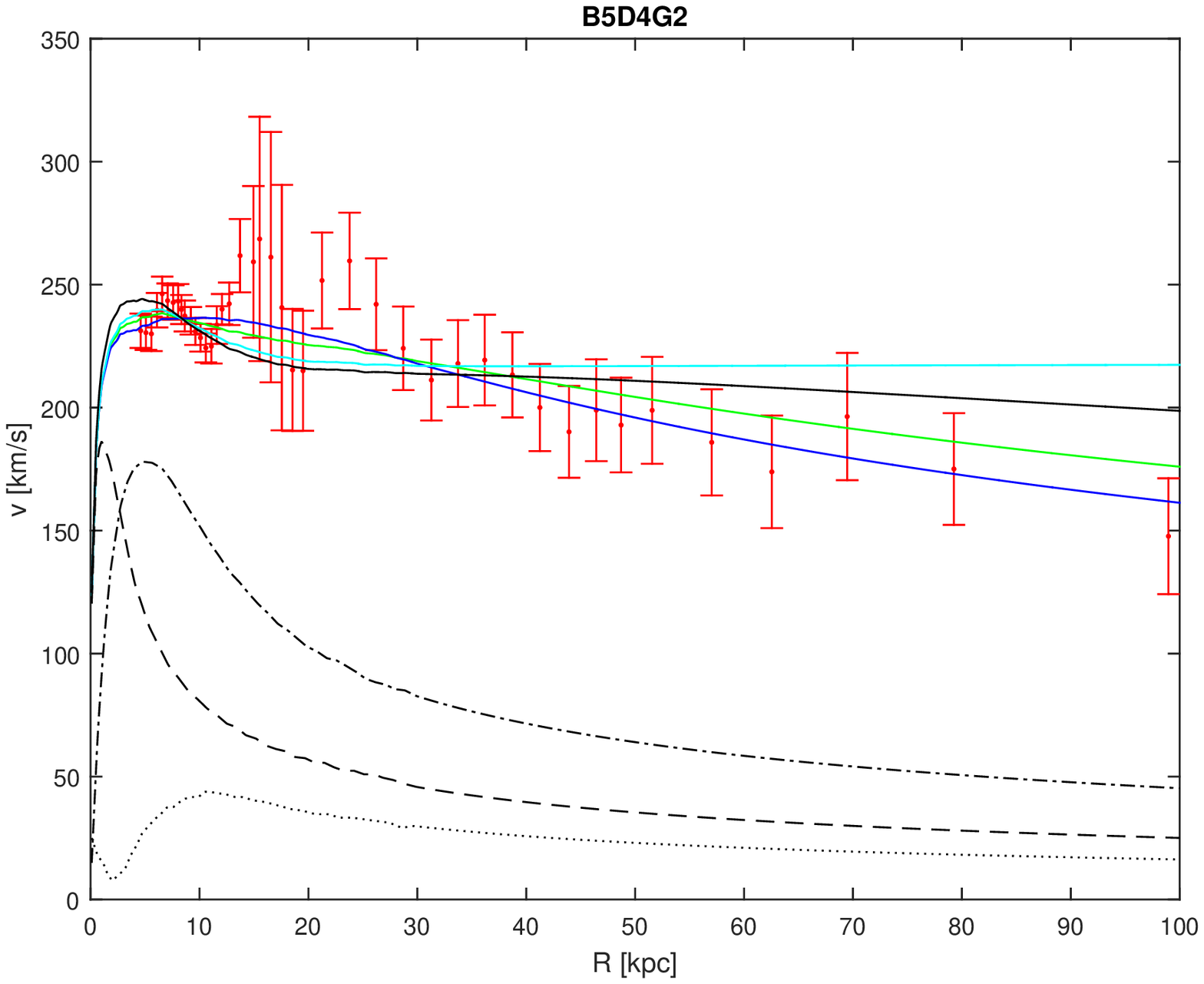}
\addtocounter{figure}{-1}
\caption{--continued}
\end{figure}

\begin{figure}
\centering
\includegraphics[width=0.4\textwidth]{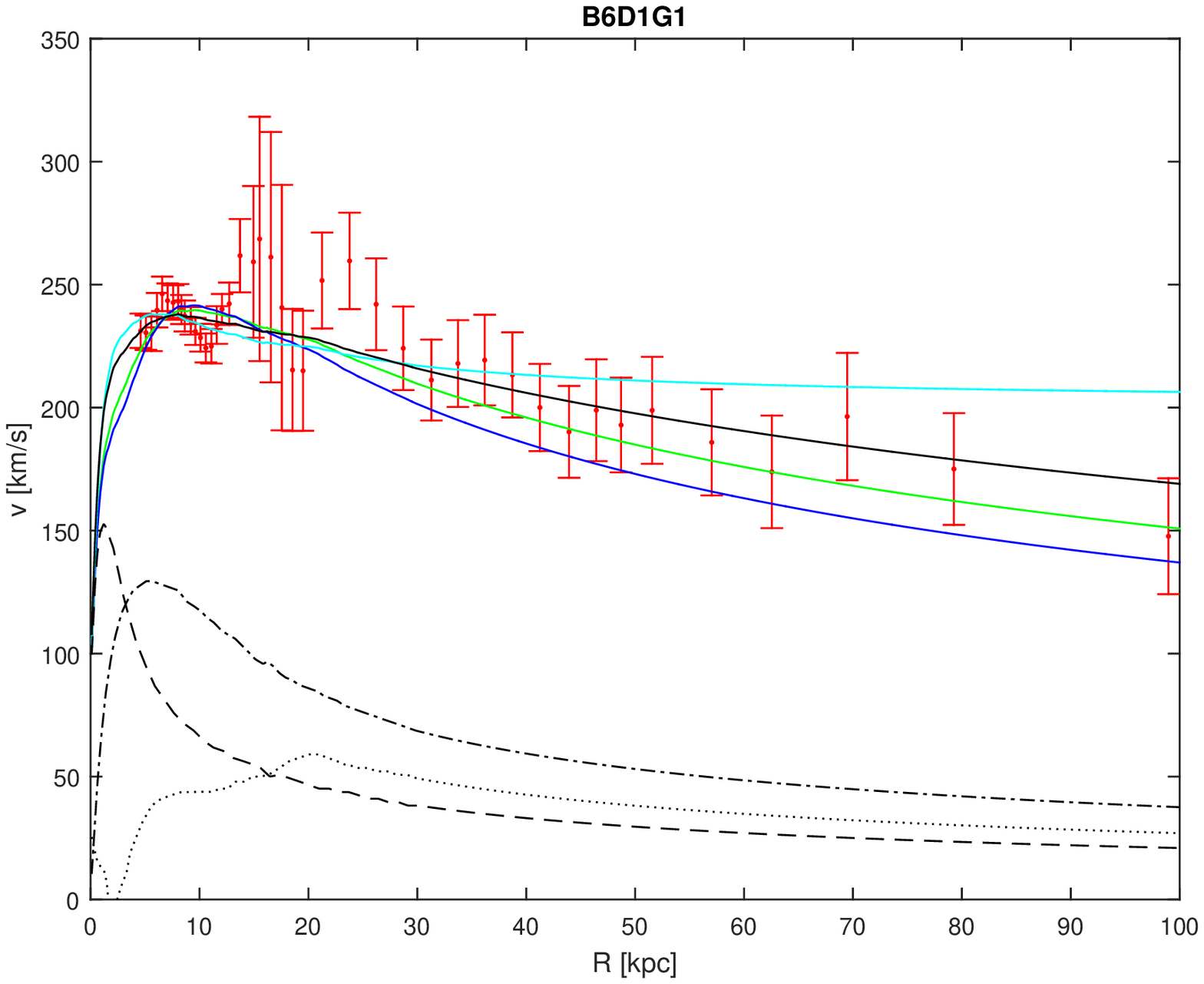}\hspace{0.5cm}
\includegraphics[width=0.4\textwidth]{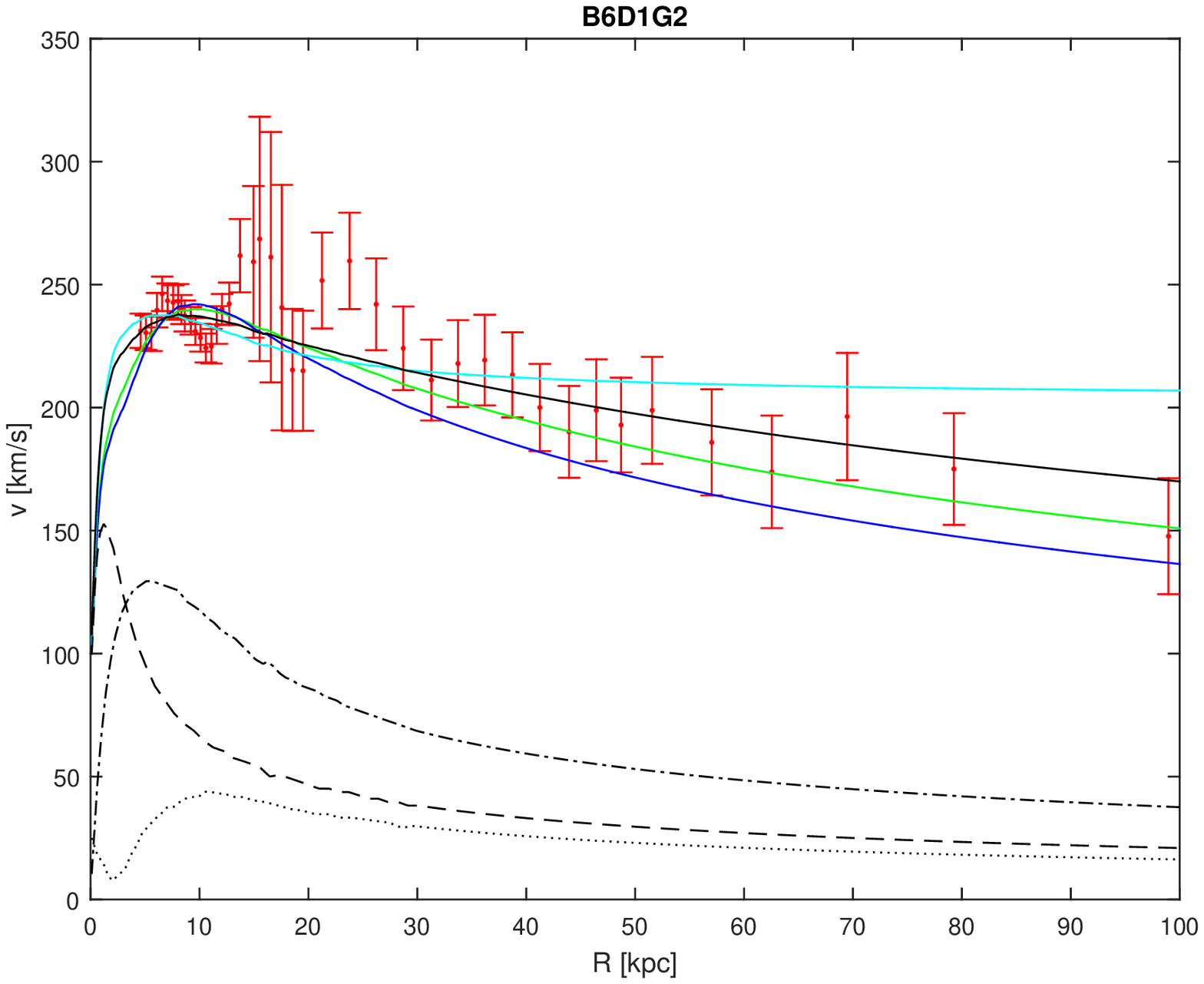}
\includegraphics[width=0.4\textwidth]{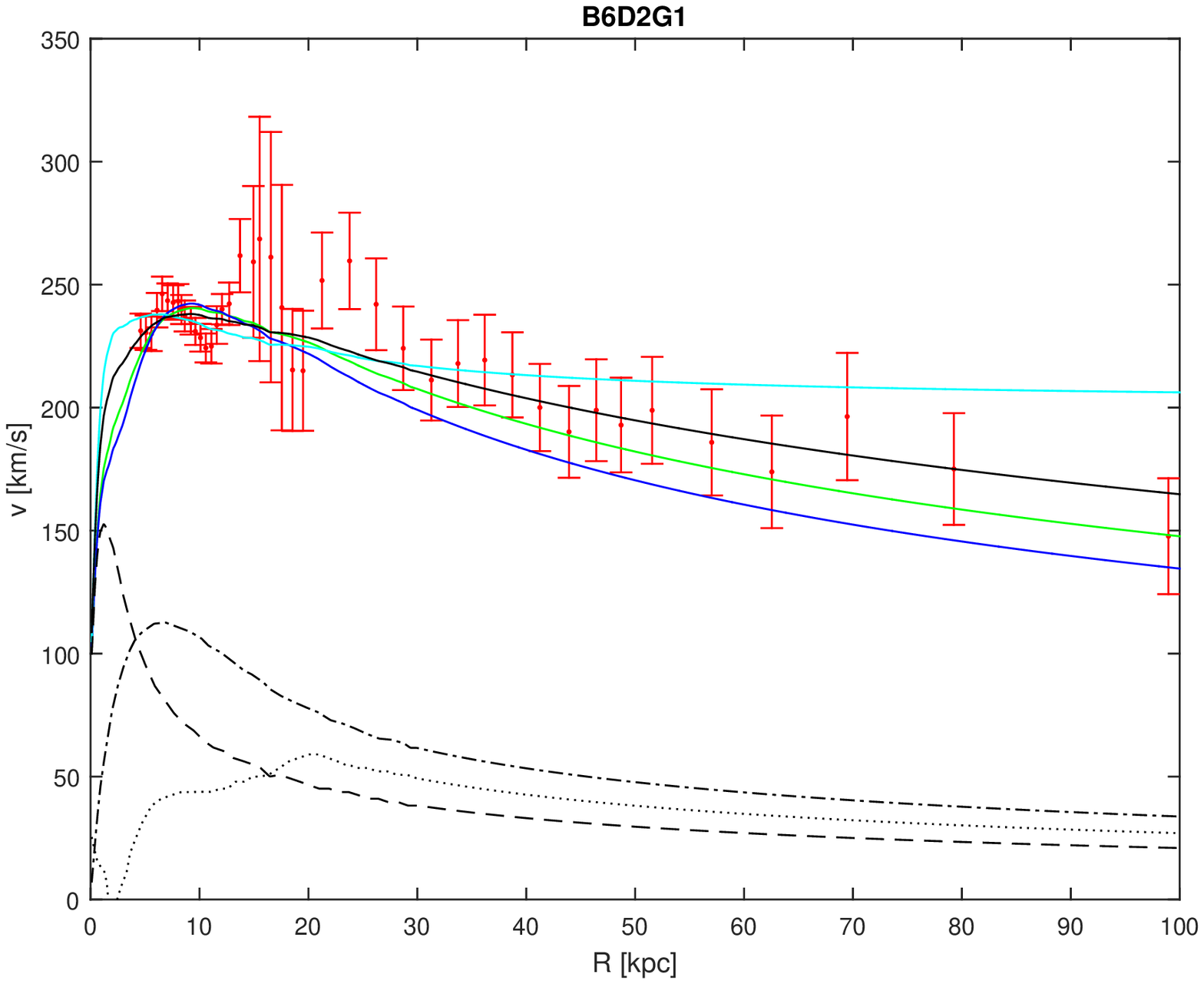}\hspace{0.5cm}
\includegraphics[width=0.4\textwidth]{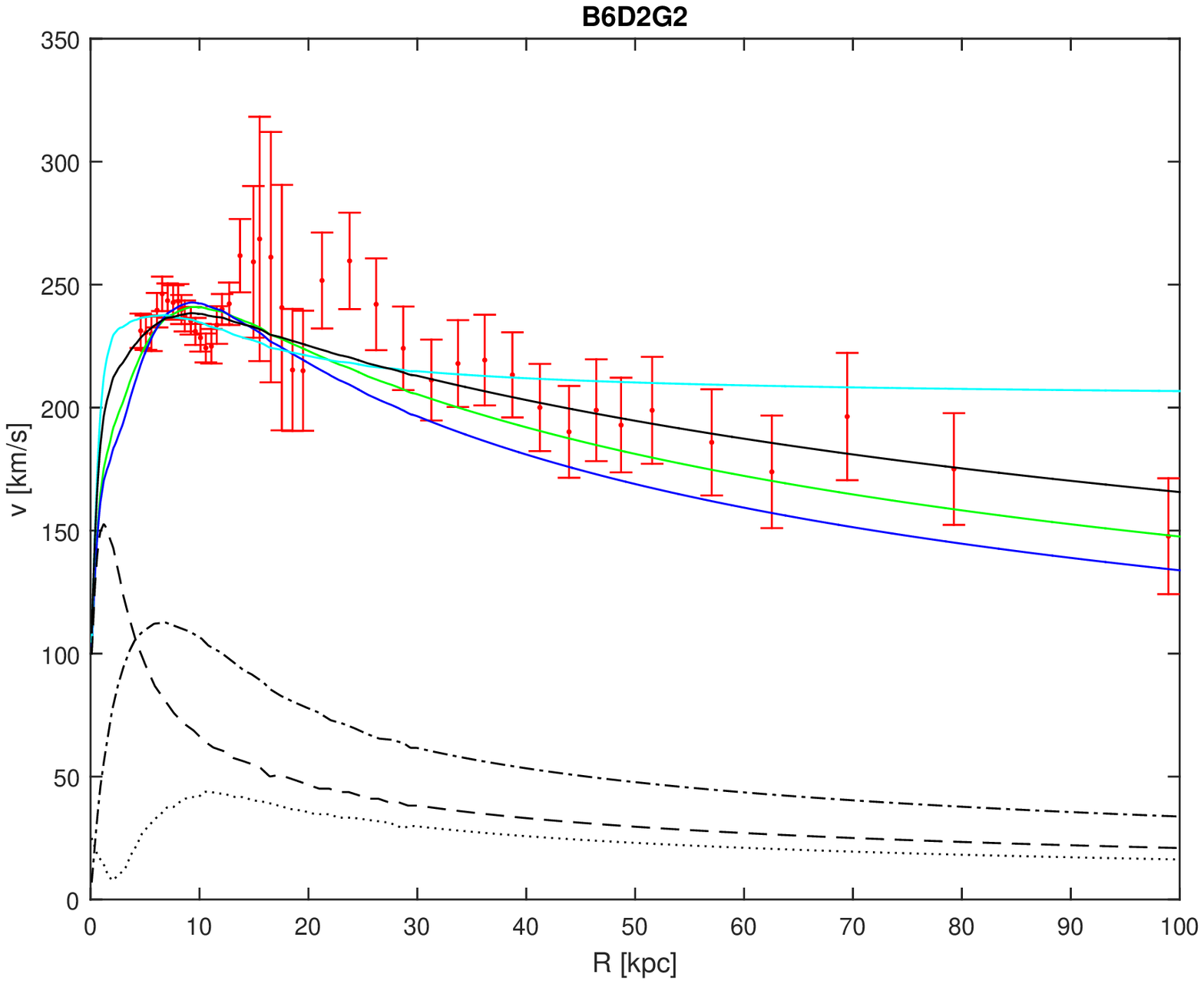}
\includegraphics[width=0.4\textwidth]{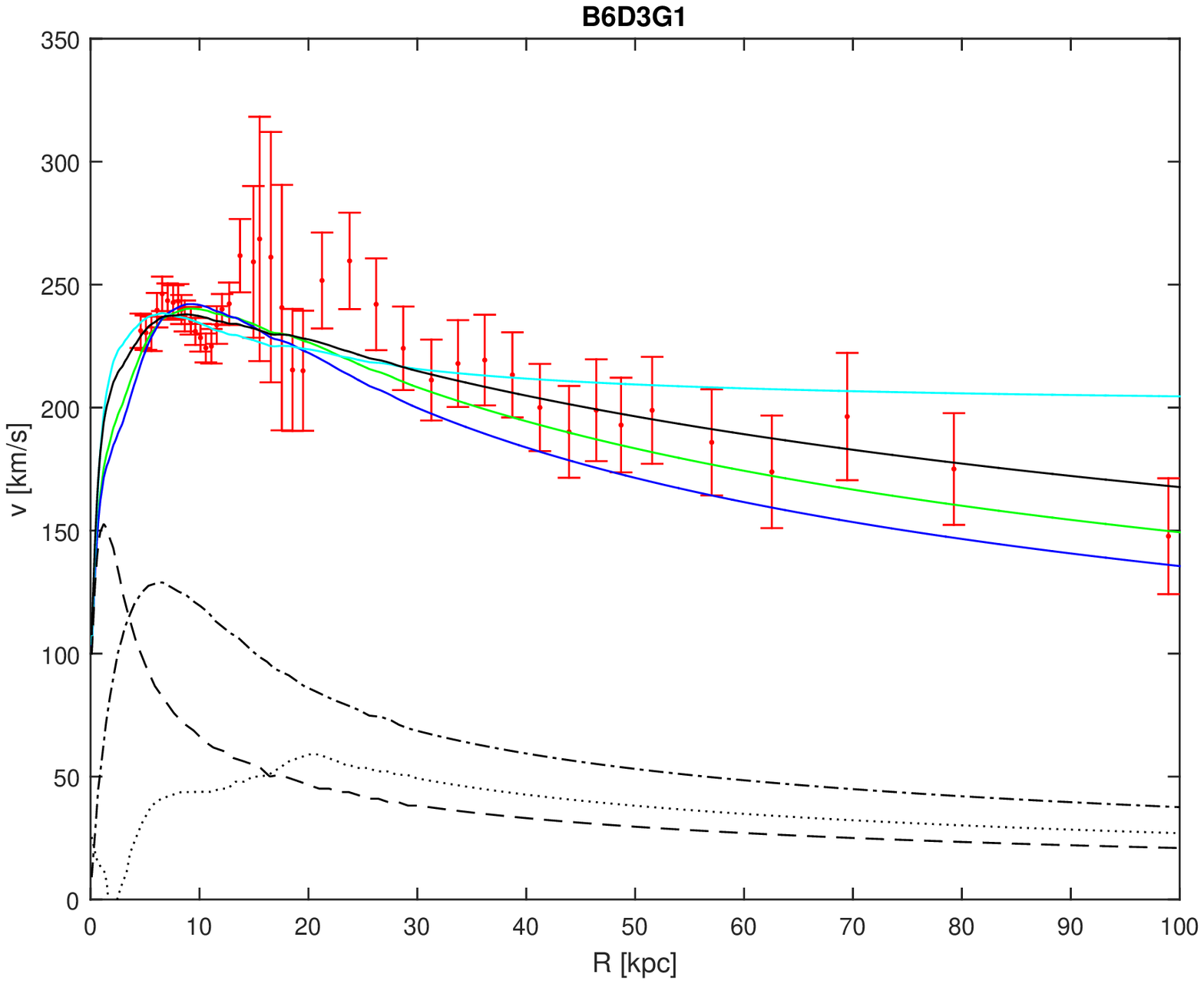}\hspace{0.5cm}
\includegraphics[width=0.4\textwidth]{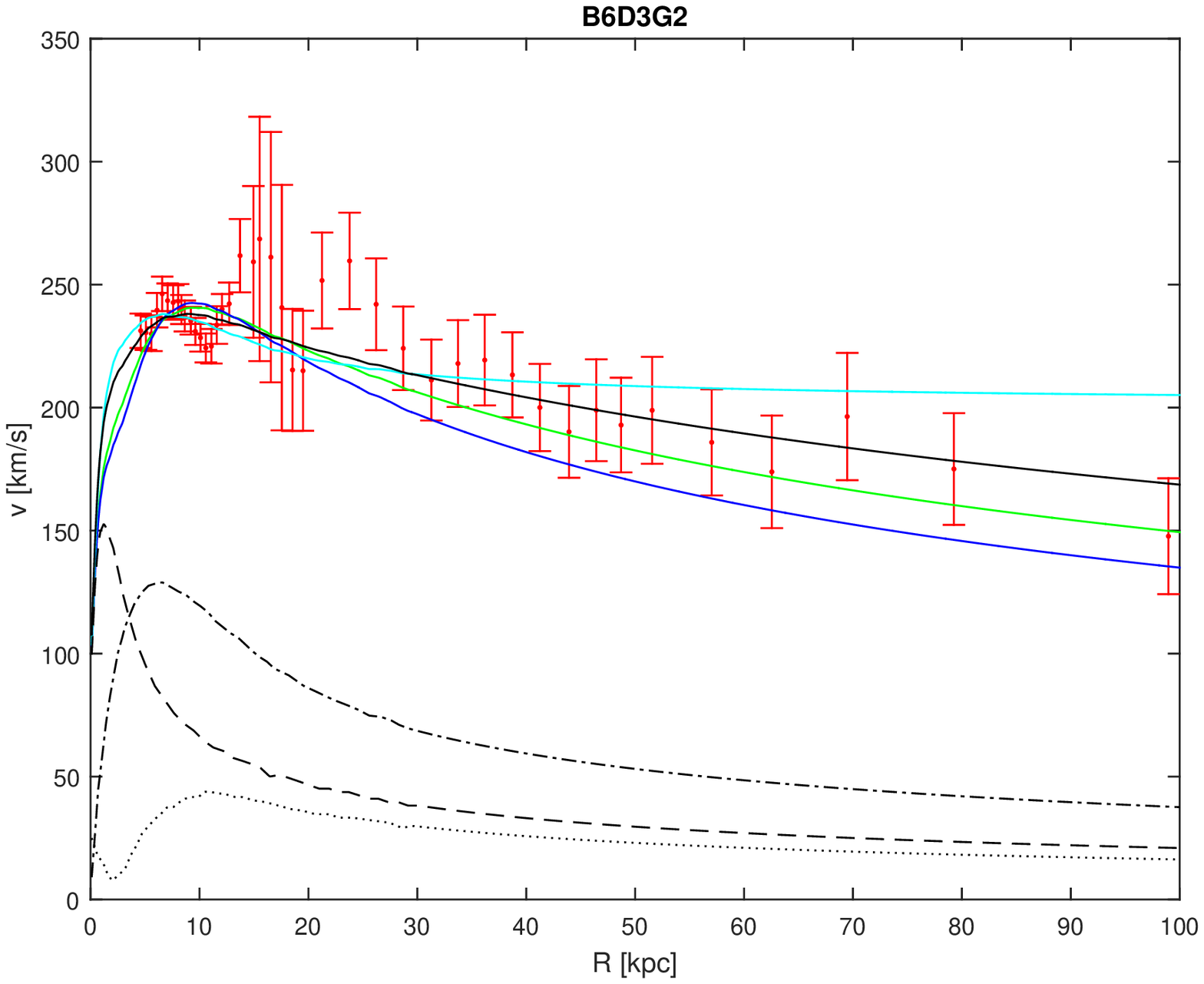}
\includegraphics[width=0.4\textwidth]{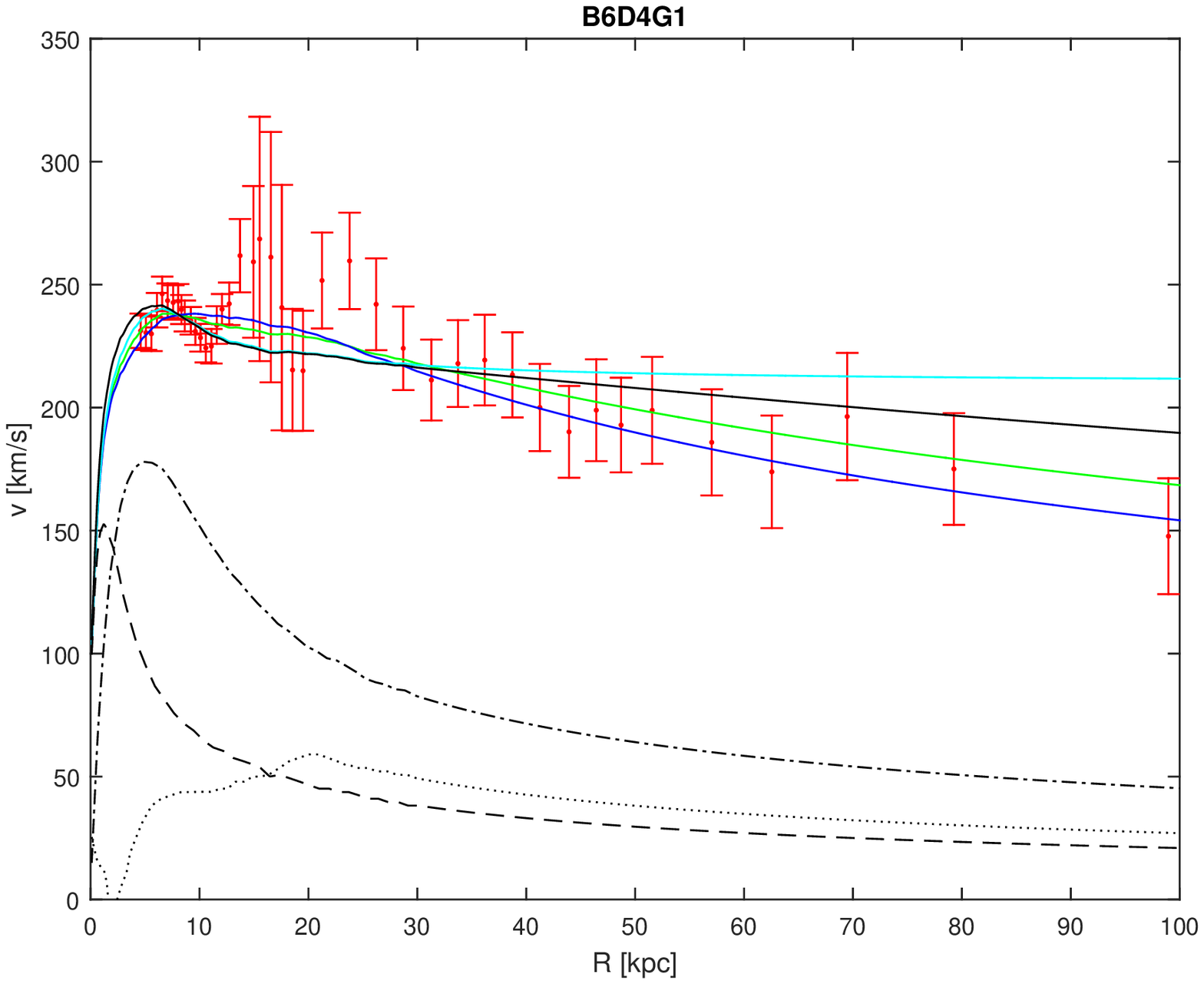}\hspace{0.5cm}
\includegraphics[width=0.4\textwidth]{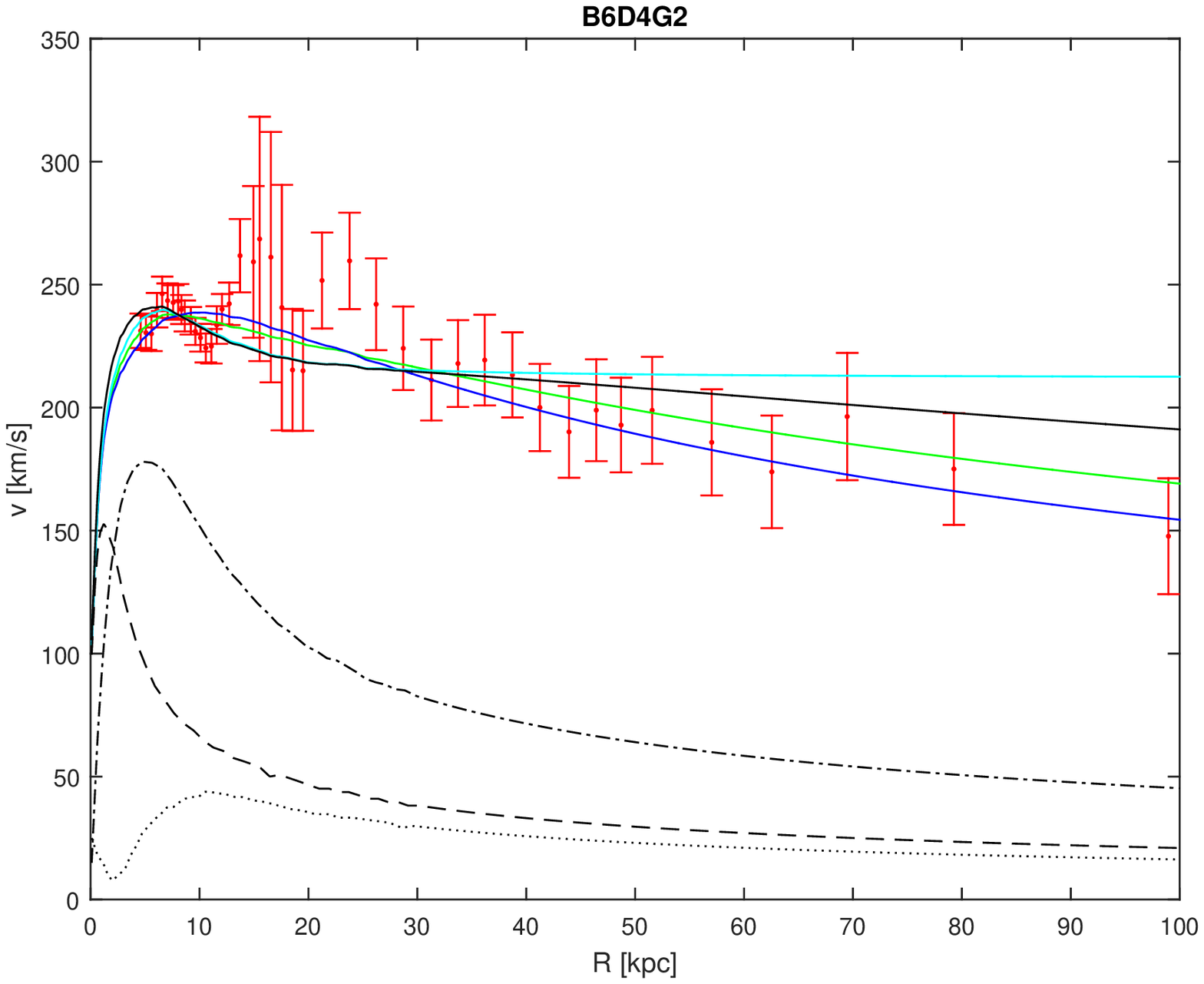}
\addtocounter{figure}{-1}
\caption{--continued}
\end{figure}

\begin{figure}
\centering
\includegraphics[width=0.4\textwidth]{B7D1G1.eps}\hspace{0.5cm}
\includegraphics[width=0.4\textwidth]{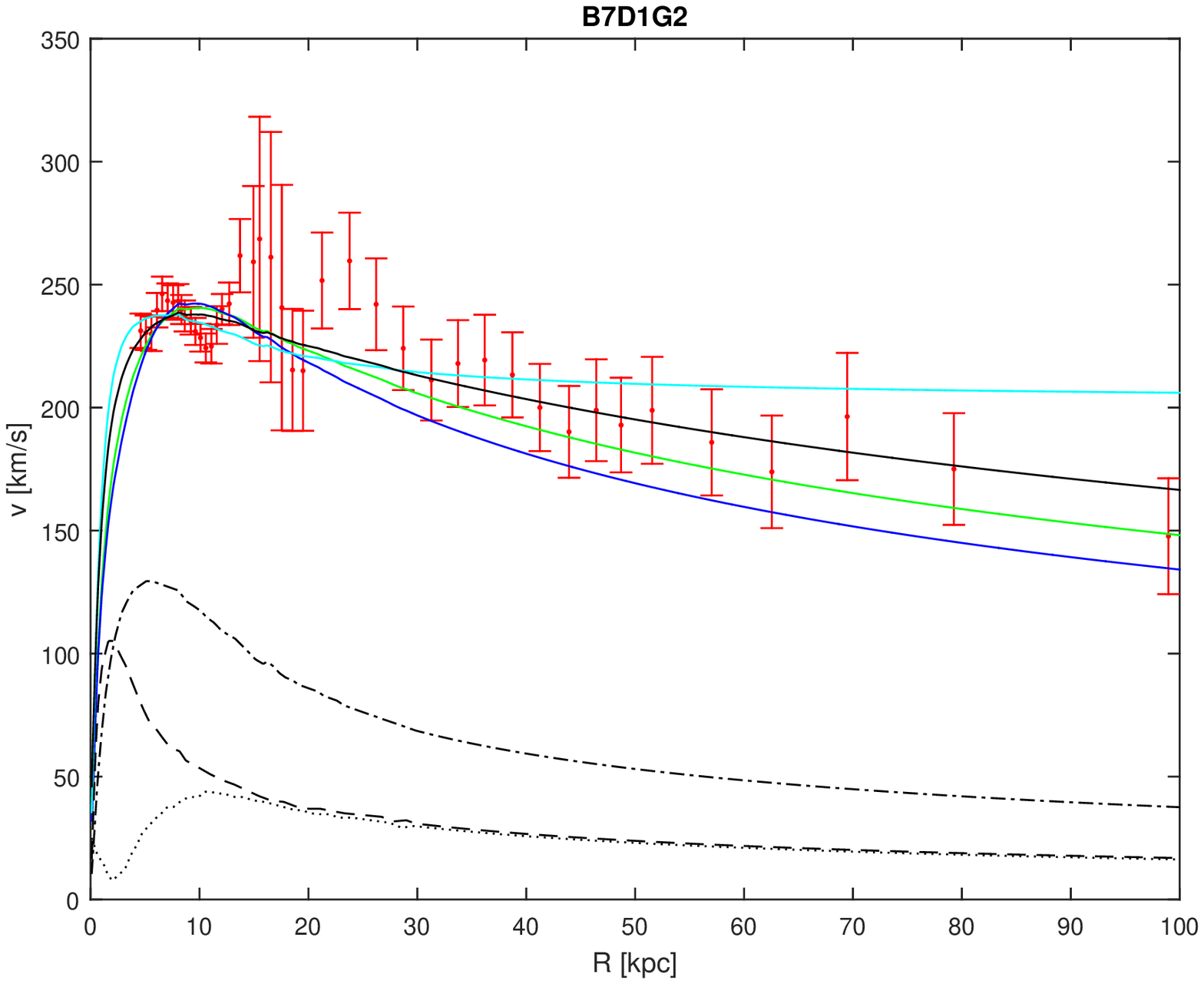}
\includegraphics[width=0.4\textwidth]{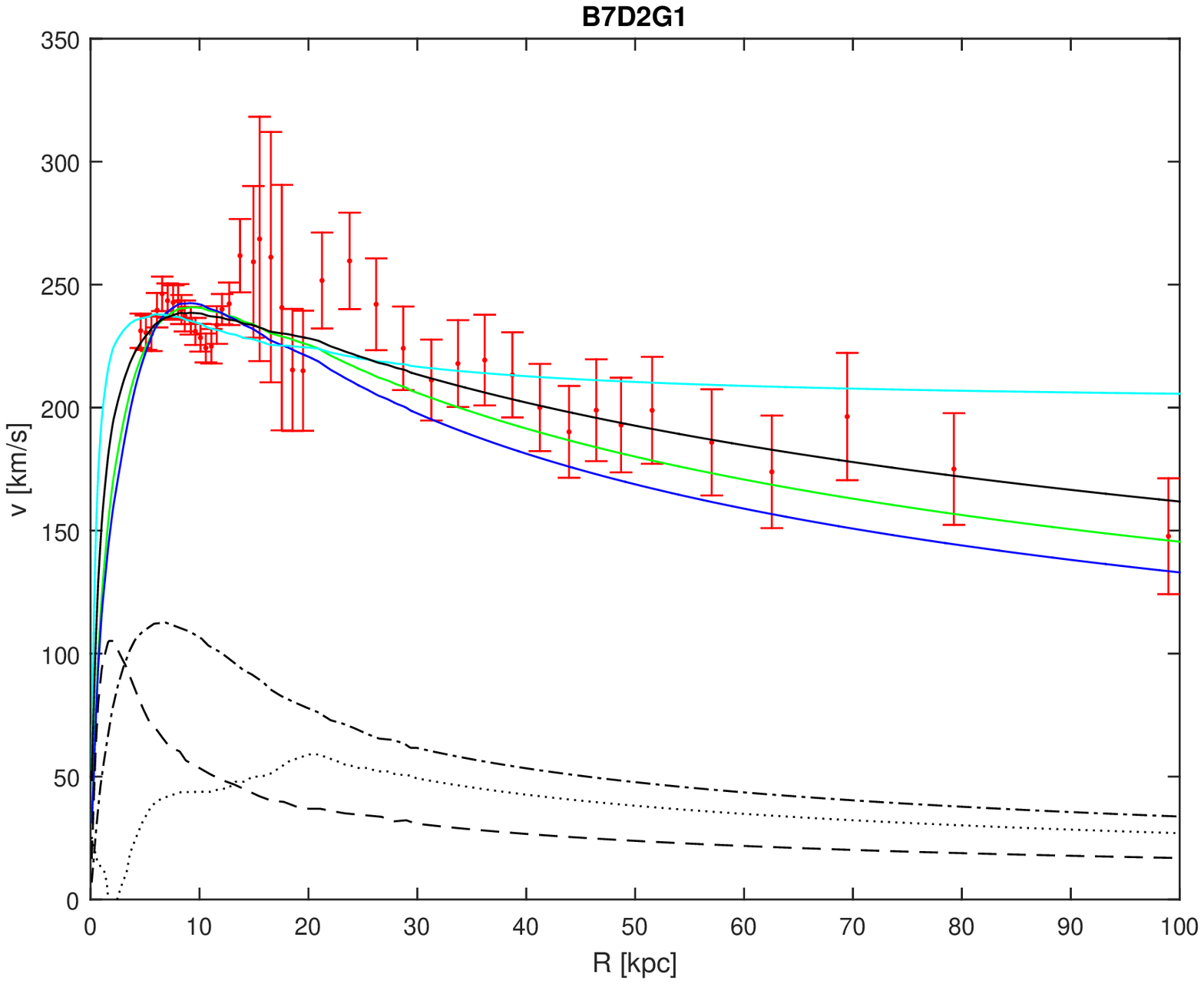}\hspace{0.5cm}
\includegraphics[width=0.4\textwidth]{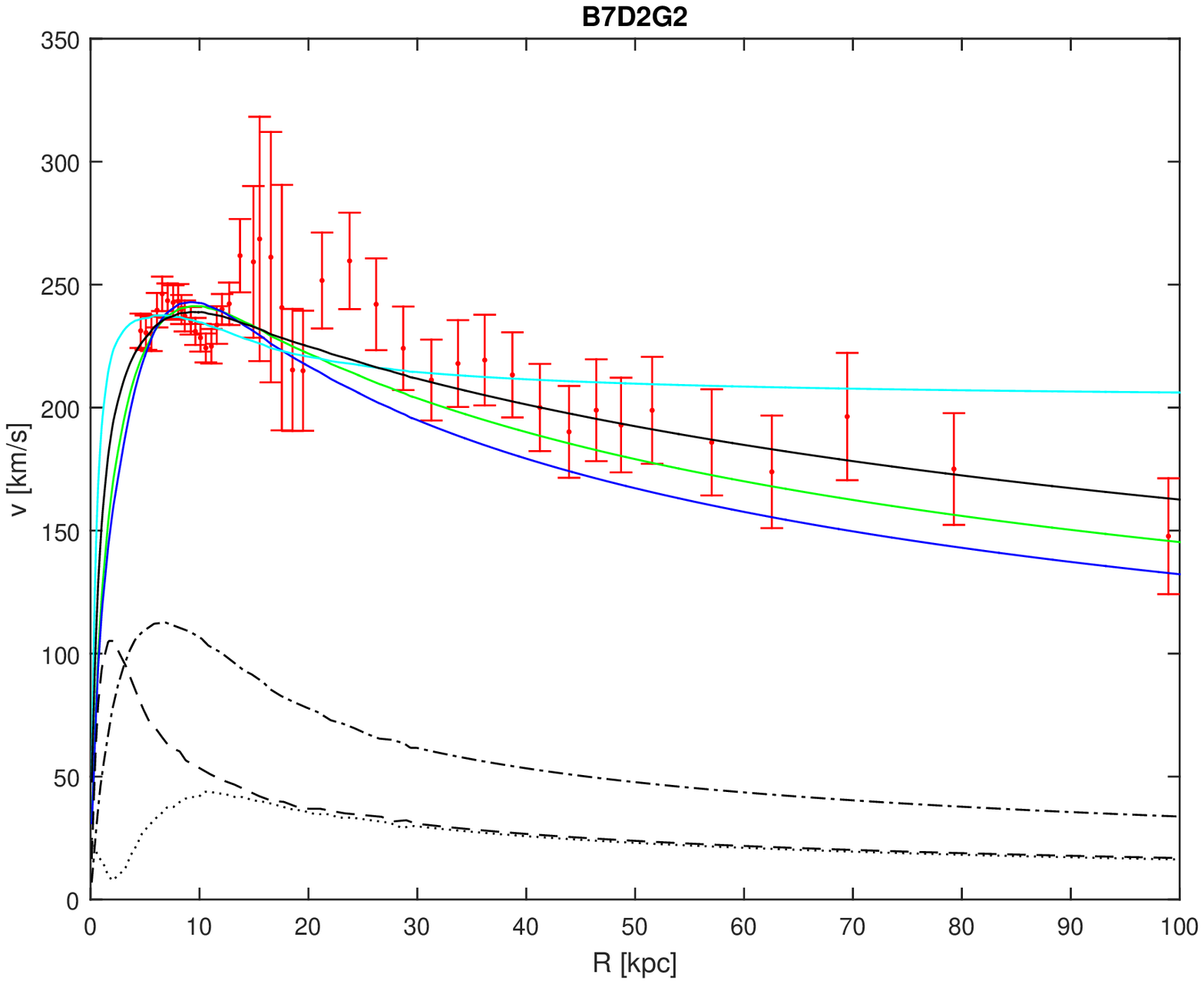}
\includegraphics[width=0.4\textwidth]{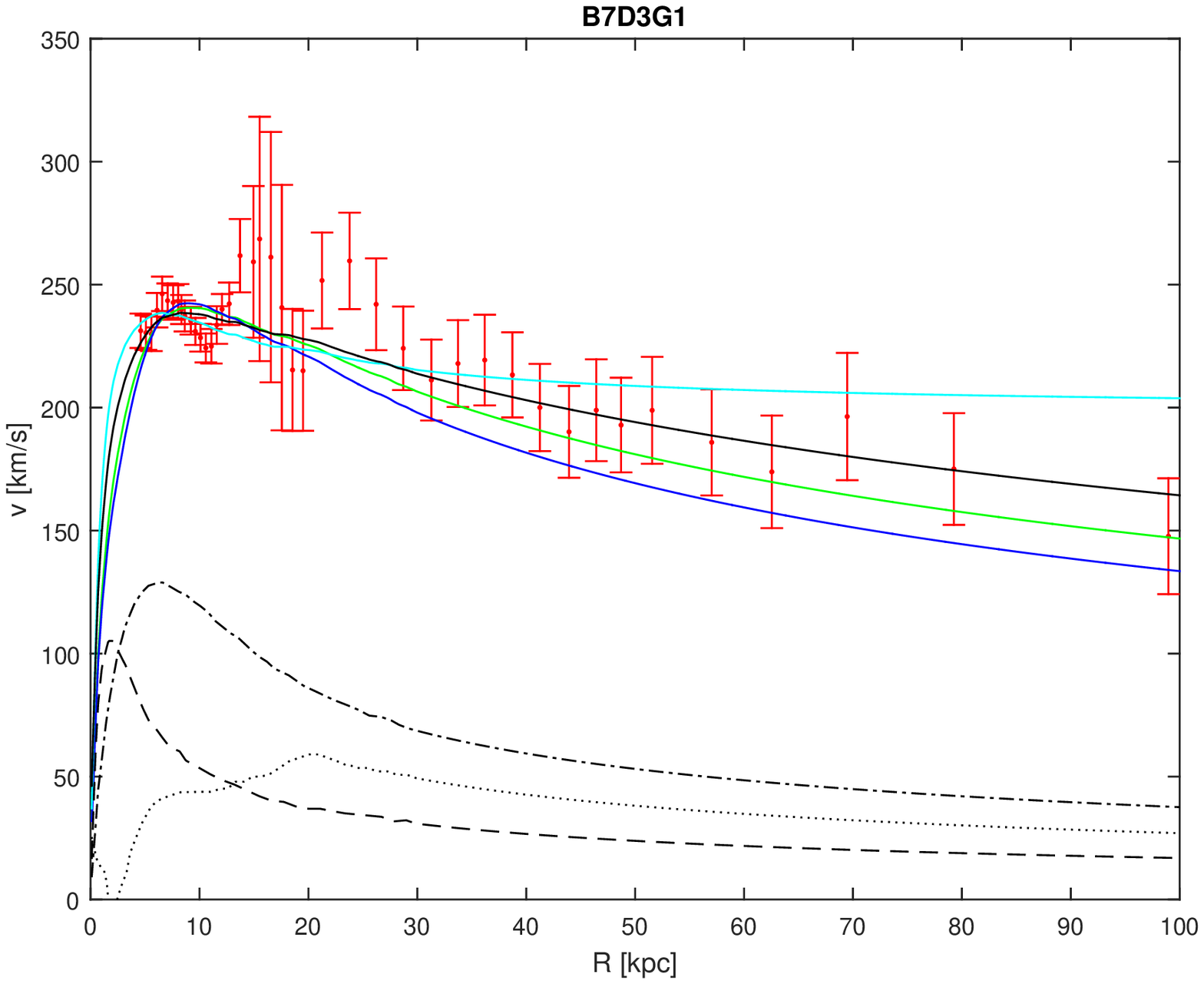}\hspace{0.5cm}
\includegraphics[width=0.4\textwidth]{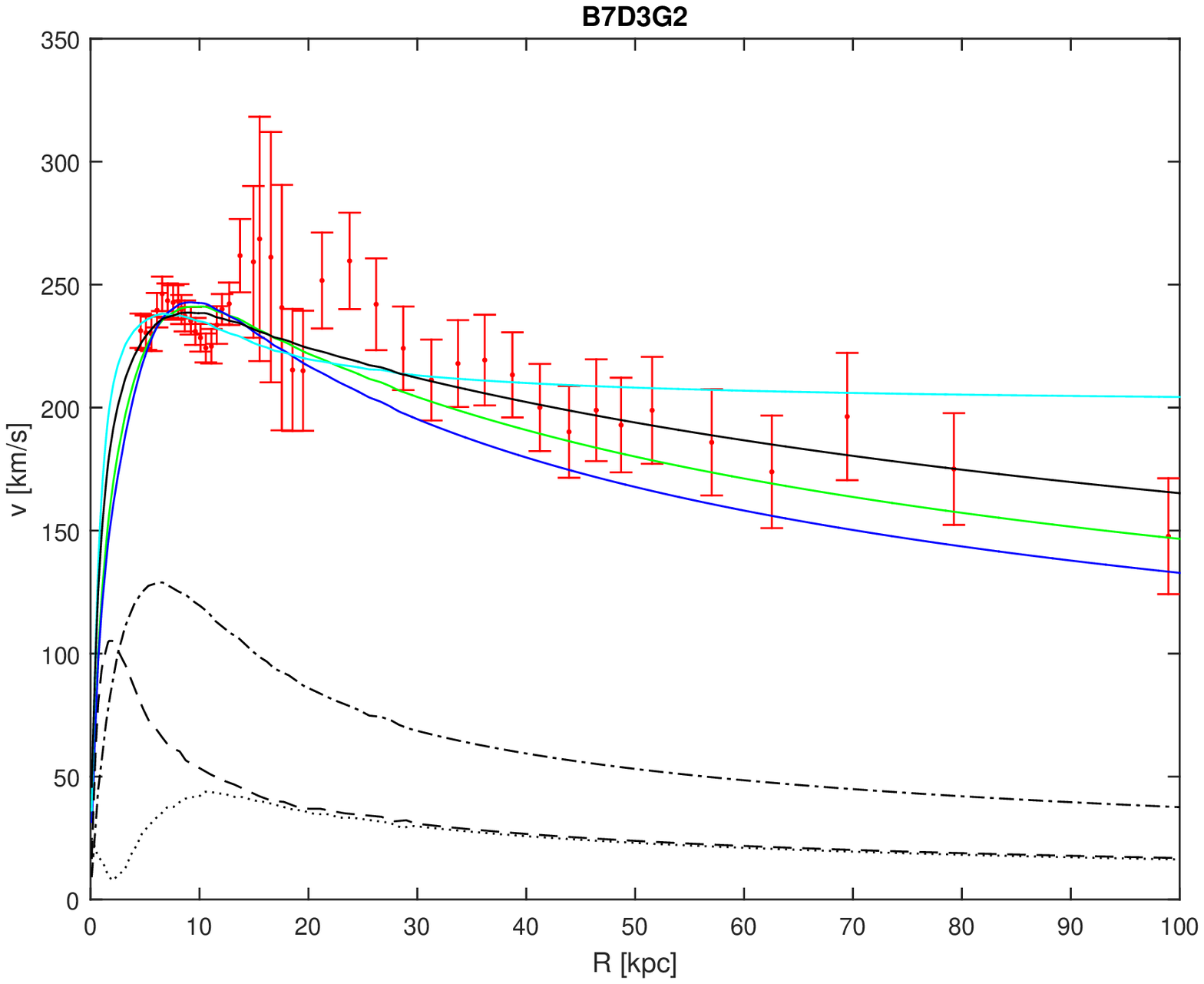}
\includegraphics[width=0.4\textwidth]{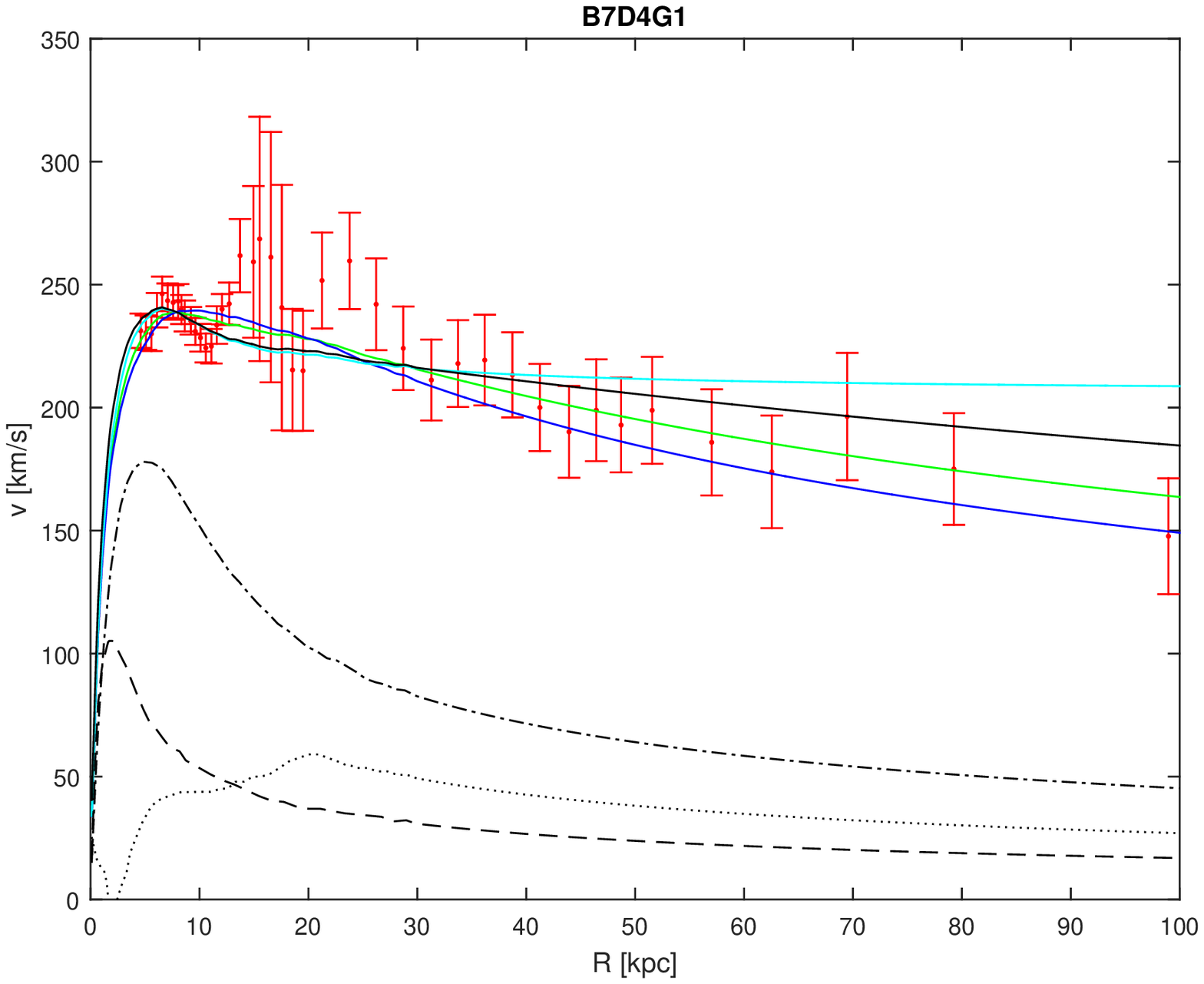}\hspace{0.5cm}
\includegraphics[width=0.4\textwidth]{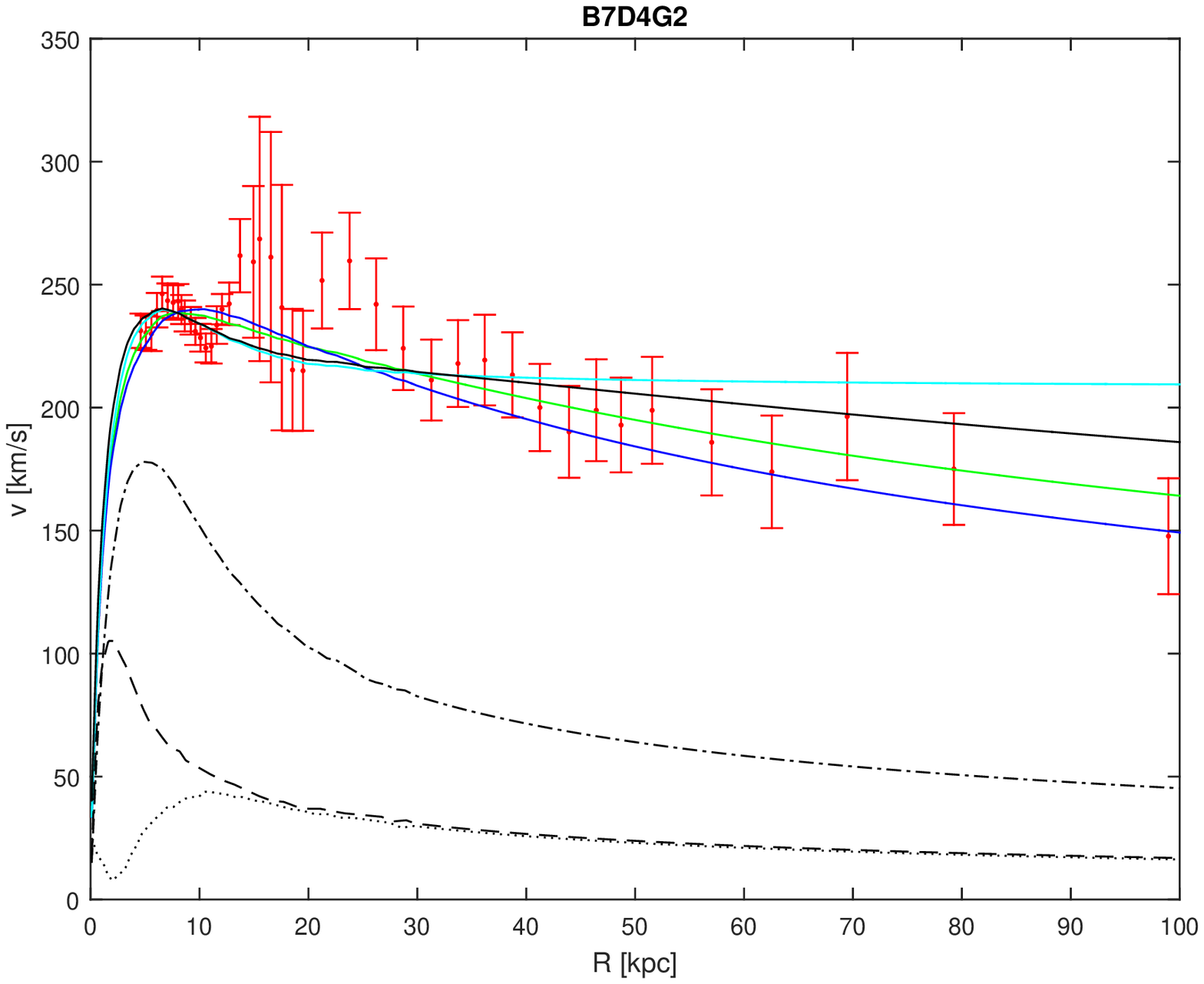}
\addtocounter{figure}{-1}
\caption{--continued}
\end{figure}